\documentclass[11pt,a4paper]{article}
\pdfoutput=1
\usepackage{jheppub}        % official JHEP class/package (loads amsmath, hyperref, graphicx, ...)

\usepackage{amssymb}
\usepackage{bm}
\usepackage{slashed}
\usepackage{braket}

\usepackage{tikz}
\usepackage{tikz-feynman}
\usetikzlibrary{positioning}

\newcommand{\cB}{\mathcal{B}}
\newcommand{\Rep}[2]{(\mathbf{#1},\mathbf{#2})}

\newcommand{\PiS}{\Pi_{S}}
\newcommand{\PiV}{\Pi_{V}}
\newcommand{\SU}[1]{\mathrm{SU}(#1)}

\usepackage{ulem}
\usepackage{color}
\definecolor{ar}{rgb}{1.0, 0.01, 0.24}
\definecolor{al}{rgb}{0.82, 0.1, 0.26}
\definecolor{ev}{rgb}{0.56, 0.0, 1.0}

\newcommand{\blueflag}[1]{{\color{blue} #1}}

\title{A quark-diquark model for parity doublet structure of baryons
}

\author[a]{Bikai Gao}
\author[b,c,d]{Masayasu Harada}

\affiliation[a]{Research Center for Nuclear Physics (RCNP), Osaka University, Osaka 567-0047, Japan}
\affiliation[b]{Kobayashi-Maskawa Institute for the Origin of Particles and the Universe, Nagoya University, Nagoya,
464-8602, Japan}
\affiliation[c]{Department of Physics, Nagoya University, Nagoya, 464-8602, Japan}
\affiliation[d]{Advanced Science Research Center, Japan Atomic Energy Agency, Tokai 319-1195, Japan}

\emailAdd{bikai@rcnp.osaka-u.ac.jp}
\emailAdd{harada@hken.phys.nagoya-u.ac.jp}

\abstract{%
The chiral invariant mass of baryons is a phenomenological input of parity doublet models, and its microscopic origin remains an open question. We
propose that the chiral invariant mass and the parity doublet structure originate from the diquarks: 
the scalar ($0^+$) and
pseudoscalar ($0^-$) diquarks form a parity doublet whose invariant mass is
generated by gluon dynamics rather than by the quark condensate. We construct a three-flavor chiral quark--diquark model in which  a quark and a diquark 
are bounded into a baryon through a  chiral-invariant 
four-body interaction  whose structure is reduced  from one-gluon exchange. 
It is shown that the quark--diquark structure automatically yields the two
chiral representations and the mirror assignment of the parity doublet model,
and the composite baryons acquire chiral invariant masses even for massless quarks. 
We find that the octet baryon spectrum and the nucleon sigma
terms are reproduced very well 
with a minimal set of parameters.
Furthermore, after chiral symmetry restoration, the model predicts a distinctive inverted mass hierarchy: the nucleon remains relatively heavy, whereas the $\Sigma$ and $\Xi$ baryon become lighter than the nucleon. This inverse mass ordering may therefore provide a novel, experimentally testable signature towards chiral symmetry restoration.
}


\begin{document}
\maketitle
\flushbottom

% ----------------------------------------------------------------------------

% ============================================================================
\section{Introduction}
\label{sec:intro}
% ============================================================================

The origin of the nucleon mass is one of the most fundamental open problems in the strong-interaction sector of the Standard Model. The Higgs mechanism fixes the current masses of the light quarks, but these account for only a few percent of the nucleon mass; the overwhelming remainder is generated dynamically by the nonperturbative dynamics of quantum chromodynamics
(QCD).
In the conventional picture this dynamics is governed by the spontaneous breaking of chiral symmetry: a nonvanishing quark
condensate $\langle\bar q q\rangle$ dresses the nearly massless current quarks into massive constituent quarks and sets the scale of the low-lying hadron
spectrum~\cite{Manohar:1983md,Kugo:1991da,Klevansky:1992qe,Roberts:1994dr,Hatsuda:1994pi,Miransky:1994vk}.
Taken literally, this mechanism ties essentially
the entire nucleon mass to the chiral order parameter, so that the mass is
expected to become almost zero as chiral symmetry is restored at high temperature or baryon
density.

The parity doublet model (PDM) offers a physically distinct
viewpoint~\cite{DeTar:1988kn,Jido:1999hd,Jido:2001nt}. In this framework, a baryon and its chiral partner are embedded in a single
chiral multiplet, which naturally accommodates a chiral invariant mass 
$m_0$ that does not originate from the quark condensate. As a consequence, only
part of the baryon mass is generated by the chiral condensate: the  chiral  invariant mass $m_0$ persists even in the chiral restored phase, where the two chiral
partners become degenerate. The existence of chiral invariant masss is supported  by the lattice QCD calculations at finite
temperature~\cite{Aarts:2015mma,Aarts:2017rrl,Aarts:2018glk} along with the sum rule calculations~\cite{Kim:2020zae,Kim:2021xyp,Lee:2023ofg} and the model has
furthermore been successfully applied to the studies of nuclear matter and neutron stars~\cite{Hatsuda:1988mv,Zschiesche:2006zj,Dexheimer:2007tn,Sasaki:2010bp,Mukherjee:2016nhb,Suenaga:2017wbb,Marczenko:2017huu,Marczenko:2018jui,Eser:2023oii,Brodie:2025nww,Negreiros:2026ode,Yamazaki:2019tuo,Minamikawa:2020jfj,Gao:2024chh,Kong:2025dwl,Gao:2025nkg,Gao:2025vdc}.

Recent observations from neutron star provide strong constraints and a relatively large value of $m_0$ is favored~\cite{Yamazaki:2019tuo,Minamikawa:2020jfj,Gao:2024chh,Kong:2025dwl,Gao:2025nkg,Gao:2025vdc}. 
More recently, the framework has been extended from two to
three flavors, placing the octet baryons and their parity partners on an equal
footing~\cite{Chen:2009sf,Chen:2010ba,Nishihara:2015fka,Sasaki:2017glk,Minamikawa:2023ypn,Gao:2024mew,Gao:2025eax,Gao:2026scv}.
In all of these realizations of the parity doublet model, the analysis is
carried out from a purely hadronic perspective, and the chiral invariant mass
$m_0$ enters as a phenomenological parameter: it is fixed by fitting the baryon spectrum or nuclear and neutron-star data, rather than being derived from more fundamental degrees of freedom. Which component of the QCD dynamics generates a baryon mass that is insensitive to the chiral condensate thus remains an open question, and clarifying the microscopic origin of the chiral invariant mass is one of the most important problems in this
context.

A natural microscopic perspective on this question is provided by the diquark
correlations inside the baryon. Lattice QCD, Dyson--Schwinger studies, and
phenomenology consistently support a picture in which two quarks form a
compact color-antitriplet diquark correlation that subsequently combines with
the remaining quark to form a color-singlet baryon, with the scalar ``good''
diquark ($J^P=0^+$) being the most strongly correlated
channel~\cite{Anselmino:1992vg,Maris:2003vk,Barabanov:2020jvn}. Since the
diquark itself is a colored object, it should not be regarded as an
asymptotic hadron, but rather as a confined substructure inside the baryon. 
Its binding and mass scale are therefore generated by the non-Abelian gluonic dynamics that is responsible for confinement in the full
color-singlet baryon. In this sense, a substantial part of the diquark mass
can be attributed to chromodynamic interactions, rather than to the quark
condensate associated with spontaneous chiral symmetry breaking. 
In addition, the gluon-mediated interaction that
binds the diquark and the remaining quark into the baryon is of chromodynamic origin and insensitive to the chiral condensate. 
This observation suggests that the chiral invariant  
interaction together with the chiral invariant mass of the diquark, both of which survive even  
when chiral symmetry is restored, generates the chiral invariant mass of baryons. The diquark sector, moreover, carries a parity doubling structure
of its own~\cite{Harada:2019udr}:
alongside the scalar ($0^+$) diquark there exists a pseudoscalar
($0^-$) diquark, and the two transform into one another under chiral
$\SU{3}_R\times\SU{3}_L$ and form  a diquark
parity doublet.
Furthermore, two diquarks are degenerated with each other to have the chiral invariant mass when the chiral symmetry is restored.
Coupling the remaining quark to this diquark doublet transfers
the doubling structure to the baryons and naturally realizes the mirror
assignment of the parity doublet model. 
In this picture, both characteristic ingredients of the parity doublet
model---the chiral invariant baryon mass and the parity doubling of
baryons---emerge from the same underlying diquark dynamics driven
predominantly by gluonic interactions.

Motivated by these considerations, we construct a three-flavor chiral
quark--diquark model to describe the baryons and their excited states
starting from the underlying quark and diquark degrees of freedom, in the
spirit of NJL-type quark--diquark descriptions of
baryons~\cite{Buck:1992wz,Ishii:1995bu}. The scalar ($0^+$) and pseudoscalar
($0^-$) diquarks are organized into a parity doublet characterized by a
chiral invariant diquark mass, while the quark and diquark are bound into a
baryon through an effective one-gluon-exchange interaction. This interaction
is formulated as a chiral symmetric four-body quark--diquark coupling
together with its crossed and  U(1)$_A$-breaking  counterparts. Employing an auxiliary-field hadronization reformulation, we introduce color-singlet
composite baryon fields and evaluate the one- and two-loop quark--diquark
self-energies, from which the baryon mass matrix---including the mixing
induced by the chiral order parameter $\Sigma$---is derived analytically.
A notable feature of this construction is that the chiral
representations of the baryons are not assigned by hand but dictated by the
quark--diquark structure: pairing 
a right-handed or left-handed 
quark with a diquark of the same
chirality produces composites in $(8,1)\oplus(1,8)$  representation 
under the SU(3)$_R \times$SU(3)$_L$ symmetry, while 
the crossed pairing
produces composites in $(3,\bar3)\oplus(\bar3,3)$  representation, 
and each comes
automatically with its mirror partner. The existence of the mirror
assignment and of the two distinct chiral representations---inputs of the
conventional hadronic parity doublet model---is thus naturally explained,
and all four multiplets are fully mixed by the chiral order parameter.
The resulting spectrum is organized by these two chiral
representations, whose mixing determines both the physical masses and the
chiral invariant mass content of each state. Within this framework, the
baryon chiral invariant masses
are no longer
phenomenological input parameters of a hadronic Lagrangian but emerge dynamically from the chiral invariant diquark mass in
combination with the four-body quark--diquark interaction: as we will  show explicitly, the composite baryon acquires a finite mass from the
quark--diquark loop dynamics even for massless quarks, with both the diquark
mass and the interaction strength setting its scale.
When chiral symmetry is restored, the constituent quark mass
generated by spontaneous chiral symmetry breaking vanishes,
and the remaining baryon mass is inherited from the chiral
invariant diquark sector ---the invariant diquark mass and the
four-body quark--diquark interaction---
thereby providing a microscopic realization of the parity-doublet mass
parameter.

We then confront the model with experimental data.
Scanning the two chiral invariant masses and fitting the remaining seven couplings, we find that the model reproduces the
masses of the octet ground states and their excited partners as well as the
nucleon sigma terms $\sigma_{\pi N}$ and $\sigma_{sN}$ remarkably well, with
the correct ordering $m_N<m_\Sigma<m_\Xi$ in both parity sectors---an economy
of description that was not achieved in the purely hadronic three-flavor
parity doublet models to the best of our knowledge. 

The remainder of the paper is organized as follows. In Sec.~\ref{sec:fields} we fix our conventions for the chiral and discrete transformations of the quark, meson, and diquark fields.
Section~\ref{sec:Leff} constructs the chiral invariant four-body quark--diquark interaction, defines the color-singlet composites together with their mirror and crossed partners, and recasts the interaction through the auxiliary-field method. 
In Sec.~\ref{sec:mass} we evaluate the one-loop quark--diquark 
ontribution to the 
two-point function  of the composite baryon. 
We derive the chiral invariant mass of baryons  even  in the absence of the chiral order parameter, exhibiting explicitly how the baryon mass is generated in the model. In Sec.~\ref{sec:sigma} we couple the quarks and diquarks to the meson field which induces the mixing among the composite baryons  
We then assemble  the $N$, $\Sigma$, and $\Xi$ mass matrices, including  the
explicit chiral symmetry breaking. Section~\ref{sec:numerics} presents the numerical analysis: fitting 
to the octet spectrum and the nucleon sigma terms; determination of chiral-invariant mass components in eigenstates and parity-doubling patterns under the restoration of chiral symmetry. 
We summarize and discuss our results in Sec.~\ref{sec:summary}. Technical details
regarding the proof of discrete symmetries
and of the loop integrals are collected in Appendices~\ref{app:PC} and~\ref{app:2pt}.

% ============================================================================
\section{Fields, chiral representations, and transformations}
\label{sec:fields}
% ============================================================================

The model is built on the chiral symmetry group $G=\SU{3}_R\times\SU{3}_L$
acting on three light quark flavors. We use the chiral projectors
$P_{R,L}=\tfrac12(1\pm\gamma_5)$ and write $q_{R,L}=P_{R,L}\,q$ for the
right- and left-handed quark fields. The dynamical degrees of freedom are the quarks,  color-antitriplet scalar ($0^+$) and pseudoscalar ($0^-$)
diquarks,  and treat the  color-singlet meson field as an external field. 
In this section we fix our conventions for the chiral and discrete-symmetry transformations of these fields.
Throughout, the first
entry of a representation label $\Rep{R}{L}$ refers to $\SU{3}_R$ and the
second to $\SU{3}_L$.

\subsection{Quarks}

The right- and left-handed quarks carry a color index $a$ (fundamental
$\mathbf 3$) and flavor indices $\alpha$ and $i$, and belong to the following representations: 
\begin{align}
q_{R}^{a,\blueflag{\alpha}}&\sim\Rep{3}{1}\,, &
q_{L}^{a,i}&\sim\Rep{1}{3}\,, &
\bar{q}_{R,a, \alpha}&\sim\Rep{\bar3}{1}\,, &
\bar{q}_{L,a,i}&\sim\Rep{1}{\bar3}\,,
\end{align}
which transform as 
\begin{align}
q_{R}^{a,\alpha}\to (U_R)^{\alpha}{}_ \beta \, q_{R}^{a,\beta}, &\qquad
q_{L}^{a,i}\to (U_L)^{i}{}_{j}\, q_{L}^{a,j}\,, \\
\bar q_{R,a,\alpha}\to \bar q_{R,a,\beta}\,(U_R^\dagger)^{\beta}{}_{\alpha}, &\qquad
\bar q_{L,a,i}\to \bar q_{L,a,j}\,(U_L^\dagger)^{j}{}_{i}\,,
\end{align}
with $U_{R,L} \in \mbox{SU(3)}_{R,L}$.
Here and throughout we use indices $(\alpha,\beta,\dots)$ for $\SU{3}_R$ and
$(i,j,\dots)$ for $\SU{3}_L$. 
Under  the parity transfomation, the chirality of the quark  
is flipped,
\begin{equation}
P:\quad q_{R}^{a}(t,\vec x)\to\gamma^0 q_{L}^{a}(t,-\vec x)\,.
\end{equation}

\subsection{Meson field}

The chiral order parameter is  expressed by 
the complex $3\times3$ matrix as 
\begin{equation}
\Sigma^{i}{}_{\alpha}=\sigma^{i}{}_{\alpha}+i\pi^{i}{}_{\alpha}\sim\Rep{\bar3}{3}\,,
\qquad
\Sigma\to U_L\,\Sigma\,U_R^\dagger\,,
\qquad
P:\ \Sigma\to\Sigma^\dagger\,,
\end{equation}
whose scalar ($\sigma^{i}{}_{\alpha}$) and pseudoscalar ($\pi^{i}{}_{\alpha}$) components describe
the scalar and pseudoscalar meson nonets.
Spontaneous chiral symmetry breaking is encoded
in the vacuum expectation value $\langle\Sigma^{i}{}_{\alpha}\rangle=f\,\delta^{i}{}_{\alpha}$ with
$f\simeq f_\pi$,
where $f_\pi$ is the pion decay constant; 
once $\SU{3}$ flavor breaking is included this generalizes to
$\langle\Sigma\rangle=\mathrm{diag}(f_\pi,f_\pi,f_s)$ with $f_s=2 f_K - f_\pi$,
where $f_K$ is the kaon decay constant. 

\subsection{Scalar and pseudoscalar diquarks}

The spin-$0$, color-antitriplet diquarks are expressed by
the local quark bilinears as 
\begin{equation}
(d_{R})_{a,\alpha}=\epsilon_{abc}\,\epsilon_{\alpha \beta \gamma}\,
\big[(q_{R}^{T})^{b,\beta}\,C\,(q_{R})^{c,\gamma}\big]\,,
\qquad
(d_{L})_{a,i}=\epsilon_{abc}\,\epsilon_{ijk}\,
\big[(q_{L}^{T})^{b,j}\,C\,(q_{L})^{c,k}\big]\,,
\end{equation}
with the charge-conjugation matrix $C=i\gamma^0\gamma^2$. The antisymmetric
color ($\epsilon_{abc}$) and flavor ($\epsilon_{\alpha \beta \gamma}, \epsilon_{ijk}$) contractions project
onto the color and flavor antitriplets, so that the diquarks form chiral
antitriplets,
\begin{align}
(d_{R})_{a,\alpha}\sim\Rep{\bar3}{1}, &\qquad
(d_{L})_{a,i}\sim\Rep{1}{\bar3},
\end{align}
transforming as
\begin{align}
(d_{R})_{a,\alpha}\to (d_{R})_{a,\beta}\,(U_R^\dagger)_{\alpha}^{\beta}, &\qquad
(d_{L})_{a,i}\to (d_{L})_{a,j}\,(U_L^\dagger)_{i}^{j}.
\end{align}
Under the parity transformation,
two chiralities are interchanged with flipping the sign as 
\begin{equation}
P:\quad d_{R}\leftrightarrow - d_{L},
\end{equation}
which follows directly from the quark parity transformation together with
$\gamma^0 C\gamma^0=-C$.% (The explicit derivation can be relegated to an appendix.)
\ The parity eigenstates are the scalar ($0^+$) and pseudoscalar ($0^-$)
combinations given by 
\begin{align}
S_{a,i}=\tfrac{1}{\sqrt2}\big[(d_{R})_{a,i}-(d_{L})_{a,i}\big]\quad(0^+),
\qquad
P_{a,i}=\tfrac{1}{\sqrt2}\big[(d_{R})_{a,i}+(d_{L})_{a,i}\big]\quad(0^-),
\end{align}
or equivalently
$(d_{R})_{a,i}=\tfrac{1}{\sqrt2}(P_{a,i}+S_{a,i})$ and $(d_{L})_{a,i}=\tfrac{1}{\sqrt2}(P_{a,i}-S_{a,i})$. 
%$(d_{R})_{a,\alpha}=\tfrac{1}{\sqrt2}(P_{a,\alpha}+S_{a,\alpha})$ and $(d_{L})_{a,\alpha}=\tfrac{1}{\sqrt2}(P_{a,\alpha}-S_{a,\alpha})$. 
We summarize the quantum numbers and chiral representations of all the fields in Table~\ref{tab:reps}. The scalar and pseudoscalar
diquarks thus constitute a parity doublet; their coupling to $\Sigma$,
introduced in the next stage of the construction, together with
$\langle\Sigma\rangle$ will split their masses.

\begin{table}[t]
  \centering
  \renewcommand{\arraystretch}{1.35}
  \begin{tabular}{lcccc}
    \hline\hline
    Field & Spin & Color & Flavor & Chiral rep.\ \\
    \hline
    $q_{R}^{a,\alpha}$          & $\tfrac12$ & $\mathbf{3}$       & $\mathbf{3}$      & $\Rep{3}{1}$    \\
    $q_{L}^{a,i}$               & $\tfrac12$ & $\mathbf{3}$       & $\mathbf{3}$      & $\Rep{1}{3}$    \\
    $\Sigma^{i}{}_{\alpha}$     & $0$        & $\mathbf{1}$       & $\bar{\mathbf{3}}\otimes\mathbf{3}$ & $\Rep{\bar3}{3}$ \\
    $(d_{R})_{a,\alpha}$        & $0$        & $\bar{\mathbf{3}}$ & $\bar{\mathbf{3}}$ & $\Rep{\bar3}{1}$ \\
    $(d_{L})_{a,i}$             & $0$        & $\bar{\mathbf{3}}$ & $\bar{\mathbf{3}}$ & $\Rep{1}{\bar3}$ \\
    \hline\hline
  \end{tabular}
  \caption{Quantum numbers and chiral representations of the quark, meson, and diquark fields.}
  \label{tab:reps}
\end{table}

% ============================================================================
\section{Effective Lagrangian of the quark-diquark model without the $\Sigma$ field
}
\label{sec:Leff}
% ============================================================================

It is instructive to first analyze the quark--diquark dynamics in the absence
of the chiral order parameter $\Sigma$; the coupling to $\Sigma$ is introduced in the next section. 
In this section we write down a chiral symmetric four-body
quark--diquark interaction 
generated by one-gluon exchenge, and 
define the color-singlet quark--diquark (baryon) composites together with their mirror
and crossed partners.

The Lagrangian of the quark-diquark model without meson field $\Sigma$ is expressed as  
\begin{align}
\mathcal{L} =& \;\bar q\,i\slashed\partial\,q
+\partial_\mu d_{R,\alpha}(\partial^\mu d_{R,\alpha})^\dagger
+\partial_\mu d_{L,i}(\partial^\mu d_{L,i})^\dagger
-m_d^2\!\left(d_{R,\alpha}d_{R,\alpha}^\dagger+d_{L,i}d_{L,i}^\dagger\right)
+{\mathcal L}_{4},
\label{eq:La1}
\end{align}
where $m_d$ is the  chiral invariant diquark mass and ${\mathcal L}_4$
collects all the four-body interactions discussed below.

\subsection{Four-body quark--diquark interactions  }
\label{sec:4body}

The chiral invariant four-body interaction, which pairs a quark and a diquark of the  same chirality is given by
\begin{equation}
\begin{aligned}
{\mathcal L}_{qd}
= & -G_{qd}\Bigg[
\left\{ \left( \bar{q}_L \right)_{c,i} \left( d_L^\dagger \right)^{c,j} \right\}
\left\{ \left( i \partial_\mu d_L \right)_{a,j} \gamma^\mu \, \left( q_L \right)^{a,i} \right\}
\\
&\quad\quad\quad+\left\{ \left( \bar{q}_L \right)_{c,i} \gamma^\mu \left( - i \partial_\mu d_L^\dagger \right)^{c,j} \right\}
\left\{ \left( d_L \right)_{a,j} \left( q_L \right)^{a,i} \right\}
\\
&\quad \qquad +
\left\{ \left( \bar{q}_R \right)_{c,\alpha} \left( d_R^\dagger \right)^{c,\beta} \right\}
\left\{ \left( i \partial_\mu d_R \right)_{a,\beta} \gamma^\mu \, \left( q_R \right)^{a,\alpha} \right\}
\\
& \qquad\quad+
\left\{ \left( \bar{q}_R \right)_{c,\alpha} \gamma^\mu \left( - i \partial_\mu d_R^\dagger \right)^{c,\beta} \right\}
\left\{ \left( d_R \right)_{a,\beta} \left( q_R \right)^{a,\alpha} \right\}
\Bigg]\ .
\label{eq:Lqd}
\end{aligned}
\end{equation}
We also allow
the crossed interaction
\begin{equation}
\begin{aligned}
{\mathcal L}_{qd}^{(2)}
= & -G_{qd}^{(2)}\Bigg[
\left\{ \left( \bar{q}_L \right)_{c,i} \left( d_R^\dagger \right)^{c,\beta} \right\}
\left\{ \left( i \partial_\mu d_R \right)_{a,\beta} \gamma^\mu \, \left( q_L \right)^{a,i} \right\}
\\&\qquad \quad+
\left\{ \left( \bar{q}_L \right)_{c,i} \gamma^\mu \left( - i \partial_\mu d_R^\dagger \right)^{c,\beta} \right\}
\left\{ \left( d_R \right)_{a,\beta} \left( q_L \right)^{a,i} \right\}
\\
& \qquad \quad +
\left\{ \left( \bar{q}_R \right)_{c,\alpha} \left( d_L^\dagger \right)^{c,j} \right\}
\left\{ \left( i \partial_\mu d_L \right)_{a,j} \gamma^\mu \, \left( q_R \right)^{a,\alpha} \right\}
\\& \qquad\quad+
\left\{ \left( \bar{q}_R \right)_{c,\alpha} \gamma^\mu \left( - i \partial_\mu d_L^\dagger \right)^{c,j} \right\}
\left\{ \left( d_L \right)_{a,j} \left( q_R \right)^{a,\alpha} \right\}
\Bigg],
\label{eq:Lqd2}
\end{aligned}
\end{equation}
which pairs a quark and a diquark of opposite chirality.

\begin{figure}[htbp]
\centering
% ---------- (a) one-gluon exchange ----------
\begin{tikzpicture}[baseline=(current bounding box.center)]
\begin{feynman}
    \vertex (qi) at (-1.9, 1.1) {\( q \)};
    \vertex (qo) at ( 1.9, 1.1) {\( q \)};
    \vertex (di) at (-1.9,-1.1) {\( d \)};
    \vertex (do) at ( 1.9,-1.1) {\( d \)};
    \vertex [dot] (v1) at (0, 1.1) {};
    \vertex [dot] (v2) at (0,-1.1) {};
    \diagram*{
        (qi) -- [fermion] (v1) -- [fermion] (qo),
        (di) -- [charged scalar] (v2) -- [charged scalar] (do),
        (v1) -- [gluon, edge label'=\( g \)] (v2),
    };
    \node at (0,-1.9) {(a)};
\end{feynman}
\end{tikzpicture}
\hspace{0.8cm}\( \Longrightarrow \)\hspace{0.8cm}
% ---------- (b) contact four-body vertex ----------
\begin{tikzpicture}[baseline=(current bounding box.center)]
\begin{feynman}
    \vertex (qi) at (-1.9, 1.1) {\( q \)};
    \vertex (qo) at ( 1.9, 1.1) {\( q \)};
    \vertex (di) at (-1.9,-1.1) {\( d \)};
    \vertex (do) at ( 1.9,-1.1) {\( d \)};
    \vertex [square dot] (v) at (0,0) {};
    \diagram*{
        (qi) -- [fermion] (v) -- [fermion] (qo),
        (di) -- [charged scalar] (v) -- [charged scalar] (do),
    };
    \node [right=4pt of v] {\( G_{qd},\ G_{qd}^{(2)} \)};
    \node at (0,-1.9) {(b)};
\end{feynman}
\end{tikzpicture}
\caption{Microscopic origin of the four-body quark--diquark interactions.
(a) One-gluon exchange between the quark (solid line) and the spin-$0$,
color-antitriplet diquark (dashed line). The upper vertex is the
chirality-conserving quark--gluon coupling $\bar q\,\gamma^\mu T^A q$, and
the lower vertex is the scalar-QCD diquark--gluon coupling
$i\,d^\dagger \bar{T}^A\overleftrightarrow{\partial^\mu} d$, 
which supplies the
single derivative acting on the diquark. (b) In the contact approximation
for the gluon propagator, a color Fierz rearrangement into the attractive
color-singlet quark--diquark channel reduces (a) to the four-body vertices
of Eqs.~\eqref{eq:Lqd} and \eqref{eq:Lqd2} (filled square), which inherit
the $\gamma^\mu$ structure and the derivative from the two currents in
(a).}
\label{fig:OGE}
\end{figure}
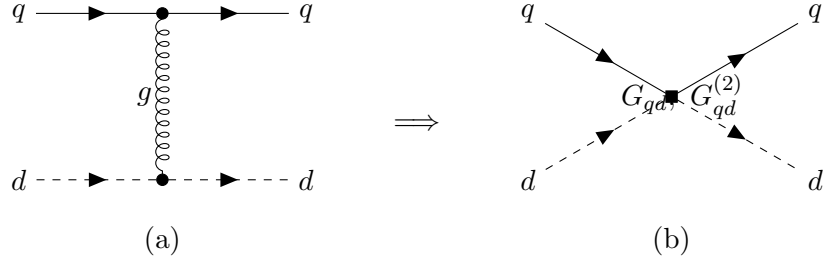

We woule like to stress that both 
interactions may be motivated by one-gluon exchange between the quark
and the diquark, as illustrated in Fig.~\ref{fig:OGE}\,(a). The gluon
couples to the quark through the color current
$\bar q\,\gamma^\mu T^A q$, which conserves the quark chirality, and to the
spin-$0$ color-antitriplet diquark through the scalar-QCD current
\begin{align*}
i\,d^\dagger \bar T^A \overleftrightarrow{\partial^\mu} d =
i d^\dagger \bar T^A \left( \partial^\mu d \right) - 
i \left( \partial^\mu d^\dagger \right) \bar T^A d 
,
\end{align*}
where $\bar T^A=-(T^A)^*$ is the generator in the
$\bar{\mathbf 3}$ representation. The diquark--gluon vertex therefore
contains a single derivative acting on the diquark line. In the contact
approximation for the gluon propagator, as is standard in NJL-type models,
the exchange reduces to the current--current interaction
\begin{align}
{\mathcal L}_{\rm OGE}
\;\propto\;
-\,g_{\rm eff}^{2}\,
\big(\bar q\,\gamma_\mu T^A q\big)
\big(i\,d^\dagger \bar T^A
\overleftrightarrow{\partial^\mu} d\big).
\label{eq:OGE}
\end{align}
Projecting this interaction onto the color-singlet quark--diquark channel,
$\mathbf 3\otimes\bar{\mathbf 3}\ni\mathbf 1$, gives the attractive color interaction.
Then, after decomposing the quark and diquark fields into their chiral components 
we obtain the structures in
Eqs.~\eqref{eq:Lqd} and \eqref{eq:Lqd2}: a chirality-conserving quark
current contracted with a single derivative acting on the spin-$0$ diquark,
with the flavor indices connected between the two color-singlet
quark--diquark brackets. At the pointlike one-gluon-exchange level the
gluon is blind to the chiral labels of both the quark and the diquark, so
one naturally expects $G_{qd}=G_{qd}^{(2)}$. In the low-energy effective
theory, however, compositeness effects, higher-dimensional gluonic
operators, and U(1)$_A$-breaking interactions can renormalize the two channels
differently. We therefore treat $G_{qd}$ and $G_{qd}^{(2)}$ as independent
phenomenological couplings.
By imposing hermiticity of the
Lagrangian together with parity and charge-conjugation invariance, one can show
that the couplings $G_{qd}$ and $G_{qd}^{(2)}$ are real (see  Appendix.~\ref{app:PC}).
% (Proof: discrete symmetries of the four-body interaction; to be given in an appendix.)

%%%%%%%%%%%%%%%%%%%%%%%%%%%%%%%
%%%%%%%%%%%%%%%%%%%%%%%%%%%%%%%

In addition to the U(1)$_A$-symmetric interactions \eqref{eq:Lqd} and
\eqref{eq:Lqd2}, we may include a four-body  interaction 
term that breaks the U(1)$_A$
symmetry,
\begin{align}
{\mathcal L}_{A} = G_A
\left[
\left\{ \left( \bar{q}_R \right)_{c,i} \left( d_R^\dagger \right)^{c,i} \right\}
\left\{ \left( d_L \right)_{a,j} \left( q_L \right)^{a,j} \right\}
+
\left\{ \left( \bar{q}_L \right)_{c,i} \left( d_L^\dagger \right)^{c,i} \right\}
\left\{ \left( d_R \right)_{a,j} \left( q_R \right)^{a,j} \right\}
\right]\ ,
\label{eq:LA}
\end{align}
which can be derived from the Kobayashi-Maskawa-'t~Hooft (KMT) type six-quark interaction.

Some comments on the interaction basis are in order. Since the diquarks
carry spin zero, a local four-body interaction in the quark--diquark channel
has the schematic form
\begin{align*}
\{\bar q\,\Gamma_1\,d^\dagger\}
\{d\,\Gamma_2\,q\},
\end{align*}
where $\Gamma_1$ and $\Gamma_2$ denote certain combinations of the gamma matrices, which 
act only on the quark fields. The Dirac structure
therefore reduces to an ordinary quark bilinear
$\bar q\,\Gamma_1\Gamma_2 q$, possibly accompanied by derivatives acting
on the diquark fields. Chirality imposes a useful selection rule. With
$q_{L,R}=P_{L,R}q$ and
$\bar q_{L,R}=\bar q P_{R,L}$, scalar bilinears connect opposite chiral
labels, whereas vector bilinears connect equal chiral labels:
\begin{align}
\bar q_L q_L=0,\qquad
\bar q_R q_R=0,\qquad
\bar q_L\gamma^\mu q_R=0,\qquad
\bar q_R\gamma^\mu q_L=0.
\end{align}
and similarly
\begin{align}
\bar q_L\sigma^{\mu\nu}q_L=0,\qquad
\bar q_R\sigma^{\mu\nu}q_R=0.
\end{align}
Consequently, at zeroth order in derivatives the nonvanishing
quark--diquark structures are chirality-flipping scalar structures. The
$\SU{3}_L\times\SU{3}_R$ invariant single-trace combination of this type is
the U(1)$_A$-breaking interaction in Eq.~\eqref{eq:LA}. In terms of the
underlying quarks it has the schematic six-quark structure
$\bar q_R^{\,3}q_L^{\,3}+\bar q_L^{\,3}q_R^{\,3}$ of the KMT-type. 
At first order in derivatives, the leading vector structure is instead the
chirality-conserving current
$\bar q_{L,R}\gamma^\mu q_{L,R}$ contracted with the scalar-diquark current
$i d^\dagger\overleftrightarrow{\partial_\mu}d$. The resulting
U(1)$_A$-symmetric single-trace operators are precisely the same-chirality
interaction \eqref{eq:Lqd} and the crossed interaction \eqref{eq:Lqd2}.
Operators in which the derivative acts on the quark field can be shifted,
up to total derivatives, into the above basis plus terms proportional to
the free quark equation of motion. Tensor structures involving
$\sigma^{\mu\nu}$ require additional Lorentz indices and therefore enter
only at higher order in the derivative expansion.

Within the single-trace form in flavor space for the 
quark--diquark exchange channel used to generate baryons, 
Eqs.~\eqref{eq:Lqd}--\eqref{eq:LA}
therefore form the minimal leading-order interaction basis adopted in this
work. Additional local operators are in principle allowed, such as
double-flavor-trace terms, in which the flavor indices are closed within
each bracket rather than between the two brackets, e.g.
\begin{align}
{\mathcal L}_{qd}^{(\rm 2tr)}
= -\tilde G_{qd}
\left\{ \left( \bar{q}_L \right)_{c,i} \left( d_L^\dagger \right)^{c,i} \right\}
\left\{ \left( i \partial_\mu d_L \right)_{a,j} \gamma^\mu \left( q_L \right)^{a,j} \right\}
+ \big( L \to R \big)\,,
\end{align}
and direct current--current interactions, in which the quark and diquark
fields are contracted into separately color-singlet bilinears, e.g.
\begin{align}
{\mathcal L}^{(\rm cc)}
= -G_{V}
\left[ \left(\bar q_L\right)_{c,i}\gamma_\mu \left(q_L\right)^{c,i} \right]
\left[\, i \left(d_L^\dagger\right)^{a,j}
\overleftrightarrow{\partial^\mu} \left(d_L\right)_{a,j} \right]
+ \cdots\,,
\end{align}
where the ellipsis stands for the flavor-octet, $(L\to R)$, and
mixed-chirality combinations such as
$(\bar q_L\gamma_\mu q_L)\,(i\,d_R^\dagger\overleftrightarrow{\partial^\mu}d_R)$.
Their effects are beyond the minimal basis considered here and may be absorbed into higher-order contact terms in the effective
theory.%

\subsection{Quark--diquark composites and their mirror and crossed partners}
\label{sec:composites}

Contracting the color indices while leaving the flavor indices free, we define
the color-singlet quark--diquark composites as 
\begin{align}
(\mathcal{Z}_{1R})^{\alpha}_{\beta} \sim (q_R)^{a,\alpha}\,(d_R)_{a,\beta}, &\qquad
(\mathcal{Z}_{1L})^{i}_{j} \sim (q_L)^{a,i}\,(d_L)_{a,j},
\label{eq:Bfields}
\end{align}
each  of which is 
a spin-$\tfrac12$ Dirac field. Under $\SU{3}_R\times\SU{3}_L$ they are decomposed as
\begin{align}
\mathcal{Z}_{1R}\sim\Rep{1}{1}\oplus\Rep{8}{1}, &\qquad
\mathcal{Z}_{1L}\sim\Rep{1}{1}\oplus\Rep{1}{8}.
\end{align}
It is convenient to introduce the baryon fields with mirror assignment as
\begin{align}
(\mathcal{Z}_{2L})^{\alpha}_{\beta} \sim (\gamma^\mu q_R)^{a,\alpha}\,(i\partial_\mu d_R)_{a,\beta}, &\qquad
(\mathcal{Z}_{2R})^{i}_{j} \sim (\gamma^\mu q_L)^{a,i}\,(i\partial_\mu d_L)_{a,j},
\label{eq:B2fields}
\end{align}
which belong to
\begin{align}
\mathcal{Z}_{2L}\sim\Rep{1}{1}\oplus\Rep{8}{1}, &\qquad
\mathcal{Z}_{2R}\sim\Rep{1}{1}\oplus\Rep{1}{8}.
\end{align}
We further decompose the $\mathcal{Z}$ field into the singlet and the octet part such as
\begin{align}
    (\mathcal{Z}_{1R})^\alpha_{\beta} = \cB^{{\rm sin}}_{1R} \,\delta^\alpha_\beta + (\cB_{1R})^{\alpha}_\beta \ .
\end{align}
Since the four-body interaction ${\mathcal L}_A$ in Eq.~(\ref{eq:LA}) contributes to the masses of ${\mathcal B}_1^{\rm sin}$ and ${\mathcal B}_2^{\rm sin}$, and we expect that the flavor-singlet baryons are heavier than the octet baryons,
in the following content, we will only focus on the octet part $\cB$. 
We should note that, at this level, 
both  $\cB_{1}$ and $\cB_{2}$ carry positive parity, with parity acting as
\begin{align}
\cB_{1L}\mathop{\longleftrightarrow}_{P}\cB_{1R}, \qquad
\cB_{2L}\mathop{\longleftrightarrow}_{P}\cB_{2R}.
\end{align}

Finally, we introduce the crossed composites, built from a quark and a diquark
of opposite chirality,
\begin{align}
(\psi_{1R})^{\alpha}{}_{i} \, \sim \, (q_R)^{a,\alpha}\,(d_L)_{a,i}, &\qquad
(\psi_{1L})^{i}{}_{\alpha} \,\sim\, (q_L)^{a,i}\,(d_R)_{a,\alpha},
\notag\\
(\psi_{2L})^{\alpha}{}_{i}\,\sim\,(\gamma^\mu q_R)^{a,\alpha}\,(i\partial_\mu d_L)_{a,i}, &\qquad
(\psi_{2R})^{i}{}_{\alpha} \,\sim\, (\gamma^\mu q_L)^{a,i}\,(i\partial_\mu d_R)_{a,\alpha},
\label{eq:psifields}
\end{align}
which correspond to the  $\Rep{3}{\bar{3}} \oplus \Rep{\bar{3}}{3}$ 
chiral representations.

\subsection{Auxiliary-field reformulation}
\label{sec:AFM}

To prepare for the loop calculation, we adopt the auxiliary-field method.
For the same-chirality sector we add
\begin{align}
{\mathcal L}_{qdAF} =& C_{qd}
\Bigg[
\left(
(\bar{\mathcal B}_{1L})_i^{j} - \frac{1}{C_{qd}}
\left\{ \left( \bar{q}_L \right)_{c,i} \left( d_L^\dagger \right)^{c,j} \right\}
\right)
\left(
\left( {\mathcal B}_{2R} \right)^i_j - G_{qd}
\left\{ \left( i \partial_\mu d_L \right)_{a,j} \gamma^\mu \, \left( q_L \right)^{a,i} \right\}
\right)
\notag\\
& \qquad +
\left(
\left( \bar{\mathcal B}_{2R} \right)_i^j - G_{qd}
\left\{ \left( \bar{q}_L \right)_{c,i} \gamma^\mu \left( - i \partial_\mu d_L^\dagger \right)^{c,j} \right\}
\right)
\left(
\left( {\mathcal B}_{1L} \right)_j^i - \frac{1}{C_{qd}}
\left\{ \left( d_L \right)_{a,j} \left( q_L \right)^{a,i} \right\}
\right)
\notag\\
& \qquad +
\left(
\left( \bar{\mathcal B}_{1R} \right)_i^j - \frac{1}{C_{qd}}
\left\{ \left( \bar{q}_R \right)_{c,i} \left( d_R^\dagger \right)^{c,j} \right\}
\right)
\left(
\left( {\mathcal B}_{2L} \right)_j^i - G_{qd}
\left\{ \left( i \partial_\mu d_R \right)_{a,j} \gamma^\mu \, \left( q_R \right)^{a,i} \right\}
\right)
\notag\\
& \qquad +
\left(
\left( \bar{\mathcal B}_{2L} \right)_i^j - G_{qd}
\left\{ \left( \bar{q}_R \right)_{c,i} \gamma^\mu \left( - i \partial_\mu d_R^\dagger \right)^{c,j} \right\}
\right)
\left(
\left( {\mathcal B}_{1R} \right)_j^i - \frac{1}{C_{qd}}
\left\{ \left( d_R \right)_{a,j} \left( q_R \right)^{a,i} \right\}
\right)
\Bigg],
\label{eq:LqdAF}
\end{align}
and, for the crossed sector,
\begin{align}
{\mathcal L}_{qdAF}^{(2)} =& \; C_{qd}^{(2)}
\Bigg[
\left(
(\bar\psi_{1L})_i{}^{\beta} - \frac{1}{C_{qd}^{(2)}}
\left\{ \left( \bar{q}_L \right)_{c,i} \left( d_R^\dagger \right)^{c,\beta} \right\}
\right)
\left(
\left( \psi_{2R} \right)^{i}{}_{\beta} - G_{qd}^{(2)}
\left\{ \left( i \partial_\mu d_R \right)_{a,\beta} \gamma^\mu \left( q_L \right)^{a,i} \right\}
\right)
\notag\\
& \quad +
\left(
(\bar\psi_{2R})_i{}^{\beta} - G_{qd}^{(2)}
\left\{ \left( \bar{q}_L \right)_{c,i} \gamma^\mu \left( -i \partial_\mu d_R^\dagger \right)^{c,\beta} \right\}
\right)
\left(
\left( \psi_{1L} \right)^{i}{}_{\beta} - \frac{1}{C_{qd}^{(2)}}
\left\{ \left( d_R \right)_{a,\beta} \left( q_L \right)^{a,i} \right\}
\right)
\notag\\
& \quad +
\left(
(\bar\psi_{1R})_\alpha{}^{j} - \frac{1}{C_{qd}^{(2)}}
\left\{ \left( \bar{q}_R \right)_{c,\alpha} \left( d_L^\dagger \right)^{c,j} \right\}
\right)
\left(
\left( \psi_{2L} \right)^{\alpha}{}_{j} - G_{qd}^{(2)}
\left\{ \left( i \partial_\mu d_L \right)_{a,j} \gamma^\mu \left( q_R \right)^{a,\alpha} \right\}
\right)
\notag\\
& \quad +
\left(
(\bar\psi_{2L})_\alpha{}^{j} - G_{qd}^{(2)}
\left\{ \left( \bar{q}_R \right)_{c,\alpha} \gamma^\mu \left( -i \partial_\mu d_L^\dagger \right)^{c,j} \right\}
\right)
\left(
\left( \psi_{1R} \right)^{\alpha}{}_{j} - \frac{1}{C_{qd}^{(2)}}
\left\{ \left( d_L \right)_{a,j} \left( q_R \right)^{a,\alpha} \right\}
\right)
\Bigg],
\label{eq:LqdAF2}
\end{align}
where $C_{qd}$ and $C_{qd}^{(2)}$ are constants of mass dimension one that
ensure the composite fields carry the canonical mass dimension $3/2$. Other
choices for the normalization of the auxiliary fields are possible; the fields can be normalized canonically at a later stage.

By construction, the equations of motion of the auxiliary fields reproduce the
original four-body interactions, so that adding \eqref{eq:LqdAF} to
\eqref{eq:Lqd} yields
\begin{align}
{\mathcal L}^\prime_{qd}
= & \;{\mathcal L}_{qd}+{\mathcal L}_{qdAF}
\notag\\
=& \;C_{qd}\,\mbox{tr}
\left[
\bar{\mathcal B}_{1L}{\mathcal B}_{2R}
+\bar{\mathcal B}_{2R}{\mathcal B}_{1L}
+\bar{\mathcal B}_{1R}{\mathcal B}_{2L}
+\bar{\mathcal B}_{2L}{\mathcal B}_{1R}
\right]
\notag\\
& -C_{qd}G_{qd}\,
(\bar{\mathcal B}_{1L})_i{}^{j}
\left\{ \left( i \partial_\mu d_L \right)_{a,j} \gamma^\mu \, \left( q_L \right)^{a,i} \right\}
-
\left\{ \left( \bar{q}_L \right)_{c,i} \left( d_L^\dagger \right)^{c,j} \right\}
\left( {\mathcal B}_{2R} \right)^i{}_j
\notag\\
& -
\left( \bar{\mathcal B}_{2R} \right)_i{}^j
\left\{ \left( d_L \right)_{a,j} \left( q_L \right)^{a,i} \right\}
-C_{qd}G_{qd}\,
\left\{ \left( \bar{q}_L \right)_{c,i} \gamma^\mu \left( - i \partial_\mu d_L^\dagger \right)^{c,j} \right\}
\left( {\mathcal B}_{1L} \right)_j{}^i
\notag\\
& -C_{qd}G_{qd}\,
\left( \bar{\mathcal B}_{1R} \right)_i{}^j
\left\{ \left( i \partial_\mu d_R \right)_{a,j} \gamma^\mu \, \left( q_R \right)^{a,i} \right\}
-
\left\{ \left( \bar{q}_R \right)_{c,i} \left( d_R^\dagger \right)^{c,j} \right\}
\left( {\mathcal B}_{2L} \right)_j{}^i
\notag\\
& -
\left( \bar{\mathcal B}_{2L} \right)_i{}^j
\left\{ \left( d_R \right)_{a,j} \left( q_R \right)^{a,i} \right\}
-C_{qd}G_{qd}\,
\left\{ \left( \bar{q}_R \right)_{c,i} \gamma^\mu \left( - i \partial_\mu d_R^\dagger \right)^{c,j} \right\}
\left( {\mathcal B}_{1R} \right)_j{}^i,
\label{eq:AFvertex}
\end{align}
and, analogously for the crossed sector,
\begin{align}
{\mathcal L}_{qd}^{\prime (2)}
=& \;{\mathcal L}_{qd}^{(2)}+{\mathcal L}_{qdAF}^{(2)}
\notag\\
=& \;C_{qd}^{(2)}\,\mbox{tr}
\left[
\bar\psi_{1L}\,\psi_{2R}
+\bar\psi_{2R}\,\psi_{1L}
+\bar\psi_{1R}\,\psi_{2L}
+\bar\psi_{2L}\,\psi_{1R}
\right]
\notag\\
& -C_{qd}^{(2)}G_{qd}^{(2)}\,
(\bar\psi_{1L})_i{}^{\beta}
\left\{ \left( i \partial_\mu d_R \right)_{a,\beta} \gamma^\mu \left( q_L \right)^{a,i} \right\}
-
\left\{ \left( \bar{q}_L \right)_{c,i} \left( d_R^\dagger \right)^{c,\beta} \right\}
\left( \psi_{2R} \right)^{i}{}_{\beta}
\notag\\
& -
(\bar\psi_{2R})_i{}^{\beta}
\left\{ \left( d_R \right)_{a,\beta} \left( q_L \right)^{a,i} \right\}
-C_{qd}^{(2)}G_{qd}^{(2)}\,
\left\{ \left( \bar{q}_L \right)_{c,i} \gamma^\mu \left( -i \partial_\mu d_R^\dagger \right)^{c,\beta} \right\}
\left( \psi_{1L} \right)^{i}{}_{\beta}
\notag\\
& -C_{qd}^{(2)}G_{qd}^{(2)}\,
(\bar\psi_{1R})_\alpha{}^{j}
\left\{ \left( i \partial_\mu d_L \right)_{a,j} \gamma^\mu \left( q_R \right)^{a,\alpha} \right\}
-
\left\{ \left( \bar{q}_R \right)_{c,\alpha} \left( d_L^\dagger \right)^{c,j} \right\}
\left( \psi_{2L} \right)^{\alpha}{}_{j}
\notag\\
& -
(\bar\psi_{2L})_\alpha{}^{j}
\left\{ \left( d_L \right)_{a,j} \left( q_R \right)^{a,\alpha} \right\}
-C_{qd}^{(2)}G_{qd}^{(2)}\,
\left\{ \left( \bar{q}_R \right)_{c,\alpha} \gamma^\mu \left( -i \partial_\mu d_L^\dagger \right)^{c,j} \right\}
\left( \psi_{1R} \right)^{\alpha}{}_{j}.
\label{eq:AFvertex2}
\end{align}
Equations \eqref{eq:AFvertex} and \eqref{eq:AFvertex2} are the forms used in
the subsequent one-loop analysis of the baryon masses.

% ============================================================================
% ============================================================================
\section{Baryon mass without the $\Sigma$ field}
\label{sec:mass}
% ============================================================================

In this section, we study the baryon mass when the chiral symmetry is restored. In such a case, 
the vanishing of the chiral order parameter removes the dynamically generated constituent quark mass.
Here, we treat the current
quark mass is zero, so that the quark propagator is $S_q(k)=\slashed k/k^2$, while the diquark
propagator is $D(P-k)=1/[(P-k)^2-m_d^2]$ with the chiral
invariant diquark
mass  $m_d$
of Eq.~\eqref{eq:La1}. Our aim is to construct 
the effective Lagrangian for
the composite baryons at leading order in the derivative expansion: the mass
terms are read off from the two-point functions at vanishing external
momentum $P$,  and the kinetic terms from their first order in $\slashed P$.

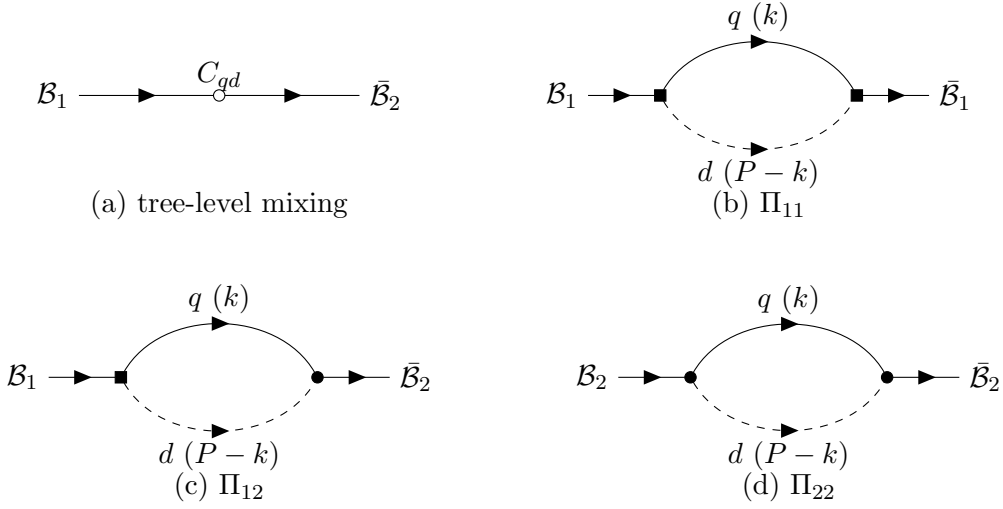
\begin{figure}[htbp]
\centering
% ---------- (a) tree-level mixing ----------
\begin{tikzpicture}
\begin{feynman}
    \vertex (i) at (-1.3,0) {\( \cB_{1} \)};
    \vertex [empty dot] (c) at (0.9,0) {};
    \vertex (o) at (3.1,0) {\( \bar{\cB}_{2} \)};
    \diagram*{
        (i) -- [fermion] (c),
        (c) -- [fermion] (o),
    };
    \node [above=7pt of c] {\( C_{qd} \)};
    \node at (0.9,-1.45) {(a) tree-level mixing};
\end{feynman}
\end{tikzpicture}
\hspace{1.4cm}
% ---------- (b) Pi_11 ----------
\begin{tikzpicture}
\begin{feynman}
    \vertex (i) at (-1.3,0) {\( \cB_{1} \)};
    \vertex [square dot] (a) at (0,0) {};
    \vertex [square dot] (b) at (2.6,0) {};
    \vertex (o) at (3.9,0) {\( \bar{\cB}_{1} \)};
    \diagram*{
        (i) -- [fermion] (a),
        (a) -- [fermion, out=60, in=120, edge label=\( q\ (k) \)] (b),
        (a) -- [charged scalar, out=-60, in=-120, edge label'=\( d\ (P-k) \)] (b),
        (b) -- [fermion] (o),
    };
    \node at (1.3,-1.45) {(b) \( \Pi_{11} \)};
\end{feynman}
\end{tikzpicture}
\\[16pt]
% ---------- (c) Pi_12 ----------
\begin{tikzpicture}
\begin{feynman}
    \vertex (i) at (-1.3,0) {\( \cB_{1} \)};
    \vertex [square dot] (a) at (0,0) {};
    \vertex [dot] (b) at (2.6,0) {};
    \vertex (o) at (3.9,0) {\( \bar{\cB}_{2} \)};
    \diagram*{
        (i) -- [fermion] (a),
        (a) -- [fermion, out=60, in=120, edge label=\( q\ (k) \)] (b),
        (a) -- [charged scalar, out=-60, in=-120, edge label'=\( d\ (P-k) \)] (b),
        (b) -- [fermion] (o),
    };
    \node at (1.3,-1.45) {(c) \( \Pi_{12} \)};
\end{feynman}
\end{tikzpicture}
\hspace{1.4cm}
% ---------- (d) Pi_22 ----------
\begin{tikzpicture}
\begin{feynman}
    \vertex (i) at (-1.3,0) {\( \cB_{2} \)};
    \vertex [dot] (a) at (0,0) {};
    \vertex [dot] (b) at (2.6,0) {};
    \vertex (o) at (3.9,0) {\( \bar{\cB}_{2} \)};
    \diagram*{
        (i) -- [fermion] (a),
        (a) -- [fermion, out=60, in=120, edge label=\( q\ (k) \)] (b),
        (a) -- [charged scalar, out=-60, in=-120, edge label'=\( d\ (P-k) \)] (b),
        (b) -- [fermion] (o),
    };
    \node at (1.3,-1.45) {(d) \( \Pi_{22} \)};
\end{feynman}
\end{tikzpicture}
\caption{Two-point functions of the composite baryon fields in the absence of
the $\Sigma$ field. Solid internal lines denote the massless quark propagator
$S_q(k)=\slashed k/k^2$ and dashed lines the diquark propagator
$D(P-k)=1/[(P-k)^2-m_d^2]$. Filled squares denote the derivative ($f_1$)
vertex of Eq.~\eqref{eq:f1} with coupling $C_{qd}G_{qd}$, filled circles the
trivial ($f_2$) vertex with unit coupling, and the open circle in (a) the
tree-level $\bar{\cB}_1\cB_2$ mixing $C_{qd}$ of Eq.~\eqref{eq:AFvertex}.
Panels (b)--(d) give the loop components $\Pi_{11}$, $\Pi_{12}$, and
$\Pi_{22}$ of Eqs.~\eqref{eq:Pi11}, \eqref{eq:Pi12}, and \eqref{eq:Pi22}: at
vanishing external momentum the diagonal components (b) and (d) generate the
kinetic normalizations $Z_1$ and $Z_3$, while the off-diagonal component (c),
combined with the tree-level mixing (a), yields the $\bar{\cB}_1\cB_2$ 
 chiral-invariant mass
term that becomes the physical baryon mass $M_\cB$ of Eq.~\eqref{eq:MB} after
canonical normalization. The identical diagrams with
$C_{qd}\to C_{qd}^{(2)}$, $G_{qd}\to G_{qd}^{(2)}$ apply to the crossed
($\psi$) sector.}
\label{fig:2pt}
\end{figure}

The reformulated Lagrangian ${\mathcal L}'_{qd}$ of Eq.~\eqref{eq:AFvertex}
contains two distinct quark--diquark vertices, so that the one-loop
quark--diquark two-point function becomes a $2\times2$ matrix in vertex space as shown in Fig.~\ref{fig:2pt}(b)-(d).  For a quark line of momentum $k$ and a diquark line of momentum $q_d=P-k$, the
two vertex factors are
\begin{align}
f_1=\slashed q_d=\slashed P-\slashed k \quad ({\rm coupling}\ C_{qd}G_{qd}),
\label{eq:f1}
\end{align}
shown by filled square in Fig.~\ref{fig:2pt}, and
\begin{align}
f_2=1 \quad ({\rm coupling}\ 1),
\label{eq:f2}
\end{align}
shown by filled circle,
each accompanied by a chiral projector $\hat P_{R,L}=\tfrac12(1\pm\gamma_5)$.
Since the massless quark propagator anticommutes with the projectors,
$\hat P_\chi\,\slashed k=\slashed k\,\hat P_{-\chi}$, the projectors merely
route the left and right components of the composite fields and can be
stripped from the loop integral. The two-point function then reduces to
\begin{align}
\Pi_{ab}(P)=i\int\frac{d^4k}{(2\pi)^4}\,
\frac{f_a\,g_a\;\slashed k\;f_b\,g_b}{k^2\left[(P-k)^2-m_d^2\right]},
\qquad a,b=1,2,
\label{eq:Piab}
\end{align}
with $g_a,g_b$ the couplings at the two vertices, $g_1 = C_{qd}G_{qd}$ and $g_2 = 1$. Only two loop functions
appear in the evaluation: the diquark tadpole $T_d$ and the vector function
$\PiV$,
\begin{align}
T_d &= i\int\frac{d^4k}{(2\pi)^4}\frac{1}{(P-k)^2-m_d^2}
%\nonumber\\&
%=\frac{1}{2\pi^2}\int^\Lambda_0 dk\,\frac{k^2}{2E_d},
\label{eq:Td}\\[2pt]
\slashed P\,\PiV &= i\int\frac{d^4k}{(2\pi)^4}
\frac{\slashed k}{k^2\left[(P-k)^2-m_d^2\right]}.
\end{align}
By taking the rest frame $P^\mu = (W, \vec{0})$ and 
introducing a three-momentum cutoff $\Lambda$, these are evaluated as
\begin{align}
T_d &= 
\frac{1}{2\pi^2}\int^\Lambda_0 dk\,\frac{k^2}{2E_d},
\\ \PiV &=-\frac{1}{2\pi^2}\int^\Lambda_0 dk\,
\frac{k^2}{2E_d\left[(k+E_d)^2-W^2\right]},
\label{eq:PiV}
\end{align}
with $E_d=\sqrt{\bm k^2+m_d^2}$
(see Appendix~\ref{app:2pt} for detail.)
Note that $T_d$ is independent of the
external momentum, and that $\PiV$ is finite and negative at  $W^2 = 0$.

%%%%%%%%%%%%%%%%%%%%%%%%%%%%%%%
%%%%%%%%%%%%%%%%%%%%%%%%%%%%%%%

\paragraph{$\Pi_{22}$ component.}
With the trivial vertex $f_2 g_2 = 1 $  
on both sides,
Eq.~\eqref{eq:Piab} directly gives
\begin{align}
\Pi_{22}=\slashed P\,\PiV.
\label{eq:Pi22}
\end{align}

\paragraph{$\Pi_{12}=\Pi_{21}$ component.}
With the derivative vertex $f_1$ on one side and the trivial vertex $f_2$ on
the other, the numerator reduces via
$\slashed q_d\,\slashed k=(\slashed P-\slashed k)\,\slashed k
=\slashed P\,\slashed k-k^2$, so that
\begin{align}
\Pi_{12}
=C_{qd}G_{qd}\left[\slashed P\,\slashed P\,\PiV-T_d\right]
=C_{qd}G_{qd}\left(P^2\,\PiV-T_d\right).
\label{eq:Pi12}
\end{align}

\paragraph{$\Pi_{11}$ component.}
With the derivative vertex $f_1$ on both sides, we write
$\slashed q_d\,\slashed k\,\slashed q_d
=(\slashed q_d\,\slashed k)\,\slashed P-k^2\,\slashed q_d$. The first piece
reproduces the integrand of $\Pi_{12}$, while the second cancels the quark
propagator and leaves $i\!\int\!d^4k/(2\pi)^4\;\slashed q_d/[q_d^2-m_d^2]$,
which vanishes: in the rest frame the shift $k^0\to k^0+W$ leaves the
three-momentum cutoff untouched, and the resulting integrand is odd in the
loop momentum. Hence
\begin{align}
\Pi_{11}
=C_{qd}G_{qd}\,\slashed P\,\Pi_{12}
=(C_{qd}G_{qd})^2\,\slashed P\left(P^2\,\PiV-T_d\right).
\label{eq:Pi11}
\end{align}

\paragraph{Effective Lagrangian and baryon mass.}
We now take the 
vanishing external momentum. Since $\PiV$ is finite at
$P^2=0$, the terms $P^2\PiV$ in Eqs.~\eqref{eq:Pi12} and \eqref{eq:Pi11} drop
out, and
\begin{align}
\Pi_{11}\to-(C_{qd}G_{qd})^2\,T_d\,\slashed P,
\qquad
\Pi_{12}\to-C_{qd}G_{qd}\,T_d,
\qquad
\Pi_{22}\to\PiV\,\slashed P .
\end{align}
The diagonal components contain no mass term. This is dictated by chiral
symmetry: with massless quarks a $\bar{\cB}_1\cB_1$ or $\bar{\cB}_2\cB_2$ mass
term would require a chirality flip along the quark line, whereas the
off-diagonal $\bar{\cB}_1\cB_2$ channel connects opposite Dirac chiralities
through the $\gamma^\mu$ of the $\cB_2$ interpolating field and is therefore
chiral invariant. The loop thus generates the kinetic terms
\begin{equation}
Z_1\,\bar{\cB}_1\,i\slashed\partial\,\cB_1,
\qquad
Z_3\,\bar{\cB}_2\,i\slashed\partial\,\cB_2,
\qquad
Z_1=\left(C_{qd}G_{qd}\right)^2 T_d,
\quad
Z_3=-\PiV>0,
\end{equation}
together with an off-diagonal mass mixing between $\cB_1$ and $\cB_2$,
\begin{align}
Z_2\,T_d\,(\bar{\cB}_1\cB_2+\bar{\cB}_2\cB_1),
\qquad Z_2=C_{qd}G_{qd},
\end{align}
where $\PiV$ is understood at $P^2=0$ hereafter. Collecting these together
with the tree-level mixing $C_{qd}$ of Eq.~\eqref{eq:AFvertex},
\begin{align}
\mathcal{L}_{\rm eff}
=Z_1\,\bar{\cB}_1\,i\slashed\partial\,\cB_1
+Z_3\,\bar{\cB}_2\,i\slashed\partial\,\cB_2
-\big(C_{qd}+Z_2T_d\big)\big(\bar{\cB}_1\cB_2+\bar{\cB}_2\cB_1\big).
\end{align}
Canonical normalization of the kinetic terms turns the off-diagonal mass into
the physical baryon mass,
\begin{align}
M_{\cB}=\frac{C_{qd}+Z_2T_d}{\sqrt{Z_1Z_3}}
=\frac{C_{qd}(1+G_{qd}T_d)}{\sqrt{(C_{qd}G_{qd})^2T_d\,(-\PiV)}}
=\frac{1+G_{qd}T_d}{G_{qd}\sqrt{-T_d\,\PiV}},
\label{eq:MB}
\end{align}
which is independent of the normalization $C_{qd}$ and finite even though the
quarks are massless. It should be noted that the baryon mass is not simply
proportional to the diquark mass: $M_\cB$ remains nonzero even in the
$m_d\to0$ limit, where the loop functions are saturated by the cutoff, and it
grows as $1/G_{qd}$ at weak coupling, signaling that a weakly bound composite
is pushed up in mass. Rather, the mass scale is generated jointly by the
invariant diquark mass $m_d$, entering through $T_d$ and $\PiV$, and the
gluon-mediated quark--diquark coupling $G_{qd}$, with the cutoff $\Lambda$
setting the scale of the underlying  interactions. The essential
point is that all of these ingredients are independent of the chiral order
parameter: the baryon mass \eqref{eq:MB} is therefore chiral invariant by
construction, originating from the gluonic dynamics of the diquark sector
rather than from the quark condensate. The identical analysis in the crossed
sector, Eq.~\eqref{eq:AFvertex2}, yields the same expression with
$G_{qd}\to G_{qd}^{(2)}$ and $C_{qd}\to C_{qd}^{(2)}$; these two masses are
the chiral invariant masses $m_{\cB0}$ and $m_{\psi0}$ of the composite
sectors introduced in the next section.

% ============================================================================
\section{ Effective  Lagrangian including the $\Sigma$ field}
\label{sec:sigma}
% ============================================================================
 
We now couple the quarks and diquarks to the  meson field $\Sigma$.
By keeping the leading order of $\Sigma$, the relevant Yukawa-type interactions are
\begin{align}
\mathcal{L}_{\Sigma}=\mathcal{L}_q+\mathcal{L}_d,
\end{align}
with 
\begin{align}
\mathcal{L}_q &=-g_q\left[\bar{q}_{L,a,i}\,\Sigma^i{}_\alpha\,q_R^{a,\alpha}
+\bar{q}_{R,a,\alpha}\,(\Sigma^\dagger)^\alpha{}_i\,q_L^{a,i}\right],
\label{eq:qmass}\\
\mathcal{L}_d &=-Y_d\left[(d_L)_{a,i}\,\Sigma^i{}_\alpha\,(d_R^\dagger)^{a,\alpha}
+(d_R)_{a,\alpha}\,(\Sigma^\dagger)^\alpha{}_i\,(d_L^\dagger)^{a,i}\right].
\label{eq:Ld}
\end{align}
The quark term $\mathcal{L}_q$ is the standard chiral invariant Yukawa
coupling and
it generates the constituent
quark mass $M_q=g_q f$. The diquark term $\mathcal{L}_d$ couples the scalar and
pseudoscalar diquarks through $\Sigma$ and, with
$\langle\Sigma\rangle=\mathrm{diag}(\alpha, \alpha, \gamma)$, 
splits the $0^+$ and
$0^-$ diquark masses; it originates from a Kobayashi--Maskawa--'t~Hooft-like
six-quark interaction and therefore breaks the U(1)$_A$ symmetry.

\subsection{Effective Lagrangian and baryon mass matrix}
\begin{figure}[h]
\centering
\begin{tikzpicture}
\begin{feynman}
    \vertex (i) at (-1.2,0) { \( (\mathcal{B}_{1L})^i_j \)};
    \vertex [square dot] (a) at (0,0) {};
    \vertex [square dot] (b) at (3.2,0) {};
    \vertex [crossed dot] (s) at (1.6,1) {};
    \vertex (o) at (4.5,0) {\( (\bar\psi_{1R})^j_\alpha\)};
    \diagram*{
        (i) -- [fermion] (a),
        (a) -- [fermion, out=60, in=180] (s),
        (s) -- [fermion, out=0, in=120] (b),
        (a) -- [charged scalar, out=-60, in=-120, edge label'=\( (d_{L})_j \)] (b),
        (b) -- [fermion] (o),
    };
    \node [above=10pt of s] {\( (\Sigma^\dagger)_i^\alpha \)};
    \node at (0.3,0.9) {$q_L^i$};
    \node at (2.9,0.9) {$q_R^\alpha$};
\end{feynman}
\end{tikzpicture}
\caption{One-loop diagram mixing $\cB_{1L}$ and $\psi_{1R}$. Both
quark--diquark vertices are of the derivative ($f_1$) type, carrying
$C_{qd}G_{qd}$ and $C_{qd}^{(2)}G_{qd}^{(2)}$, respectively. This diagram
generates the induced interaction of Eq.~\eqref{eq:effB1Lpsi1R}.}
\label{fig:sig1}
\end{figure}
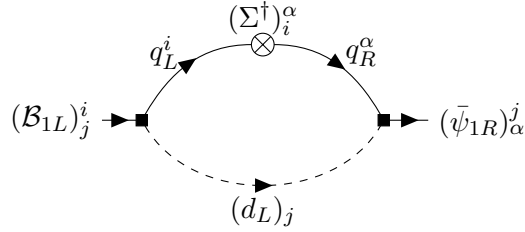

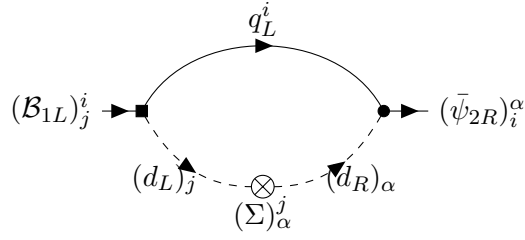
\begin{figure}[h]
\centering
\begin{tikzpicture}
\begin{feynman}
    \vertex (i) at (-1.2,0) {\( (\mathcal{B}_{1L})^i_j\)};
    \vertex [square dot] (a) at (0,0) {};
    \vertex [dot] (b) at (3.2,0) {};
    \vertex [crossed dot] (s) at (1.6,-1.0) {};
    \vertex (o) at (4.5,0) {\( (\bar\psi_{2R})^\alpha_i \)};
    \diagram*{
        (i) -- [fermion] (a),
        (a) -- [fermion, out=60, in=120, edge label=\( q_L^i \)] (b),
        (a) -- [charged scalar, out=-60, in=180] (s),
        (s) -- [charged scalar, out=0, in=-120] (b),
        (b) -- [fermion] (o),
    };
    \node [below=10pt of s] {\( (\Sigma)^j_\alpha \)};
    \node at (2.9,-0.9) {$ (d_{R})_\alpha $};
    \node at (0.3,-0.9) {$ (d_{L})_j $};
\end{feynman}
\end{tikzpicture}
\caption{One-loop diagram mixing $\cB_{1L}$ and $\psi_{2R}$. The $\cB_{1L}$ vertex is of the derivative ($f_1$) type and the
$\psi_{2R}$ vertex is of the trivial ($f_2$) type. This diagram generates the
induced interaction of Eq.~\eqref{eq:effB1Lpsi2R}.}
\label{fig:sig2}
\end{figure}

\begin{figure}[h]
\centering
\begin{tikzpicture}
\begin{feynman}
    \vertex (i) at (-1.2,0) {\( (\cB_{2L})^{\alpha}_{\beta} \)};
    \vertex [dot] (a) at (0,0) {};
    \vertex [dot] (b) at (3.2,0) {};
    \vertex [crossed dot] (s) at (1.6,1) {};
    \vertex (o) at (4.8,0) {\( (\bar\psi_{2R})_{i}^{\beta} \)};
    \diagram*{
        (i) -- [fermion] (a),
        (a) -- [fermion, out=60, in=180] (s),
        (s) -- [fermion, out=0, in=120] (b),
        (a) -- [charged scalar, out=-60, in=-120, edge label'=\( (d_{R})_\beta \)] (b),
        (b) -- [fermion] (o),
    };
    \node [above=10pt of s] {\( (\Sigma)_\alpha^i\)};
    \node at (0.3,0.9) {$q_R^{\alpha}$};
    \node at (2.9,0.9) {$q_L^{i}$};
\end{feynman}
\end{tikzpicture}
\caption{One-loop diagram mixing $\cB_{2L}$ and $\psi_{2R}$. Both quark--diquark vertices are of the trivial ($f_2$) type. This diagram generates the induced interaction of
Eq.~\eqref{eq:effB2Lpsi2R}.}
\label{fig:sig3}
\end{figure}
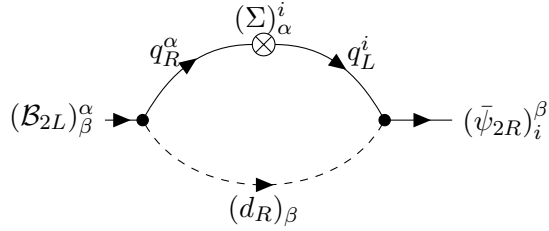

\begin{figure}[h]
\centering
\begin{tikzpicture}
\begin{feynman}
    \vertex (i) at (-1.2,0) {\( (\cB_{2L})^{\alpha}_{\beta} \)};
    \vertex [dot] (a) at (0,0) {};
    \vertex [dot] (b) at (3.2,0) {};
    \vertex [crossed dot] (s) at (1.6,-1.0) {};
    \vertex (o) at (4.8,0) {\( (\bar\psi_{1R})_{\alpha}^{j} \)};
    \diagram*{
        (i) -- [fermion] (a),
        (a) -- [fermion, out=60, in=120, edge label=\( q_R^{\alpha} \)] (b),
        (a) -- [charged scalar, out=-60, in=180] (s),
        (s) -- [charged scalar, out=0, in=-120] (b),
        (b) -- [fermion] (o),
    };
    \node [below=10pt of s] {\( (\Sigma)^\beta_j \)};
    \node at (0.3,-0.9) {$(d_{R})_\beta$};
    \node at (2.9,-0.9) {$ (d_{L})_j$};
\end{feynman}
\end{tikzpicture}
\caption{One-loop diagram mixing $\cB_{2L}$ and $\psi_{1R}$.  The $\psi_{1R}$ vertex is of the derivative ($f_1$) type and the
$\cB_{2L}$ vertex is of the trivial ($f_2$) type. This diagram generates the
induced interaction of Eq.~\eqref{eq:effB2Lpsi1R}.}
\label{fig:sig4}
\end{figure}
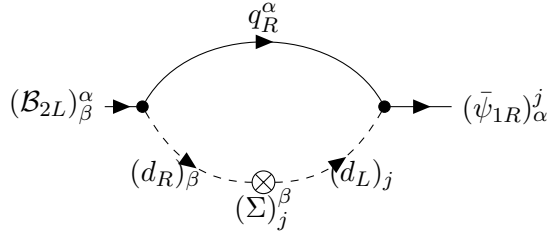

Once $\Sigma$ acquires its vacuum expectation value, the composites
$\cB_{1,2}$ mix with the crossed composites $\psi_{1,2}$ through a chirality
flip along either the quark or the diquark line. There are four such one-loop
mixing diagrams, shown in Figs.~\ref{fig:sig1}--\ref{fig:sig4}; in each case the
$\Sigma$ insertion is marked by a crossed vertex.

\paragraph{$\cB_{1L}$--$\psi_{1R}$ mixing (Fig.~\ref{fig:sig1}).}
The chirality flips along the quark line, $q_L\to q_R$. In the massless-quark
limit,
\begin{align}
\Pi^{(1)}
&= i\int\frac{d^4k}{(2\pi)^4}
\left[C_{qd}^{(2)}G_{qd}^{(2)}\slashed k\frac{1+\gamma_5}{2}\right]
\frac{\slashed k}{k^2}\frac{\slashed k}{k^2}
\left[C_{qd}G_{qd}\slashed k\frac{1-\gamma_5}{2}\right]
\frac{1}{k^2-m_d^2}\nonumber\\
&= \left(C_{qd}^{(2)}G_{qd}^{(2)}\,C_{qd}G_{qd}\right)T_d\,\frac{1-\gamma_5}{2},
\end{align}
which corresponds to the induced interaction
\begin{align}
\mathcal{L}_{\rm inter}^{(1)}
=-g_q\,(Z_1^{\cB})^{1/2}(Z_1^{\psi})^{1/2}\,
(\bar{\psi}_{1R})^j{}_\alpha\,(\Sigma^\dagger)^\alpha{}_i\,(\cB_{1L})^i{}_j,
\label{eq:effB1Lpsi1R}
\end{align}
with
\begin{align}
Z_1^{\cB}=(C_{qd}G_{qd})^2 T_d, \qquad
Z_1^{\psi}=\left(C_{qd}^{(2)}G_{qd}^{(2)}\right)^2 T_d.
\end{align}
 
\paragraph{$\cB_{1L}$--$\psi_{2R}$ mixing (Fig.~\ref{fig:sig2}).}
Here the chirality flips along the diquark line, $d_L\to d_R$, so that
%\begin{align}
%\Pi^{(2)}
%&= i\int\frac{d^4k}{(2\pi)^4}\,\frac{1-\gamma_5}{2}\,\frac{\slashed k}{k^2}
%\left(C_{qd}G_{qd}\,\slashed k\,\frac{1-\gamma_5}{2}\right)
%\left(\frac{1}{k^2-m_d^2}\right)^2\nonumber\\
%&= (C_{qd}G_{qd})\,i\int\frac{d^4k}{(2\pi)^4}
%\left(\frac{1}{k^2-m_d^2}\right)^2\frac{1-\gamma_5}{2}
%= (C_{qd}G_{qd})\,L_d\,\frac{1-\gamma_5}{2},
%\end{align}
\begin{align}
\Pi^{(2)}
&= i\int\frac{d^4k}{(2\pi)^4}\,\frac{1-\gamma_5}{2}\,\frac{\slashed k}{k^2}
\left(C_{qd}G_{qd}\,\slashed k\,\frac{1-\gamma_5}{2}\right)
\left(\frac{1}{k^2-m_d^2}\right)^2%\nonumber\\
%&
= (C_{qd}G_{qd})\,L_d\,\frac{1-\gamma_5}{2},
\end{align}
with
\begin{align}
L_d=i\int\frac{d^4k}{(2\pi)^4}\left(\frac{1}{k^2-m_d^2}\right)^2.
\label{eq:Ld_loop}
\end{align}
The induced interaction is
\begin{align}
\mathcal{L}_{\rm inter}^{(2)}
=-Y_d\,(C_{qd}G_{qd})\,L_d\,
(\bar{\psi}_{2R})^\alpha{}_i\,\Sigma^j{}_\alpha\,(\cB_{1L})^i{}_j.
\label{eq:effB1Lpsi2R}
\end{align}
 
\paragraph{$\cB_{2L}$--$\psi_{2R}$ mixing (Fig.~\ref{fig:sig3}).}
The chirality flips along the quark line, $q_R\to q_L$, and both vertices are
of the trivial ($f_2$) type:
%\begin{align}
%\Pi^{(3)}
%&= i\int\frac{d^4k}{(2\pi)^4}\,
%\frac{1-\gamma_5}{2}\,\frac{\slashed k}{k^2}\,\frac{\slashed k}{k^2}\,
%\frac{1-\gamma_5}{2}\,\frac{1}{k^2-m_d^2}
%= \left(i\int\frac{d^4k}{(2\pi)^4}\frac{1}{k^2\,[k^2-m_d^2]}\right)\frac{1-\gamma_5}{2}
%\nonumber\\
%&= \frac{1}{m_d^2}\left[i\int\frac{d^4k}{(2\pi)^4}\frac{1}{k^2-m_d^2}
%-i\int\frac{d^4k}{(2\pi)^4}\frac{1}{k^2}\right]\frac{1-\gamma_5}{2}
%= \frac{T_d-T_q^{M_q=0}}{m_d^2}\,\frac{1-\gamma_5}{2}.
%\end{align}
\begin{align}
\Pi^{(3)}
&= i\int\frac{d^4k}{(2\pi)^4}\,
\frac{1-\gamma_5}{2}\,\frac{\slashed k}{k^2}\,\frac{\slashed k}{k^2}\,\frac{1-\gamma_5}{2}\,\frac{1}{k^2-m_d^2}
= \frac{T_d-T_q}{m_d^2}\,\frac{1-\gamma_5}{2}\ , 
\end{align}
where
\begin{align}
T_q = \frac{1}{4\pi^2} \int_0^\Lambda d k k = \frac{\Lambda^2}{8 \pi^2} \ .
\end{align}
The induced interaction is
%\begin{align}
%\mathcal{L}_{\rm inter}^{(3)}
%=-g_q\,\frac{T_d-T_q^{M_q=0}}{m_d^2}\,
%(\bar{\psi}_{2R})^\beta{}_i\,\Sigma^i{}_\alpha\,(\cB_{2L})^\alpha{}_\beta.
%\label{eq:effB2Lpsi2R}
%\end{align}
\begin{align}
\mathcal{L}_{\rm inter}^{(3)}
=-g_q\,\frac{T_d-T_q}{m_d^2}\,
(\bar{\psi}_{2R})^\beta{}_i\,\Sigma^i{}_\alpha\,(\cB_{2L})^\alpha{}_\beta.
\label{eq:effB2Lpsi2R}
\end{align}
 
\paragraph{$\cB_{2L}$--$\psi_{1R}$ mixing (Fig.~\ref{fig:sig4}).}
The chirality flips along the diquark line, $d_R\to d_L$, giving
%\begin{align}
%\Pi^{(4)}
%&= i\int\frac{d^4k}{(2\pi)^4}
%\left(C^{(2)}_{qd}G^{(2)}_{qd}\,\slashed k\,\frac{1+\gamma_5}{2}\right)
%\frac{\slashed k}{k^2}\,\frac{1-\gamma_5}{2}\,
%\left(\frac{1}{k^2-m_d^2}\right)^2\nonumber\\
%&= C^{(2)}_{qd}G^{(2)}_{qd}
%\left(i\int\frac{d^4k}{(2\pi)^4}\Big(\frac{1}{k^2-m_d^2}\Big)^2\right)\frac{1-\gamma_5}{2}
%= C^{(2)}_{qd}G^{(2)}_{qd}\,L_d\,\frac{1-\gamma_5}{2}.
%\end{align}
\begin{align}
\Pi^{(4)}
&= i\int\frac{d^4k}{(2\pi)^4}
\left(C^{(2)}_{qd}G^{(2)}_{qd}\,\slashed k\,\frac{1+\gamma_5}{2}\right)
\frac{\slashed k}{k^2}\,\frac{1-\gamma_5}{2}\,
\left(\frac{1}{k^2-m_d^2}\right)^2%\nonumber\\
%&
= C^{(2)}_{qd}G^{(2)}_{qd}\,L_d\,\frac{1-\gamma_5}{2}.
\end{align}
The induced interaction is
\begin{align}
\mathcal{L}_{\rm inter}^{(4)}
=-Y_d\,C_{qd}^{(2)}G_{qd}^{(2)}\,L_d\,
(\bar{\psi}_{1R})^j{}_\alpha\,(\Sigma)^\beta{}_j\,(\cB_{2L})^\alpha{}_\beta.
\label{eq:effB2Lpsi1R}
\end{align}
 
We note that the two diagrams in which the chirality flips along the diquark line, Eqs.~\eqref{eq:effB1Lpsi2R} and
\eqref{eq:effB2Lpsi1R}, share the same coefficient form $\propto Y_d\,L_d$,
whereas the two diagrams flipping the chirality along the quark line,
Eqs.~\eqref{eq:effB1Lpsi1R} and \eqref{eq:effB2Lpsi2R}, carry different
coefficients ($\propto g_q T_d$ and 
$\propto g_q(T_d-T_q)/m_d^2$,
respectively). After canonical normalization of the $\cB$ and $\psi$ fields the
former two reduce to a single mixing coupling, while the latter two yield two
distinct couplings.

Collecting the tree-level mixings and the induced terms, the mass matrix for the nucleons 
in the basis $(\cB_1,\cB_2,\psi_1,\psi_2)$ takes the form
\begin{align}
\begin{pmatrix}
\bar{\cB}_1 & \bar{\cB}_2 & \bar{\psi}_1 & \bar{\psi}_2
\end{pmatrix}
\begin{pmatrix}
0        & m_{\cB 0} & g_1\alpha & y_1\gamma \\
m_{\cB 0}& 0         & y_1\gamma & g_2\alpha \\
g_1\alpha& y_1\gamma & 0         & m_{\psi 0}\\
y_1\gamma& g_2\alpha & m_{\psi 0}& 0
\end{pmatrix}
\begin{pmatrix}
\cB_1 \\ \cB_2 \\ \psi_1 \\ \psi_2
\end{pmatrix},
\label{eq:massmat0}
\end{align}
where $m_{\cB 0}$ and $m_{\psi 0}$ are the chiral invariant masses of the $\cB$
and $\psi$ sectors, $\alpha$ and $\gamma$ are the condensate parameters
associated with the light ($f_\pi$) and strange ($f_s$) components of
$\langle\Sigma\rangle$, and $g_{1,2}$, $y_1$ are the normalized mixing couplings
induced by the diagrams above.
 
\begin{figure}[h]
\centering
\begin{tikzpicture}
\begin{feynman}    
\vertex (i) at (-1.6,0) {\( (\psi_{1L})^{i}_{\alpha} \)};    
\vertex [dot] (a) at (0,0) {};    
\vertex [dot] (c) at (1.3,-0.9) {};    \vertex [crossed dot] (s) at (2.85,-0.15) {};    
\vertex [dot] (d) at (4.4,0.6) {};    \vertex [dot] (b) at (5.6,0) {};    \vertex (o) at (7.2,0) {\( (\bar\psi_{1R})^{k}_{\gamma} \)};    \diagram*{        (i) -- [fermion] (a),        (a) -- [fermion, out=60, in=160, edge label=\( q_L^{i} \)] (d),        (a) -- [charged scalar, out=-60, in=150, edge label'=\( (d_{R})_{\alpha} \)] (c),        (c) -- [fermion, edge label=\( q_R^{\beta} \)] (s),        (s) -- [fermion, edge label=\( q_L^{j} \)] (d),        (c) -- [fermion, out=-30, in=-120, edge label'=\( q_R^{\gamma} \)] (b),        (d) -- [charged scalar, out=-20, in=120, edge label=\( (d_{L})_{k} \)] (b),        (b) -- [fermion] (o),    };    \node [below=8pt of s] {\( (\Sigma)^{j}_{\beta} \)};
\end{feynman}
\end{tikzpicture}
\caption{Two-loop diagram generating the diagonal entry $g_1^a$ in the $\psi$ sector ($\psi_{1L}$--$\psi_{1R}$ mixing). The diquark is resolved into its constituent quarks, and a single $\Sigma$ insertion (crossed vertex), combined with the quark rearrangement, flips the chirality of both the quark and the diquark lines, $q_L\,d_R\to q_R\,d_L$. The same topology with the derivative-type ($f_1$) vertices yields $g_2^a$.}
\label{fig:sig5}
\end{figure}
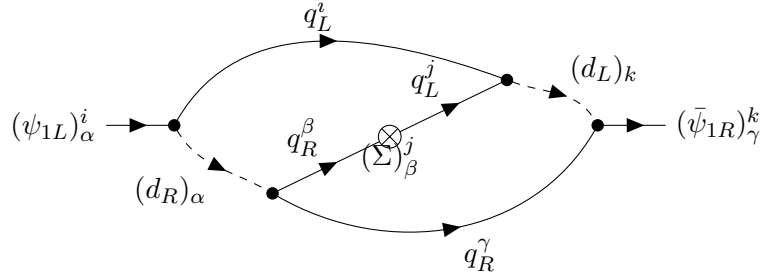

A complete evaluation must also include the two-loop diagrams in which the
chirality flips on both the quark and the diquark lines, connecting
$\psi_{1L,2L}$ to $\psi_{1R,2R}$ and thereby generating diagonal entries in
the $\psi$ sector. 
The relevant chiral invariant quark--quark--diquark Yukawa coupling is 
\begin{equation}
\begin{aligned}
\mathcal{L}_{qqd}
=& G_d\Big[
(d_L^\dagger)^{a,i}\,\epsilon_{abc}\,\epsilon_{ijk}\,
(q_L^T)^{b,j}\,C\,(q_L)^{c,k}
\\ & \qquad+(d_R^\dagger)^{a,\alpha}\,\epsilon_{abc}\,\epsilon_{\alpha\beta\gamma}\,
(q_R^T)^{b,\beta}\,C\,(q_R)^{c,\gamma}
+\mathrm{h.c.}\Big].
\label{eq:Lqqd}
\end{aligned}
\end{equation}
The antisymmetric color and flavor contractions match the quantum numbers of
the diquark fields of Sec.~\ref{sec:fields}, and the same argument as in
Appendix~\ref{app:PC} shows that parity and charge-conjugation invariance
force $G_d$ to be real. Since $\mathcal{L}_{qqd}$ is chiral invariant, its
$\Sigma$-independent loop effects only renormalize the diquark mass and the
wave-function normalizations already parametrized in Sec.~\ref{sec:mass}, and
we absorb them accordingly; its role here is to enable the double chirality
flip. A representative diagram is shown in Fig.~\ref{fig:sig5}: the diquark
dissolves through one $G_d$ vertex, the $\Sigma$ insertion flips one of the
constituent quarks with strength $g_q$, and the flipped quark recombines with
the spectator into the opposite-chirality diquark through a second $G_d$
vertex. With derivative ($f_1$) composite vertices at both ends the diagram
generates the $\psi_1$-sector diagonal entry, and with trivial ($f_2$)
vertices the $\psi_2$-sector one; rather than evaluating the two-loop
integrals explicitly, we parametrize the canonically normalized strengths of
these entries by the couplings $g_1^a$ and $g_2^a$, treated as independent
parameters in the fit below. 
In addition, mixed diagrams with one derivative ($f_1$) and one trivial
($f_2$) composite vertex connect $\psi_1$ and $\psi_2$; as discussed in
Sec.~\ref{sec:summary}, they generate no mass term but only a chiral
invariant kinetic mixing, which we neglect as a higher-order correction.

%%%%%%%%%%%%%%%%%%%%%%%%%%%%%%%
%%%%%%%%%%%%%%%%%%%%%%%%%%%%%%%
Assembling the results obtained so far, the total effective Lagrangian for the
normalized composite baryon fields reads
\begin{align}
\mathcal{L}_{\rm eff}
=&\;\sum_{n=1,2}\Big[\,\bar{\cB}_n\,i\slashed\partial\,\cB_n
                     +\bar{\psi}_n\,i\slashed\partial\,\psi_n\Big]
 -m_{\cB0}\big(\bar{\cB}_1\cB_2+\bar{\cB}_2\cB_1\big)
 -m_{\psi0}\big(\bar{\psi}_1\psi_2+\bar{\psi}_2\psi_1\big)
\notag\\
&-g_1^s\,(\bar\psi_{1R})^{j}{}_{\alpha}\,(\Sigma^\dagger)^{\alpha}{}_{i}\,(\cB_{1L})^{i}{}_{j}
 \;-\;g_2^s\,(\bar\psi_{2R})^{\beta}{}_{i}\,\Sigma^{i}{}_{\alpha}\,(\cB_{2L})^{\alpha}{}_{\beta}
\notag\\
&-y_1\Big[(\bar\psi_{2R})^{\alpha}{}_{i}\,\Sigma^{j}{}_{\alpha}\,(\cB_{1L})^{i}{}_{j}
         +(\bar\psi_{1R})^{j}{}_{\alpha}\,\Sigma^{\beta}{}_{j}\,(\cB_{2L})^{\alpha}{}_{\beta}\Big]
\notag\\
&+\;
- g_1^a \left( \bar{\psi}_{1R} \right)^i{}_\alpha \, \left( \Sigma \right)^j{}_\beta \, \left( \psi_{1L} \right)^k{}_\gamma\, \epsilon_{ijk} \epsilon^{\alpha\beta\gamma} 
- g_2^a \left( \bar{\psi}_{2L} \right)^i{}_\alpha \, \left( \Sigma \right)^j{}_\beta \, \left( \psi_{2R} \right)^k{}_\gamma\, \epsilon_{ijk} \epsilon^{\alpha\beta\gamma} 
\notag\\
& +\;
%\mathcal{L}_{\psi}^{a} 
+ \big(L\leftrightarrow R,  {\rm h.c.} \big) \,,
\label{eq:Leff_full}
\end{align}
where the terms proportional to $g_1^a$ or $g_2^a$ are obtained from 
too-loop contribution as shown above.
Distinguishing these diagonal couplings from the off-diagonal mixing couplings ($g^s$), and resolving the flavor (strangeness)
content of each octet member through the diagonal structure of
$\langle\Sigma\rangle$, the mass matrices in the $N$, $\Sigma$, and $\Xi$
sectors read
\begin{align}
{\mathcal M}_N &=
\begin{pmatrix}
0          & m_{\cB 0}  & g_1^s\alpha & y_1\gamma \\
m_{\cB 0}  & 0          & y_1\gamma   & g_2^s\alpha\\
g_1^s\alpha& y_1\gamma  & g_1^a\alpha & m_{\psi 0}\\
y_1\gamma  & g_2^s\alpha& m_{\psi 0}  & g_2^a\alpha
\end{pmatrix},
\label{eq:massN}\\[4pt]
{\mathcal M}_\Sigma &=
\begin{pmatrix}
0          & m_{\cB 0}  & g_1^s\alpha & y_1\alpha \\
m_{\cB 0}  & 0          & y_1\alpha   & g_2^s\alpha\\
g_1^s\alpha& y_1\alpha  & g_1^a\gamma & m_{\psi 0}\\
y_1\alpha  & g_2^s\alpha& m_{\psi 0}  & g_2^a\gamma
\end{pmatrix},
\label{eq:massSigma}\\[4pt]
{\mathcal M}_\Xi &=
\begin{pmatrix}
0          & m_{\cB 0}  & g_1^s\gamma & y_1\alpha \\
m_{\cB 0}  & 0          & y_1\alpha   & g_2^s\gamma\\
g_1^s\gamma& y_1\alpha  & g_1^a\alpha & m_{\psi 0}\\
y_1\alpha  & g_2^s\gamma& m_{\psi 0}  & g_2^a\alpha
\end{pmatrix}.
\label{eq:massXi}
\end{align}

We should note that, since we include the flavor-symmetry breaking
contribution generated by the difference between $\alpha$ and $\gamma$, the
iso-singlet members ($\Lambda$ baryons) of the octet baryons mix with the
flavor-singlet baryons, which are omitted in the present analysis. We
therefore concentrate on the mass spectra of the nucleons, $\Sigma$, and
$\Xi$ baryons in the following analysis.

\subsection{Explicit chiral symmetry breaking}
\label{sec:explicit}
 
Explicit chiral symmetry breaking from the current quark masses is incorporated
through the quark Yukawa term \eqref{eq:qmass} with bare quark mass matrix,
\begin{align}
\mathcal{L}_q =-g_q\left[\bar{q}_{L,a,i}\,\Sigma^i{}_j\,q_R^{a,j}+\mathrm{h.c.}\right]
-\left[\bar{q}_{L,a,i}\,({\mathcal M}_q)^i{}_j\,q_R^{a,j}
+\bar{q}_{R,a,j}\,({\mathcal M}_q^\dagger)^j{}_i\,q_L^{a,i}\right],
\label{eq:qmassmod}
\end{align}
with
\begin{equation}
{\mathcal M}_q=\mathrm{diag}(m_u,m_d,m_s)\ ,
\end{equation}
where $m_u$, $m_d$, $m_s$ are the current quark masses.
In the following analysis, we take the isospin limit $m_d=m_u$ and use $m_u$ only. 
Equation~(\ref{eq:qmassmod}) is rewritten as  
\begin{align}
\mathcal{L}_q =-g_q\left[\bar{q}_{L,a,i}\,(\Sigma+c_1\,{\mathcal M}_q)^i{}_j\,q_R^{a,j}
+\bar{q}_{R,a,j}\,(\Sigma^\dagger+c_1\,{\mathcal M}_q^\dagger)^j{}_i\,q_L^{a,i}\right],
\qquad c_1=\frac{1}{g_q},
\label{eq:qmassmod2}
\end{align}
so that, in the mass matrices
\eqref{eq:massN}--\eqref{eq:massXi}, the condensate parameters entering the
$g_1^s$-, $g_2^s$-, $g_1^a$-, and $g_2^a$-terms are replaced by
\begin{equation}
\alpha\to\alpha+c_1\,m_u, \qquad \gamma\to\gamma+c_1\,m_s.
\end{equation}
An analogous bare-mass term in $\mathcal{L}_d$ produces, in the $y_1$
entries,\footnote{The bare-mass term in $\mathcal{L}_d$ is tied to the
U(1)$_A$-breaking (Kobayashi--Maskawa--'t~Hooft-like) origin of $\mathcal{L}_d$
itself; its strength $c_2$ is treated as an independent parameter.}
\begin{equation}
\alpha\to\alpha+c_2\,m_u, \qquad \gamma\to\gamma+c_2\,m_s.
\end{equation}

Before proceeding, we point out an exact relation among the mass matrices
\eqref{eq:massN}--\eqref{eq:massXi}. The chiral invariant masses $m_{\cB0}$,
$m_{\psi0}$ and the mixing couplings $g_{1,2}^{s}$, $y_1$ enter only
off-diagonal entries, so the trace of each matrix is only the function of $g_{1,2}^{a}$,, yielding
\begin{align}
\mathrm{Tr}\,{\mathcal M}_N
=\mathrm{Tr}\,{\mathcal M}_\Xi
=\big(g_1^a+g_2^a\big)\,\alpha\,,
\label{eq:trace_relation}
\end{align}
and the equality $\mathrm{Tr}\,{\mathcal M}_N=\mathrm{Tr}\,{\mathcal M}_\Xi$
holds for arbitrary values of the condensates; it is unaffected by the
explicit breaking introduced in Sec.~\ref{sec:explicit}, which replaces
$\alpha\to\alpha+c\,m_u$ and $\gamma\to\gamma+c\,m_s$ in
Eq.~\eqref{eq:trace_relation}.

% ============================================================================
\section{Mass spectra and sigma terms}
\label{sec:numerics}
% ============================================================================

Having fixed the structure of the baryon mass matrices in
Sec.~\ref{sec:sigma}, we now define the nucleon sigma terms used in our
analysis and confront the model with the octet baryon spectrum. We first
derive the sigma terms from the quark-mass dependence of the nucleon mass
matrix, and then describe the fitting procedure and the corresponding
numerical results.

% ----------------------------------------------------------------------------
\subsection{Nucleon sigma terms}
\label{sec:sigma_terms}
% ----------------------------------------------------------------------------

The nucleon sigma terms are defined by
\begin{align}
\sigma_{sN}
&=m_s\,\langle N|\bar{s}s|N\rangle
=m_s\,\frac{\partial m_N}{\partial m_s},
\qquad
\sigma_{\pi N}
=m_u\,\langle N|\bar{u}u|N\rangle
=m_u\,\frac{\partial m_N}{\partial m_u},
\label{eq:sigma_def}
\end{align}
and are evaluated in the present model using the Feynman--Hellmann theorem.
The nucleon mass matrix, obtained from Eq.~\eqref{eq:massN} together with the
explicit-breaking replacements of Sec.~\ref{sec:explicit}, is expressed as 
\begin{align}
{\mathcal M}_N =
\left(\begin{matrix}
 0 & m_{\cB0} & g_1^s(\alpha+c_1m_u) & y_1(\gamma+c_2m_s) \\
 m_{\cB0} & 0 & y_1(\gamma+c_2m_s) & g_2^s(\alpha+c_1m_u) \\
 g_1^s(\alpha+c_1m_u) & y_1(\gamma+c_2m_s) & g_1^a(\alpha+c_1m_u) & m_{\psi0} \\
 y_1(\gamma+c_2m_s) & g_2^s(\alpha+c_1m_u) & m_{\psi0} & g_2^a(\alpha+c_1m_u)
\end{matrix}\right).
\label{eq:massmat_N}
\end{align}
It is diagonalized by an orthogonal matrix $\hat{\mathcal O}$ according to
\begin{align}
\hat{\mathcal O}\,{\mathcal M}_N\,\hat{\mathcal O}^T
= {\rm diag}\big(m_N,\,m_{N(1440)},\,m_{N(1535)},\,m_{N(1650)}\big).
\end{align}
The Feynman--Hellmann theorem then gives
\begin{align}
\sigma_{sN}
&=\left[
\hat{\mathcal O}\,
m_s\frac{\partial{\mathcal M}_N}{\partial m_s}\,
\hat{\mathcal O}^T
\right]_{11},
\qquad
\sigma_{\pi N}
=\left[
\hat{\mathcal O}\,
m_u\frac{\partial{\mathcal M}_N}{\partial m_u}\,
\hat{\mathcal O}^T
\right]_{11},
\label{eq:FH}
\end{align}
where $[X]_{11}$ denotes the $(1,1)$ component associated with the nucleon
ground state.

To evaluate the derivatives, we assume that the couplings
$g_{1,2}^{s}$, $g_{1,2}^{a}$, $y_1$, $c_1$, and $c_2$ are independent of the
current quark masses, and that the quark-mass dependence of the condensate
sector is carried entirely by $\alpha$ and $\gamma$. We take the SU(3)-breaking 
splitting between the two condensates to be linear in
$m_s-m_u$,
\begin{align}
\gamma-\alpha=(m_s-m_u)\,C,
\label{eq:gamma_alpha}
\end{align}
where $C$ is a constant. The chiral-limit value of $\alpha$ is fixed by the pion decay constant in the chiral limit~\cite{Gasser:1983yg},
\begin{align}
\alpha=f_0+m_u\,C_u,
\qquad
f_0=88~{\rm MeV},
\label{eq:alpha_f0}
\end{align}
where $C_u$ is another constant.

\paragraph{Strangeness sigma term.}
At fixed $m_u$, only the entries proportional to $\gamma+c_2m_s$ depend on
$m_s$. Equation~\eqref{eq:gamma_alpha} gives
\begin{align}
m_s\frac{\partial\gamma}{\partial m_s}
=m_s\frac{\partial(\gamma-\alpha)}{\partial m_s}
=\frac{m_s}{m_s-m_u}(\gamma-\alpha),
\end{align}
and hence
\begin{align}
m_s\frac{\partial{\mathcal M}_N}{\partial m_s}
&=y_1\,\tilde\gamma
\left(\begin{matrix}
0&0&0&1\\
0&0&1&0\\
0&1&0&0\\
1&0&0&0
\end{matrix}\right),
\nonumber\\
\tilde\gamma
&=\frac{m_s}{m_s-m_u}(\gamma-\alpha)+c_2m_s.
\label{eq:dMds}
\end{align}

\paragraph{Pion--nucleon sigma term.}
At fixed $m_s$, both the explicit terms proportional to $c_1m_u$ and the
implicit quark-mass dependence of the condensates contribute. From
Eq.~\eqref{eq:alpha_f0}, one obtains
\begin{align}
m_u\frac{\partial\alpha}{\partial m_u}=\alpha-f_0.
\end{align}
Combining this relation with Eq.~\eqref{eq:gamma_alpha} gives
\begin{align}
m_u\frac{\partial\gamma}{\partial m_u}
=(\alpha-f_0)-\frac{m_u}{m_s-m_u}(\gamma-\alpha).
\end{align}
Therefore,
\begin{align}
m_u\frac{\partial{\mathcal M}_N}{\partial m_u}
=
\left(\begin{matrix}
0 & 0 & g_1^s\tilde\alpha & y_1\hat\gamma \\
0 & 0 & y_1\hat\gamma & g_2^s\tilde\alpha \\
g_1^s\tilde\alpha & y_1\hat\gamma & g_1^a\tilde\alpha & 0 \\
y_1\hat\gamma & g_2^s\tilde\alpha & 0 & g_2^a\tilde\alpha
\end{matrix}\right),
\label{eq:dMdu}
\end{align}
where
\begin{align}
\tilde\alpha&=(\alpha-f_0)+c_1m_u,
\qquad
\hat\gamma
=(\alpha-f_0)-\frac{m_u}{m_s-m_u}(\gamma-\alpha).
\label{eq:cond-cont}
\end{align}
The terms proportional to $\alpha-f_0$ describe the response of the chiral
order parameter, equivalently the pion decay constant, to $m_u$. They thus
represent the meson-sector, or condensate, contribution to
$\sigma_{\pi N}$. The numerical importance of this contribution will be
discussed below.

% ----------------------------------------------------------------------------
\subsection{Fitting procedure}
\label{sec:fitting}
% ----------------------------------------------------------------------------

The model contains seven coupling constants,
\begin{align}
g_1^s,\quad g_2^s,\quad g_1^a,\quad g_2^a,\quad y_1,\quad c_1,\quad c_2,
\end{align}
in addition to the two chiral invariant mass parameters $m_{\cB0}$ and
$m_{\psi0}$. We treat $(m_{\cB0},m_{\psi0})$ as external parameters and scan
them over a two-dimensional grid. At each grid point, 
we minimize the reduced chi-squared,
\begin{align}
\chi^2
=\frac{1}{N_{\rm dof}}\left[
\sum_i
\left(\frac{m_i^{\rm fit}-m_i^{\rm exp}}{\delta m_i}\right)^2
+\left(
\frac{\sigma_{sN}^{\rm fit}-\sigma_{sN}^{\rm exp}}
{\delta\sigma_{sN}}
\right)^2
+\left(
\frac{\sigma_{\pi N}^{\rm fit}-\sigma_{\pi N}^{\rm exp}}
{\delta\sigma_{\pi N}}
\right)^2
\right]\ ,
\label{eq:chi2}
\end{align}
to determine 
the seven couplings 
The mass inputs are listed in Table~\ref{tab-mass-inputs}. We assign
$\delta m_i=10$~MeV to the ground states and $\delta m_i=100$~MeV to the
excited states. For the strangeness sigma term, we use the FLAG\,2024
average~\cite{FlavourLatticeAveragingGroupFLAG:2024oxs},
\begin{align}
\sigma_{sN}^{\rm exp}=44.9\pm6.4~{\rm MeV}.
\end{align}
Lattice-QCD calculations of the pion--nucleon sigma
term~\cite{Durr:2015dna,Yang:2015uis,Abdel-Rehim:2016won,Bali:2016lvx,Yamanaka:2018uud,Gubler:2018ctz}
typically give values in the range
$\sigma_{\pi N}\simeq30$--$46~{\rm MeV}$, whereas an analysis of $\pi N$
scattering data gives
$\sigma_{\pi N}=59.1\pm3.5~{\rm MeV}$~\cite{Hoferichter:2015dsa}.
To accommodate most of these determinations, we use
\begin{align}
\sigma_{\pi N}^{\rm exp}=45\pm15~{\rm MeV}.
\end{align}
The current quark masses are fixed to
$(m_u,m_s)=(3.5,95)$~MeV.
%
%%%%%%%%%%%%%%%%%%%%%%%%%%%%%%%%%
\begin{table*}
\caption{
Baryon mass inputs used in the fit. Each row corresponds to one SU(3)-flavor 
octet, of which only the $N$, $\Sigma$, and $\Xi$ members are
used here. Masses are given in MeV; entries marked $(?)$ are not included in the
fit.
}
\label{tab-mass-inputs}
\centering
\begin{tabular}{c||c|c|c}
 & \multicolumn{3}{c}{Mass inputs for octet members [MeV]} \\
\hline\hline
$J^P$ & $N$  & $\Sigma$ & $\Xi$  \\
\hline
$m_1:\ 1/2^+$ (G.S.) &
$N(939):\ 939$  &
$\Sigma(1193):\ 1193$ &
$\Xi(1318):\ 1318$  \\
$m_2:\ 1/2^+$ &
$N(1440):\ 1440$  &
$\Sigma(1660):\ 1660$ &
$\Xi(?)$    \\
$m_3:\ 1/2^-$ &
$N(1535):\ 1530$  &
$\Sigma(1750):\ 1750$  &
$\Xi(?)$     \\
$m_4:\ 1/2^-$ &
$N(1650):\ 1650$  &
$\Sigma(?)$  &
$\Xi(?)$     \\
\hline
\end{tabular}
\end{table*}
%%%%%%%%%%%%%%%%%%%%%%%%%%%%%%%%%
%
%\begin{align}
%\chi^2
%=\frac{1}{N_{\rm dof}}\left[
%\sum_i
%\left(\frac{m_i^{\rm fit}-m_i^{\rm exp}}{\delta m_i}\right)^2
%+\left(
%\frac{\sigma_{sN}^{\rm fit}-\sigma_{sN}^{\rm exp}}
%{\delta\sigma_{sN}}
%\right)^2
%+\left(
%\frac{\sigma_{\pi N}^{\rm fit}-\sigma_{\pi N}^{\rm exp}}
%{\delta\sigma_{\pi N}}
%\right)^2
%\right].
%\label{eq:chi2}
%\end{align}
We note that the fit contains eight mass inputs and two sigma-term inputs. With seven
couplings determined at each grid point, the number of degrees of freedom is
therefore
\begin{align}
N_{\rm dof}=10-7=3.
\end{align}

% ----------------------------------------------------------------------------
\subsection{Mass-spectrum and sigma-term results}
\label{sec:fit_results}
% ----------------------------------------------------------------------------

\begin{figure}[htbp]
\centering
\includegraphics[width=0.7\hsize]{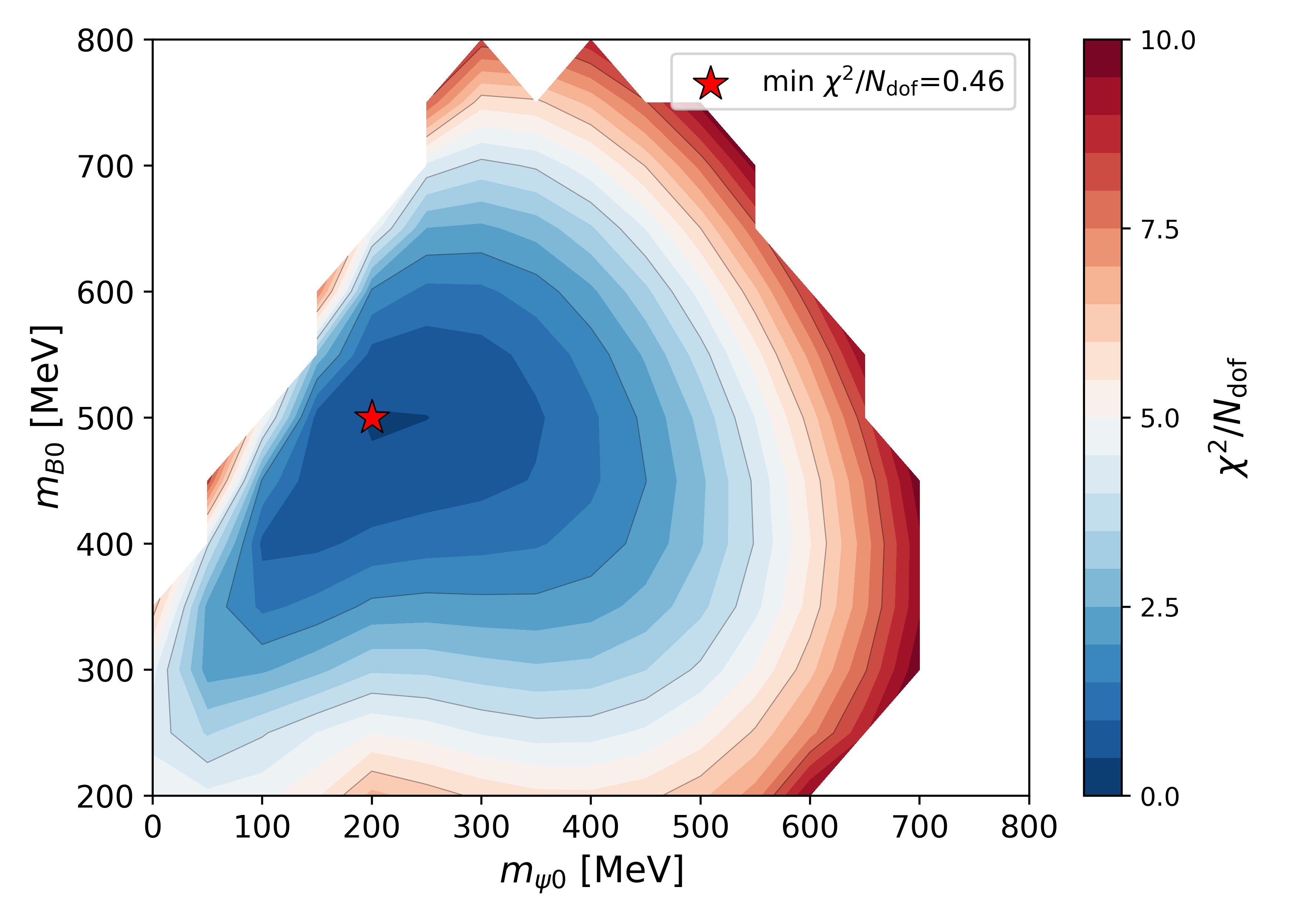}
\caption{
Reduced chi-squared, $\chi^2/N_{\rm dof}$, from the seven-coupling fit in the
$(m_{\psi0},m_{\cB0})$ plane. At each point, $\chi^2$ is minimized over the
seven couplings. The uncolored region corresponds to
$\chi^2/N_{\rm dof}>10$. The star marks the global minimum at
$(m_{\cB0},m_{\psi0})=(500,200)$~MeV.
}
\label{fig_chi2}
\end{figure}

The result of the scan is shown in Fig.~\ref{fig_chi2}. The global minimum,
\begin{align}
\frac{\chi^2}{N_{\rm dof}}=0.46,
\end{align}
is located at $(m_{\cB0},m_{\psi0})=(500,200)$~MeV. In the following, we adopt
this best-fit point as our reference. The corresponding couplings, condensate
parameters, and sigma terms are collected in Table~\ref{tab:parameters}.

\begin{table}[htbp]
\caption{
Fitted couplings, condensate parameters, and predicted sigma terms at the
best-fit point $(m_{\cB0},m_{\psi0})=(500,200)$~MeV.
}
\label{tab:parameters}
\centering
\begin{tabular}{l|c}
\hline\hline
~$(m_{\cB0},m_{\psi0})$ [MeV] ~&~~ (500, 200) ~~~~~\\
\hline
~~~~~~~~$g_1^s$ & $14.70$   \\
~~~~~~~~$g_2^s$ & $-14.30$  \\
~~~~~~~~$g_1^a$ & $-4.84$  \\
~~~~~~~~$g_2^a$ & $-5.53$  \\
~~~~~~~~$y_1$ & $-11.26$   \\
~~~~~~~~$c_1$ & $-0.29$   \\
~~~~~~~~$c_2$ & $-1.51$   \\
\hline
~~~~~~~~$\alpha$ [MeV] & $92.4$  \\
~~~~~~~~$\gamma$ [MeV] & $127.6$  \\
~~~~~~~~$f_0$ [MeV] & $88$  \\
~~~~~~~~$\alpha + c_1 m_u$ [MeV] & $91.38$  \\
~~~~~~~~$\alpha + c_2 m_u$ [MeV] & $87.11$  \\
~~~~~~~~$\gamma + c_1 m_s$  [MeV] & $100.05$ \\
~~~~~~~~$\gamma + c_2 m_s$  [MeV] & $-15.85$  \\
\hline
~~~~~~~~$\sigma_{sN}$ [MeV] & $45$  \\
~~~~~~~~$\sigma_{\pi N}$ [MeV] & $38$  \\
\hline\hline
\end{tabular}
\end{table}

\begin{figure}[htbp]
\centering
\includegraphics[width=1\hsize]{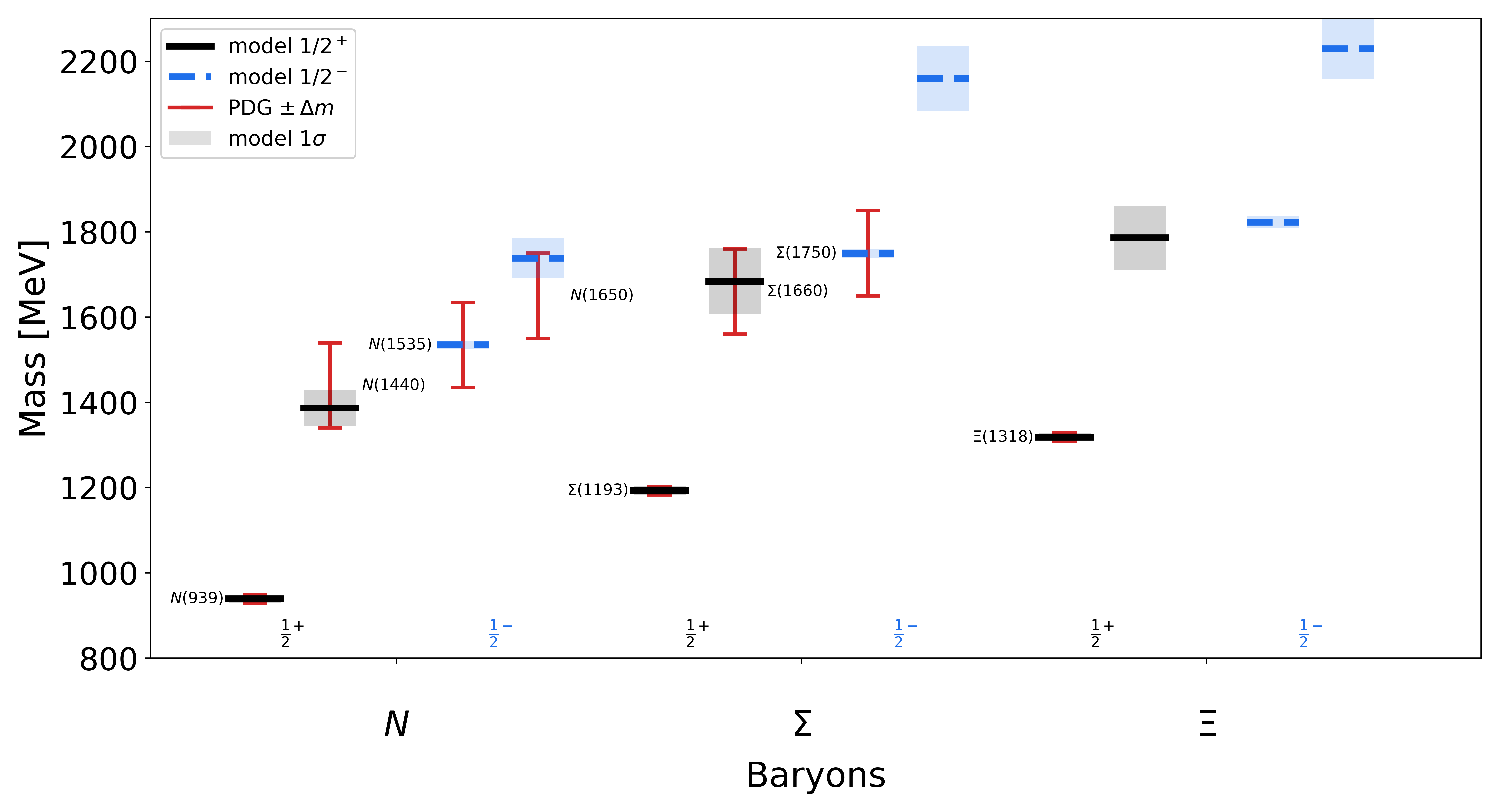}
\caption{
Octet baryon mass spectra at the best-fit point
$(m_{\cB0},m_{\psi0})=(500,200)$~MeV. Solid black and dashed blue bars denote
the model $1/2^+$ and $1/2^-$ states, respectively, while red bars show the
experimental inputs with their assigned uncertainties. The shaded bands are
the $1\sigma$ uncertainties propagated from the coupling fit.
}
\label{fig_baryon_mass}
\end{figure}

The corresponding mass spectra are displayed in
Fig.~\ref{fig_baryon_mass}. With only seven couplings, the model reproduces
all eight mass inputs and correctly yields the hierarchy
$m_N<m_\Sigma<m_\Xi$ for  all 
the ground and excited states. Achieving this with such a minimal parameter set was not possible in the purely hadronic
description of our previous analyses~\cite{Minamikawa:2023ypn,Gao:2024mew,Gao:2025eax}.
We attribute the improvement to the additional dynamical structure supplied
by the quark--diquark construction.
%%%%%%%%%%%%%%%%%%%%%%%%%%%%%%%
\begin{figure}[htbp]
\centering
\includegraphics[width=1\hsize]{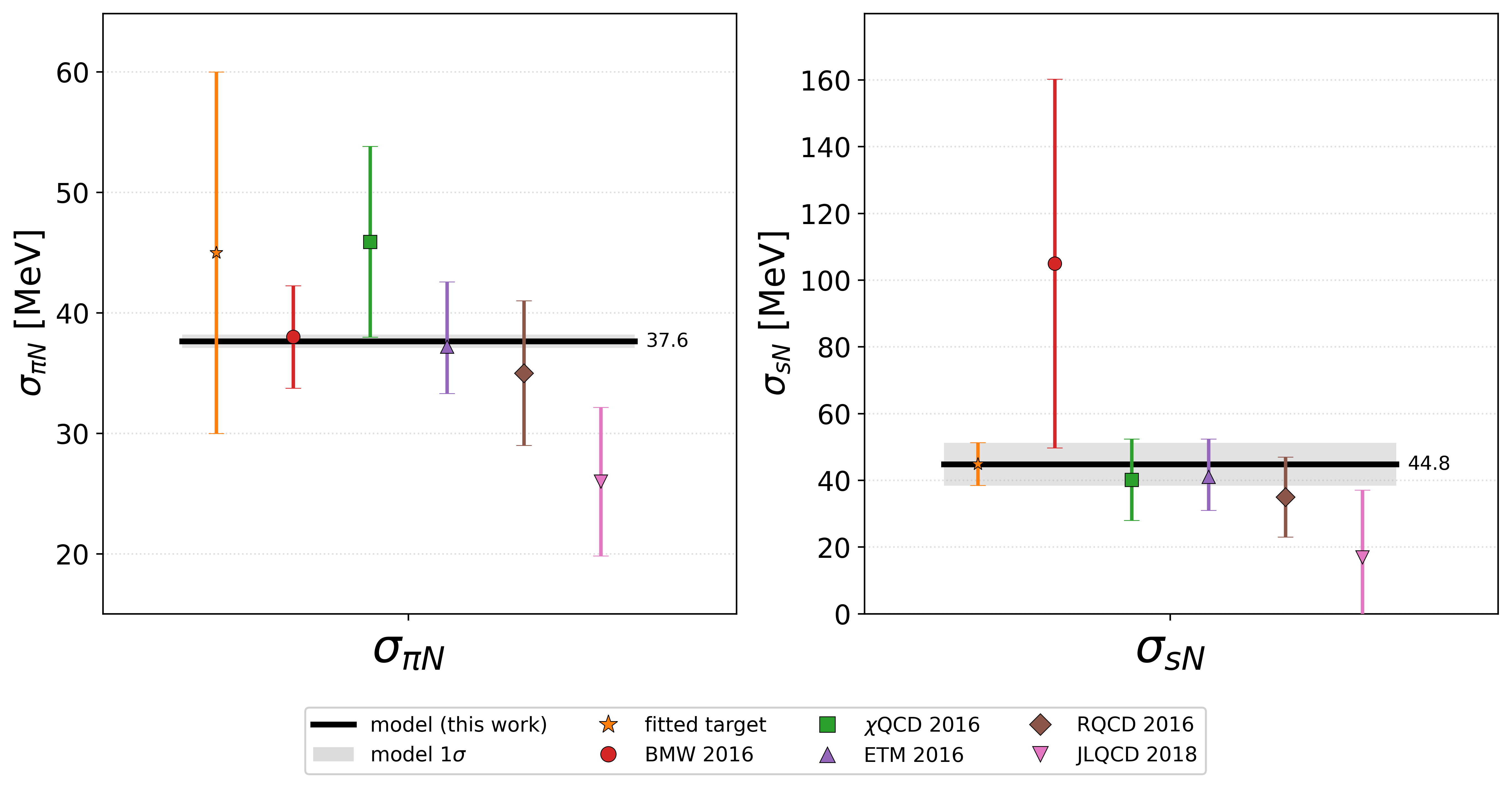}
\caption{
Nucleon sigma terms at the best-fit point
$(m_{\cB0},m_{\psi0})=(500,200)$~MeV. The model predictions, shown by their
central values and $1\sigma$ bands from the coupling fit, are compared with
lattice-QCD calculations represented by the colored error bars:
red~\cite{Durr:2015dna},
green~\cite{Yang:2015uis},
purple~\cite{Abdel-Rehim:2016won},
brown~\cite{Bali:2016lvx}, and
pink~\cite{Yamanaka:2018uud}.
}
\label{fig_sigma_spec}
\end{figure}
%%%%%%%%%%%%%%%%%%%%%%%%%%%%%%
The predicted sigma terms at the best-fit point are
$\sigma_{sN}=45$~MeV and $\sigma_{\pi N}=38$~MeV, and are compared with
lattice-QCD results in Fig.~\ref{fig_sigma_spec}. Both values are consistent
with the corresponding fit inputs and with most lattice determinations. The
condensate contribution identified 
in Eq.~(\ref{eq:cond-cont})
is essential for
the pion--nucleon sigma term: omitting the terms proportional to
$\alpha-f_0$ gives the unphysical value
$\sigma_{\pi N}=-11.4$~MeV, whereas including them yields
$\sigma_{\pi N}\simeq38$~MeV.

% ----------------------------------------------------------------------------
\subsection{Mixing structure and effective chiral invariant mass}
\label{sec:mixing}
% ----------------------------------------------------------------------------

\begin{figure*}[htbp]
\centering
\includegraphics[width=1\hsize]{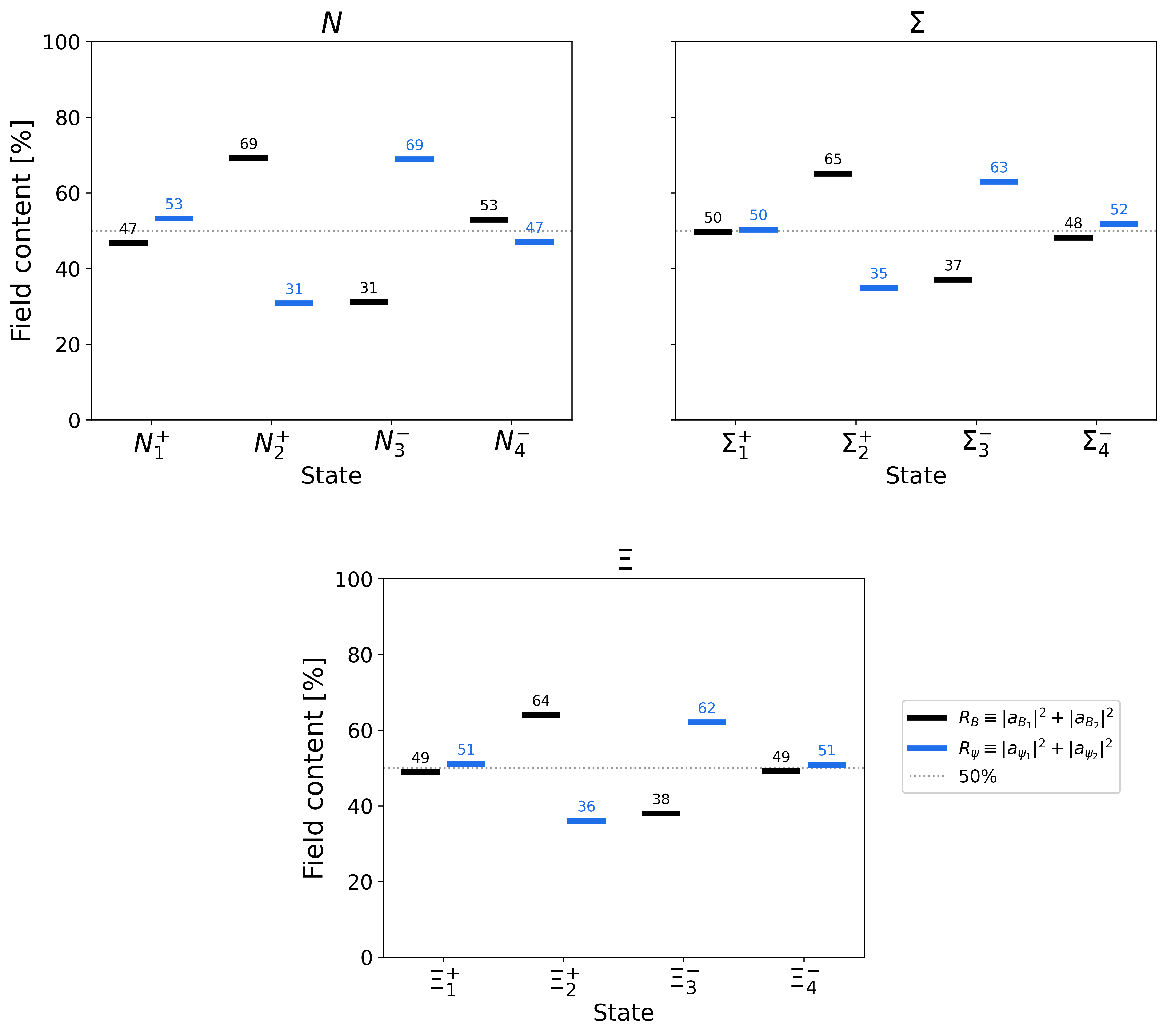}
\caption{
Field content of the octet-baryon eigenstates at the best-fit point
$(m_{\cB0},m_{\psi0})=(500,200)$~MeV. For each state, we plot the weights of
the two interpolating fields, $|\langle\mathcal{B}|\,i\rangle|^2$ and
$|\langle\psi|\,i\rangle|^2$, which sum to unity and give the fractions of the
naive $\mathcal{B}$-type and partner $\psi$-type components in the physical
baryon. States are ordered by parity and mass as
$1/2^+_1,1/2^+_2,1/2^-_1,1/2^-_2$. The shaded bands are the $1\sigma$
uncertainties propagated from the coupling fit.
}
\label{fig_field_ratio}
\end{figure*}

Each physical baryon eigenstate is a linear combination of the four basis
fields. For the nucleon,
\begin{align}
|N\rangle
=a_{\cB_1}|\cB_1\rangle+a_{\cB_2}|\cB_2\rangle
+a_{\psi_1}|\psi_1\rangle+a_{\psi_2}|\psi_2\rangle,
\label{eq:mixing}
\end{align}
with the normalization
$a_{\cB_1}^2+a_{\cB_2}^2+a_{\psi_1}^2+a_{\psi_2}^2=1$, and analogously for
the $\Sigma$ and $\Xi$ channels. Since the $\cB$ fields are assigned to the
$(8,1)\oplus(1,8)$ chiral representation and the $\psi$ fields to
$(3,\bar3)\oplus(\bar3,3)$, it is natural to group the amplitudes into the
total field contents
\begin{align}
R_B&\equiv a_{\cB_1}^2+a_{\cB_2}^2,
\qquad
R_\psi\equiv a_{\psi_1}^2+a_{\psi_2}^2,
\qquad
R_B+R_\psi=1.
\end{align}
These quantities measure the probability with which each chiral
representation is realized in a given state. The mixing at the best-fit point
is collected in Table~\ref{tab:mixing}, and its evolution across the spectrum
is shown in Fig.~\ref{fig_field_ratio}.

\begin{table}[htbp]
\centering
\caption{
Mass spectrum and eigenvector mixing structure of the octet baryons at
$(m_{\cB 0},m_{\psi 0})=(500,200)\,\mathrm{MeV}$.
The last two columns give the total $\cB$- and $\psi$-field contents,
$R_B \equiv a_{\cB_1}^2+a_{\cB_2}^2$ and
$R_\psi \equiv a_{\psi_1}^2+a_{\psi_2}^2$.
}
\label{tab:mixing}
\begin{tabular}{l|cc||rr}
\hline\hline
State & $M$ [MeV] & $P$
      & $R_B$ [\%] & $R_\psi$ [\%] \\
\hline
$N_1$ & 938.88  & $+$ & 46.73 & 53.26 \\
$N_2$ & 1386.42 & $+$ & 69.18 & 30.81 \\
$N_3$ & 1534.83 & $-$ & 31.13 & 68.86 \\
$N_4$ & 1738.18 & $-$ & 52.93 & 47.06 \\
\hline
$\Sigma_1$ & 1192.99 & $+$ & 49.67 & 50.32 \\
$\Sigma_2$ & 1683.94 & $+$ & 65.09 & 34.90 \\
$\Sigma_3$ & 1749.49 & $-$ & 37.03 & 62.96 \\
$\Sigma_4$ & 2160.28 & $-$ & 48.18 & 51.81 \\
\hline
$\Xi_1$ & 1318.17 & $+$ & 48.92 & 51.07 \\
$\Xi_2$ & 1786.12 & $+$ & 63.96 & 36.03 \\
$\Xi_3$ & 1822.94 & $-$ & 37.95 & 62.04 \\
$\Xi_4$ & 2229.05 & $-$ & 49.14 & 50.85 \\
\hline\hline
\end{tabular}
\end{table}

For every octet ground state, we find $R_B\simeq R_\psi$: the $\Rep{8}{1}\oplus\Rep{1}{8}$ and $\Rep{3}{\bar{3}} \oplus \Rep{\bar{3}}{3}$ 
components enter with
almost equal weight. This mixing pattern agrees with the purely hadronic
analysis of Ref.~\cite{Minamikawa:2023ypn} and shows that the ground-state
octet is not dominated by a single chiral representation. A direct
consequence is that the chiral invariant mass of each ground state is shared
between the two assignments. In particular, the effective invariant mass
\begin{align}
m_0^{\rm eff}\simeq R_B\,m_{\cB0}+ R_\psi\,m_{\psi0}
\end{align}
lies close to the mean $(m_{\cB0}+m_{\psi0})/2$, so neither $m_{\cB0}$ nor
$m_{\psi0}$ alone characterizes the origin of the nucleon mass.
%%%%%%%%%%%%%%%%%%%%%%%%%%%%%%%
\begin{figure}[htbp]
\centering
\includegraphics[width=1\hsize]{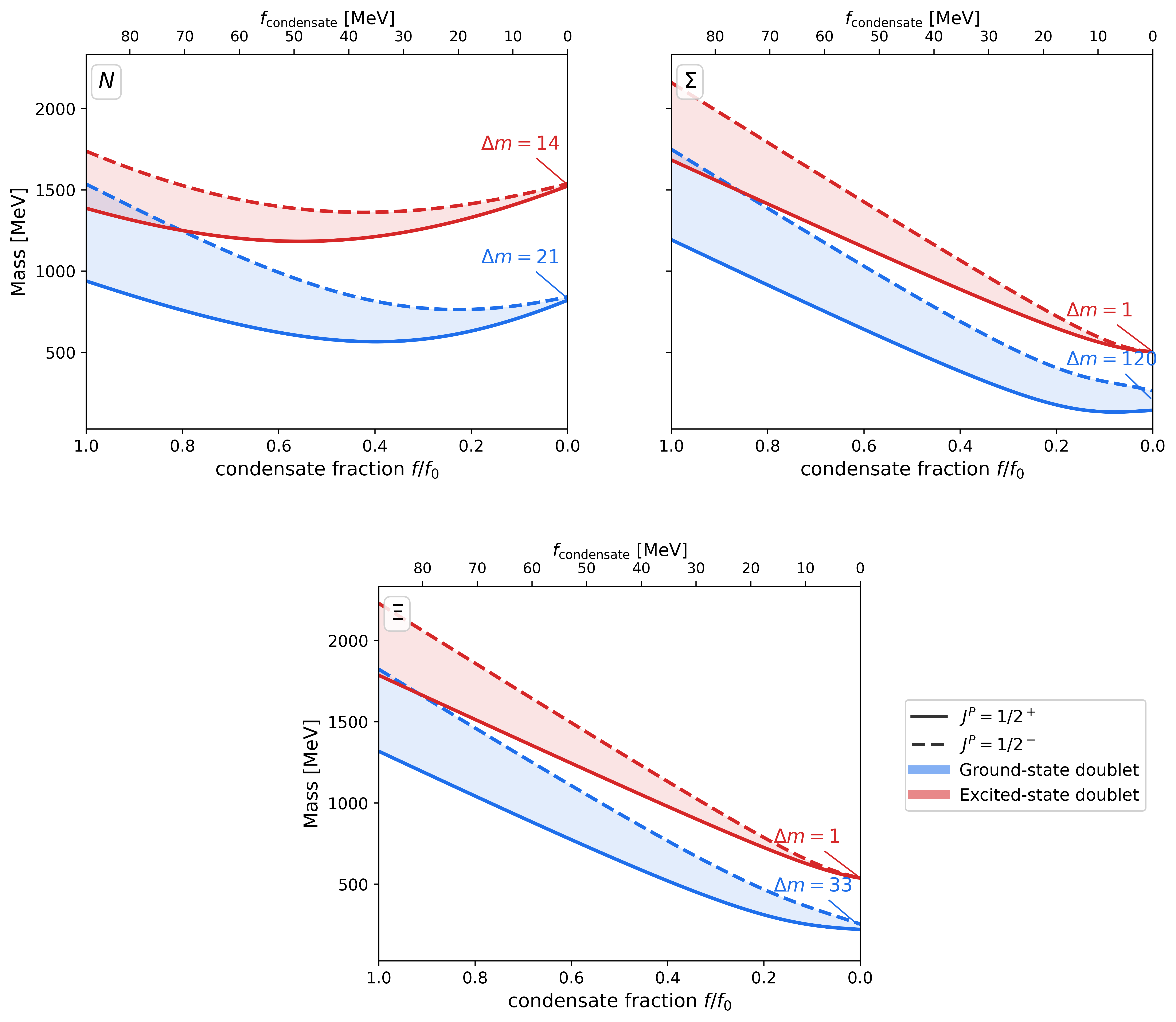}
\caption{
Masses of the four $N$ (top left), $\Sigma$ (top right), and $\Xi$ (bottom)
eigenstates at the best-fit point $(m_{\cB0},m_{\psi0})=(500,200)$~MeV as
functions of the condensate fraction $x=f/f_0$, obtained from the scaling
\eqref{eq:scaling_cond} with the current quark masses held at their physical
values; $x=1$ is the vacuum and $x=0$ the chiral-restored phase, and the
upper axes give $f=x f_0$ in MeV. Solid (dashed) curves denote the positive-
(negative-) parity states, and the blue (red) bands show the parity splitting
$\Delta m=|m_+-m_-|$ of the ground-state (excited-state) doublet; the quoted
numbers give $\Delta m$ at $x=0$ in MeV.
}
\label{fig_mass_behavior}
\end{figure}
%%%%%%%%%%%%%%%%%%%%%%%%%%%%%%%%%
The excited states are more polarized and follow a clear pattern. The first
$1/2^+$ excitation is $\cB$ dominated ($R_B>R_\psi$), the lowest $1/2^-$ state
is $\psi$ dominated ($R_\psi>R_B$), and the second $1/2^-$ state returns to a
nearly equal mixture ($R_B\simeq R_\psi$). The excited spectrum therefore carries
a more definite chiral-representation character than the ground state. This
distinguishes these levels dynamically rather than merely by their masses.
We stress that this field content is a vacuum quantity. Because the mixing is
generated by the same symmetry-breaking condensates that enter
$\mathcal{M}_N$, it depends on the condensate, and the eigenstates reorganize
into the pure $\cB$ and $\psi$ doublets as $f /f_0\to0$.

% ----------------------------------------------------------------------------
\subsection{Chiral restoration and parity doubling}
\label{sec:parity_doubling}
% ----------------------------------------------------------------------------

A distinctive property of the parity-doublet framework is that the
chiral invariant masses $m_{\cB0}$ and $m_{\psi0}$ survive in the
chiral-restored phase, where the positive- and negative-parity baryons become
degenerate. To exhibit this explicitly, we keep the best-fit point
$(m_{\cB0},m_{\psi0})=(500,200)$~MeV and the seven couplings fixed, and follow
the four eigenstates of each of ${\mathcal M}_N$, ${\mathcal M}_\Sigma$, and
${\mathcal M}_\Xi$ in Eqs.~\eqref{eq:massN}--\eqref{eq:massXi} as the chiral
order parameter is reduced from its vacuum value to zero.

Every entry of the mass matrices is proportional to one of the four
symmetry-breaking weights  $\alpha + c_i m_u$, $\gamma + c_i m_s$ ($i=1,2$). 
In the chiral limit ($m_u = m_s = 0$), all decay
constants collapse to the common value $f_0$, so each weight separates into a spontaneous (condensate) piece $f_0$ and an explicit-breaking remainder,
\begin{align}
\alpha+c_i\,m_u &= f_0+\big[(\alpha-f_0)+c_i\,m_u\big]
   \equiv f_0+\bar\alpha_{i}\,,
\label{eq:split_light}\\
\gamma+c_i\,m_s &= f_0+\big[(\gamma-f_0)+c_i\,m_s\big]
   \equiv f_0+\bar\gamma_{i}\,,
\qquad i \in\{1, 2\}\,,
\label{eq:split_strange}
\end{align}
where the strange remainders  $\bar{\gamma}_i$ 
are enhanced by $m_s\gg m_u$
[cf.~Eqs.~\eqref{eq:gamma_alpha} and \eqref{eq:alpha_f0}]. Numerically, from
Table~\ref{tab:parameters},
\begin{align}
\bar\alpha_{1}\simeq3.3 ~{\rm MeV},\qquad
\bar\alpha_{2}\simeq-0.9~{\rm MeV},\qquad
\bar\gamma_{1}\simeq12.1~{\rm MeV},\qquad
\bar\gamma_{2}\simeq-103.9~{\rm MeV}.
\label{eq:remainders}
\end{align}
In the following we model chiral restoration by scaling only the spontaneous part with the
fraction $x\equiv f/f_0\in[0,1]$, keeping the current quark masses---and
hence the explicit remainders---at their physical values,
\begin{equation}
\begin{aligned}
\alpha+c_i\,m_u &\;\longrightarrow\; x\,f_0+\bar\alpha_{i}\,,
\\
\gamma+c_i\,m_s &\;\longrightarrow\; x\,f_0+\bar\gamma_{i}\,,
\label{eq:scaling_cond}
\end{aligned}
\end{equation}
so that $x=1$ corresponds to the vacuum and $x=0$ to the chiral-restored
phase.

The key structural point is that the symmetry breaking effects are differently distributed in the mass matrices for three baryons $N$, $\Sigma$ and $\Xi$, 
as summarized in Table~\ref{tab:restored_weights}: 
the effect from the current strange quark $\bar{\gamma}_i$ enters the
$N$ channel through the $\cB$--$\psi$ mixing coupling $y_1$, the $\Sigma$
channel only through the diagonal $g_{1,2}^a$ entries of the $\psi$ sector,
and the $\Xi$ channel through the $g_{1,2}^s$ mixing entries. 
Since only the
remainders 
$\bar{\alpha}_i$ and $\bar{\gamma}_i$ in Eq.~ 
\eqref{eq:remainders} survive at $x=0$, with
$|\bar\gamma_{2}|\gg|\bar\gamma_{1}|\gg|\bar\alpha_{1}|,|\bar\alpha_{2}|$,
this assignment controls both where the restored doublets sit and how strongly
they remain mixed.

%%%%%%%%%%%%%%%%%%%%%%%%%%%%%%%
\begin{table}[htbp]
\centering
\caption{
Remnants of the explicit symmetry breaking existing 
in the $N$, $\Sigma$, and
$\Xi$ mass matrices, Eqs.~\eqref{eq:massN}--\eqref{eq:massXi}. The numbers in
parentheses give the explicit remainders (in MeV) surviving in the
chiral-restored phase, $x=0$.
}
\label{tab:restored_weights}
\renewcommand{\arraystretch}{1.3}
\begin{tabular}{l|ccc}
\hline\hline
Coupling & $N$ & $\Sigma$ & $\Xi$ \\
\hline
$g_{1,2}^{s}$ ($\cB$--$\psi$ mixing) &
$\bar\alpha_{1}~(3.3)$ & $\bar\alpha_{1}~(3.3)$ & $\bar\gamma_{1}~(12.1)$ \\
$y_1$ ($\cB$--$\psi$ mixing) &
$\bar\gamma_{2}~(-103.9)$ & $\bar\alpha_{2}~(-0.9)$ & $\bar\alpha_{2}~(-0.9)$ \\
$g_{1,2}^{a}$ ($\psi$ diagonal) &
$\bar\alpha_{1}~(3.3)$ & $\bar\gamma_{1}~(12.1)$ & $\bar\alpha_{1}~(3.3)$ \\
\hline\hline
\end{tabular}
\end{table}
%%%%%%%%%%%%%%%%%%%%%%%%%%%%%%%

Figure~\ref{fig_mass_behavior} shows the evolution of the twelve eigenvalues
with $x$. For all baryon species, the parity splittings decrease monotonically, and parity doubling becomes nearly exact at $x=0$ even at physical quark
masses. The restored masses, however, differ markedly among the different baryons, following directly from Table~\ref{tab:restored_weights}.
In the $N$ channel, the restored doublets sit at $\simeq830$ and
$\simeq1530$~MeV, far from the invariant masses $m_{\psi0}$ and $m_{\cB0}$.
The reason is that the $y_1$ entry carries the explicit symmetry-breaking effect from the strange quark, 
so the mixing
element $Y_1\equiv|y_1\,\bar\gamma_{2}|\simeq1.17$~GeV survives in the
restored phase. Keeping only this entry among the symmetry-breaking terms,
the $4 \times 4$ matrix ${\mathcal M}_N$ is partitioned into two $2 \times 2$ blocks based on 
the basis
$\left\{ (\cB_1+\cB_2)/\sqrt2 , (\psi_1+\psi_2)/\sqrt2\right\}$ 
and 
$\left\{ (\cB_1-\cB_2)/\sqrt2 , (\psi_1-\psi_2)/\sqrt2\right\}$.
As a result, there are two exactly degenerate parity doublet with masses given by
\begin{align}
m_{N}^{\rm rest,\mp}
=\sqrt{\,Y_1^{2}+\tfrac14\big(m_{\cB0}-m_{\psi0}\big)^{2}\,}
\;\mp\;\tfrac12\big(m_{\cB0}+m_{\psi0}\big)
\;\simeq\;0.83~\text{and}~1.53~{\rm GeV}\,,
\label{eq:mN_restored}
\end{align}
which are separated by exactly $m_{\cB0}+m_{\psi0}=700$~MeV. Since $Y_1$ connects the
$\cB$ and $\psi$ sectors symmetrically in parity, it shifts the doublets
without splitting them; the small residual splittings, $\Delta m\simeq21$ and
$14$~MeV, are generated by the light entries $\propto\bar\alpha_{1}$. The
restored ground-state doublet, built from $N(939)$ and $N(1535)$, thus remains
at $\simeq830$~MeV, consistent with the chiral invariant masses favored by
neutron-star
observations~\cite{Minamikawa:2020jfj,Gao:2024chh,Kong:2025dwl,Gao:2025nkg,Gao:2025vdc}
and with the gravitational-form-factor perspective of
Ref.~\cite{Kawaguchi:2025cuf}.

In the $\Sigma$ channel, on the other hand, all the entries generating $\cB$--$\psi$ mixing are propotional to $\bar{\alpha}_i$ which  
effectively vanishes at $x=0$, so ${\mathcal M}_\Sigma$ becomes block diagonal. The $\cB$ doublet settles at the invariant mass,
$m_{\cB0}=500$~MeV, with $\Delta m\simeq1$~MeV. The $\psi$ block retains the strange-quark contribution
on its diagonal and yields masses
$\simeq m_{\psi0}\pm\tfrac12(g_1^a+g_2^a)\bar\gamma_{1}\simeq 137$ 
and $262$~MeV: a doublet centered at $m_{\psi0}=200$~MeV with splitting
$\Delta m\simeq(g_1^a+g_2^a)\bar\gamma_{1}\simeq125$~MeV. 
The restored
$\Sigma$ spectrum thus exposes both chiral invariant masses directly.
In $\Xi$ channel, only the mixing entries
$g_{1,2}^s\,\bar\gamma_{1}\simeq\mp0.17$~GeV survive. Their level repulsion
lifts both doublets by
$\simeq(g^s\bar\gamma_{1})^2/(m_{\cB0}+m_{\psi0})\simeq40$~MeV, so the
restored doublets settle slightly above the invariant masses, at $\simeq540$
and $\simeq240$~MeV. The excited doublet is again almost degenerate
($\Delta m\simeq1$~MeV), while the ground-state splitting,
$\Delta m\simeq(g_1^a+g_2^a)\bar\alpha_{1}\simeq33$~MeV, comes from the diagonal entries.

The flavor hierarchy is therefore inverted in the restored phase: the hyperon
doublets drop to the vicinity of the invariant masses, while the nucleon
doublets are held up by the strange explicit breaking. This counterintuitive
pattern---an $m_s$ effect largest in the non-strange channel---reflects the U(1)$_A$ breaking coupling of the diquarks to the condensate as well as the current quark masses, similarly to the inverted hierarchy pointed in Ref.~\cite{Harada:2019udr}:
in the $y_1$ coupling, the
$\epsilon$-tensor contraction ties the light $[ud]$ diquark inside the
nucleon to the strange component of $\langle\Sigma\rangle$, whereas for the
$\Sigma$ and $\Xi$ channels it picks out the light components.

\begin{figure}[htbp]
\centering
\includegraphics[width=1\hsize]
{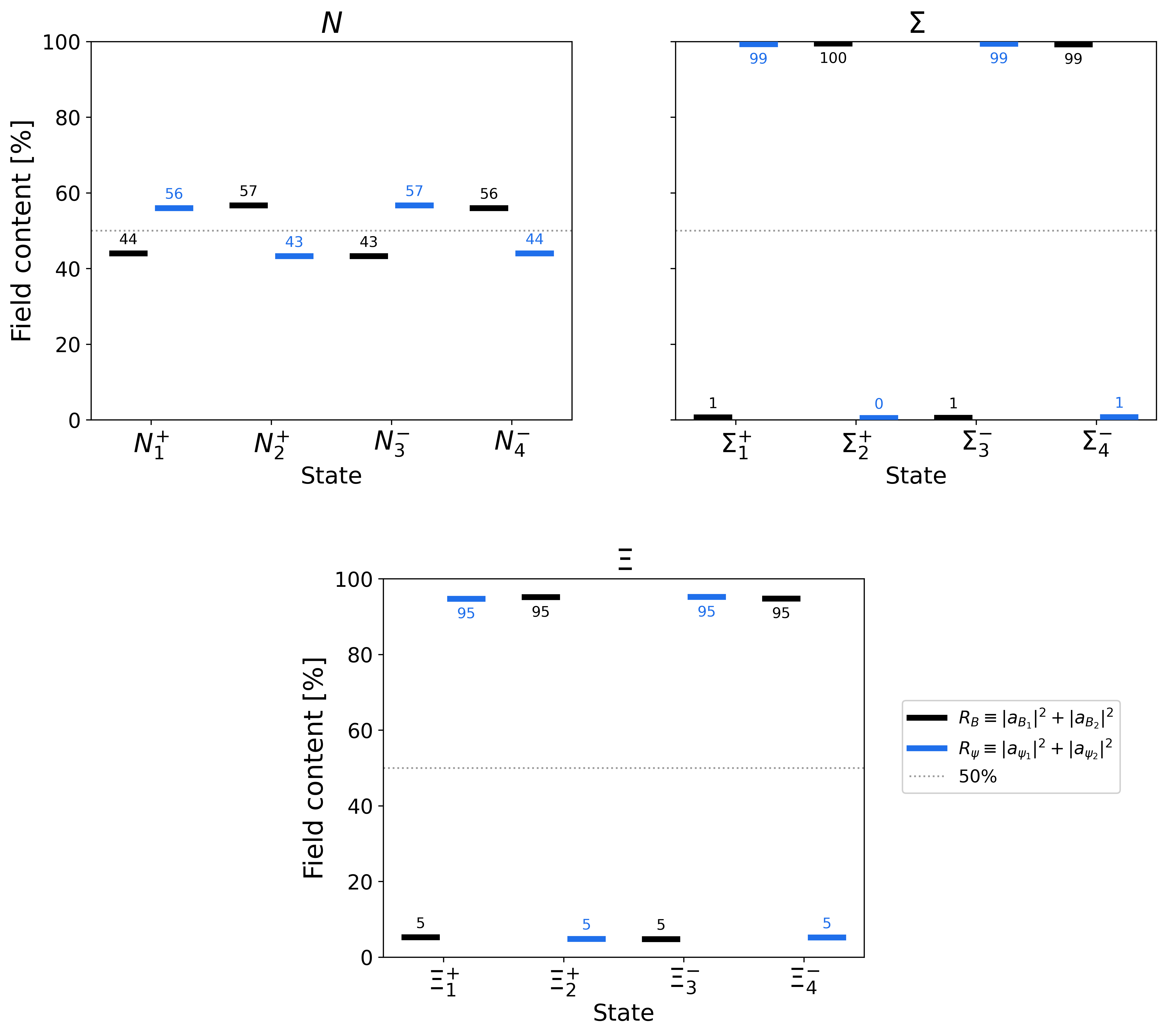}
\caption{
Chiral representation content of the $N$, $\Sigma$, and $\Xi$ eigenstates in
the chiral-restored phase ($x=0$) at the best-fit point
$(m_{\cB0},m_{\psi0})=(500,200)$~MeV. For each state, the fractions
$R_{\cB}$ and $R_\psi=1-R_{\cB}$ give the weights of the
$\Rep{8}{1}\oplus\Rep{1}{8}$ and $\Rep{3}{\bar3}\oplus\Rep{\bar3}{3}$
components.
}
\label{fig_ratio_restored}
\end{figure}

To clarify the internal structure of the restored states, we also examine the chiral representation content of the eigenvectors at $x=0$, shown in
Fig.~\ref{fig_ratio_restored}.
%%%%%%%%%%%%%%%%%%%%%%%%%%%%%%%
%%%%%%%%%%%%%%%%%%%%%%%%%%%%%%%
The approach to parity doubling is accompanied by a reorganization of the
eigenvectors: in every flavor channel, the two members of a doublet acquire
almost identical chiral-representation fractions,
\begin{equation}
 R_{\cB}(F_1^+)\simeq R_{\cB}(F_3^-),
 \qquad
 R_{\psi}(F_2^+)\simeq R_{\psi}(F_4^-),
 \qquad (F=N,\,\Sigma,\,\Xi)\,,
 \label{eq:equal_restored_content}
\end{equation}
so that the restored partners share not only nearly degenerate masses but
also a common internal chiral structure.

The degree of residual mixing again mirrors
Table~\ref{tab:restored_weights}. For the nucleon, the surviving mixing
$Y_1\propto m_s$ is non-negligible, and both doublets remain close to maximal
mixing between $\Rep{8}{1}\oplus\Rep{1}{8}$ and
$\Rep{3}{\bar3}\oplus\Rep{\bar3}{3}$,
\begin{equation}
 R_{\cB}(N_1^+)\simeq 0.44,\quad
 R_{\cB}(N_3^-)\simeq 0.43;
 \qquad
 R_{\cB}(N_2^+)\simeq 0.57,\quad
 R_{\cB}(N_4^-)\simeq 0.56.
\end{equation}
For the $\Sigma$, all mixing entries are proportional to the light remainders
$\propto m_u$, and the restored eigenstates separate cleanly into an almost
pure $\psi$ ground-state doublet ($R_\psi \simeq 0.99$)  
and an almost pure $\cB$ excited doublet ($R_{\cB} \simeq 0.99$). 
The $\Xi$ lies in between:
the surviving mixing $g_{1,2}^s\bar\gamma_{1}\propto c_1\,m_s$ is numerically
modest because the small $c_1$ compensates the large $m_s$, leaving a residual
admixture of only about $5\%$ in each doublet.

Finally, in the exact chiral limit $m_u=m_s=0$, all remainders in
Eq.~\eqref{eq:remainders} vanish, and at $x=0$ the mass matrix of every
flavor channel collapses to the common block-diagonal form
\begin{equation}
{\mathcal M}_F(x\!=\!0)\Big|_{m_u=m_s=0}=
\begin{pmatrix}
0 & m_{\cB0} & 0 & 0\\
m_{\cB0} & 0 & 0 & 0\\
0 & 0 & 0 & m_{\psi0}\\
0 & 0 & m_{\psi0} & 0
\end{pmatrix},
\qquad F=N,\,\Sigma,\,\Xi\,,
\label{eq:Mrestored}
\end{equation}
with eigenvalues $\{\pm m_{\cB0},\pm m_{\psi0}\}$. 
All twelve  baryon 
states
then collapse onto two universal parity doublets located exactly at the
chiral invariant masses, one purely of $\cB$ type and one purely of $\psi$
type. The channel-dependent offsets, splittings, and mixings found above are
therefore entirely a footprint of the explicit chiral symmetry
breaking---dominated by the strange quark mass---superimposed on this
universal parity-doubled structure.

%%%%%%%%%%%%%%%%%%%%%%%%%%%%%%%%%%
\section{Summary and discussion}
\label{sec:summary}
% ============================================================================

In this paper we have proposed a microscopic realization of the parity-doublet
structure of baryons, in which the chiral invariant mass---an input parameter
of the conventional hadronic parity doublet model---is traced to the chiral invariant quark-diquark interaction generated by the gluon dynamics combined with the chiral  
invariant mass carried by the diquark. 
Furthermore, the existence of scalar and pseudoscalar diquarks naturally leads to the emergence of a parity-doublet structure in the resulting baryons.
We would like to stress that, because 
the binding of the colored diquark is controlled by non-Abelian
gluon dynamics rather than by the spontaneous breaking of chiral symmetry,
this invariant mass is naturally insensitive to the quark condensate and
persists in the chiral restored phase.

To turn this picture into a quantitative model, we constructed a three-flavor
chiral quark--diquark model
from the underlying quark and diquark degrees of
freedom in Sec.~\ref{sec:Leff}. 
The model includes two indendent 
chiral symmetric four-body quark--diquark  interactions which are obtainable from one-gluon exchange, 
and  U(1)$_A$  breaking one.
Using the auxiliary-field method we introduced the color-singlet composite baryons: the direct composites $\cB$, belonging to $\Rep{8}{1}\oplus\Rep{1}{8}$, and their crossed partners $\psi$, belonging to
$\Rep{3}{\bar3}\oplus\Rep{\bar3}{3}$. Evaluating the one- and two-loop
quark--diquark self-energies, we derived the baryon mass matrices analytically
and showed in Sec.~\ref{sec:mass}. We showed  that a finite baryon mass, Eq.~\eqref{eq:MB},
is generated even for massless quarks, jointly by the invariant diquark mass
and the four-body quark--diquark interaction. The chiral invariant masses $m_{\cB0}$ and $m_{\psi0}$ of the two composite sectors are therefore no
longer phenomenological inputs of a hadronic Lagrangian but outputs of the
quark--diquark dynamics. Coupling the quarks and diquarks to the chiral order
parameter $\Sigma$ then induces the mixing between $\cB$ and $\psi$ and
assembles the $N$, $\Sigma$, and $\Xi$ mass matrices.

Confronting the model with data, we treated $(m_{\cB0},m_{\psi0})$ as external
parameters and fixed the seven remaining couplings at each grid point by
fitting the  mass spectra of $N$, $\Sigma$ and $\Xi$ baryons 
together with the two nucleon sigma terms
in Sec.~\ref{sec:numerics}.
The best fit, $\chi^2/N_{\rm dof}=0.46$, is reached at $(m_{\cB0},m_{\psi0})=(500,200)~\mathrm{MeV}$, where the model reproduces all
eight mass inputs and correctly orders the ground and excited states as
$m_N<m_\Sigma<m_\Xi$. Achieving this within such a minimal parameter set was
not possible in our earlier purely hadronic
analyses~\cite{Minamikawa:2023ypn,Gao:2024mew,Gao:2025eax}, and we attribute
the improvement to the additional dynamical structure of the quark--diquark
construction. At the same point the predicted sigma terms,
$\sigma_{sN}\simeq45~\mathrm{MeV}$ and $\sigma_{\pi N}\simeq38~\mathrm{MeV}$,
are consistent with recent lattice determinations; we emphasized that a
physical $\sigma_{\pi N}$ requires the response of the chiral condensate
itself (the $\alpha-f_0$ term
in Eq.~(\ref{eq:alpha_f0})
), without which the value is strongly
underestimated.

The eigenvector analysis reveals a clear organizing pattern. Each 
ground state carries almost equal weight in the two chiral representations,
$R_{\cB}\simeq R_\psi\simeq1/2$, so that its chiral invariant mass is shared
between the two assignments: the effective invariant mass sits close to the
mean $(m_{\cB0}+m_{\psi0})/2$, and neither $m_{\cB0}$ nor $m_{\psi0}$ alone
characterizes the origin of the nucleon mass. The excited states are more
polarized in chiral-representation content and thereby acquire a definite
character beyond their masses.

Dialing the chiral order parameter to zero with the current quark masses held
at their physical values, we found in Sec.~\ref{sec:parity_doubling} that
parity doubling becomes nearly exact in all three flavor channels, while the
explicit breaking leaves a characteristic, flavor-resolved imprint on the
restored spectrum. The pattern is fixed by where the current mass of the strange quark enters into the mass matrices (Table~\ref{tab:restored_weights}). In the nucleon channel it
enters the $\cB$--$\psi$ mixing, so the restored doublets are lifted to
$\simeq830$ and $\simeq1530~\mathrm{MeV}$---separated by exactly
$m_{\cB0}+m_{\psi0}$, Eq.~\eqref{eq:mN_restored}---and remain almost maximally
mixed between the two chiral representations. 
In the $\Sigma$ channel, on the other hand, 
all mixing entries do not include the strange-quark mass and thus small,
so the restored spectrum exposes the two
invariant masses directly: an almost pure $\cB$ doublet at
$m_{\cB0}=500~\mathrm{MeV}$ and an almost pure $\psi$ doublet centered at
$m_{\psi0}=200~\mathrm{MeV}$. The $\Xi$ channel is intermediate, with nearly
pure doublets settling slightly above the invariant masses. The flavor
hierarchy is thus inverted in the restored phase---all of $\Sigma$ and $\Xi$ baryons become lighter than four nucleons
---a pattern that traces back to the inverse hierarchy of the diqurak,
which ties the $[ud]$ diquark inside the nucleon to the current strange-quark mass.
Only in the exact chiral
limit do all twelve octet states collapse onto two universal parity doublets
located precisely at $m_{\psi0}=200$ and $m_{\cB0}=500~\mathrm{MeV}$,
isolating the two chiral invariant masses [Eq.~\eqref{eq:Mrestored}]. The
restored nucleon ground-state doublet at $\simeq830~\mathrm{MeV}$ is
compatible with the chiral invariant masses favored by recent neutron-star
observations~\cite{Minamikawa:2020jfj,Gao:2024chh,Kong:2025dwl,Gao:2025nkg,Gao:2025vdc},
while the distinct doubling patterns of the $N$, $\Sigma$, and $\Xi$ channels
constitute a prediction that can in principle be confronted with lattice QCD
studies of nucleon and hyperon parity doubling at finite
temperature~\cite{Aarts:2015mma,Aarts:2017rrl,Aarts:2018glk}.

Before closing, we comment on an approximation made in the two-loop analysis
of Sec.~\ref{sec:sigma}. There, we retained only the diagrams carrying the
same composite vertex type---derivative ($f_1$) or trivial ($f_2$)---at both
ends, which generate the diagonal mass entries $g_1^a$ and $g_2^a$ in the
$\psi$ sector. Mixed diagrams, with one $f_1$ and one $f_2$ composite vertex,
are also allowed and connect $\psi_1$ to $\psi_2$. A counting of Dirac
matrices shows, however, that they cannot generate a mass term: with massless
quarks each internal quark propagator supplies one $\gamma$ matrix and the
charge-conjugation matrices of the $G_d$ vertices drop out of the counting, so
the mixed diagrams carry an odd number of $\gamma$ matrices between the
external Dirac indices. The resulting structure is proportional to
$\slashed P$, vanishes at zero external momentum, and pairs composites of the
same Dirac chirality; its leading term in the derivative expansion is the
chiral invariant kinetic mixing
\begin{align}
\mathcal{L}^{\rm kin}_{\rm mix}
=\varepsilon_{\rm kin}\,
\epsilon_{ijk}\,\epsilon^{\alpha\beta\gamma}\,(\Sigma)^{k}{}_{\gamma}\,
(\bar\psi_{2L})_{\alpha}{}^{j}\,i\slashed\partial\,(\psi_{1L})^{i}{}_{\beta}
+\big(L\leftrightarrow R,\ \Sigma\to\Sigma^\dagger\big)+\mathrm{h.c.},
\label{eq:kinmix}
\end{align}
with $\varepsilon_{\rm kin}$ a two-loop coefficient. Once $\Sigma$ acquires
its expectation value, Eq.~\eqref{eq:kinmix} induces an off-diagonal kinetic
mixing between $\psi_1$ and $\psi_2$ that cannot be absorbed into the
wave-function normalizations of the individual fields: a consistent treatment
would require diagonalizing the kinetic matrix first and only then
rediagonalizing the mass matrix in the resulting basis. Moreover, owing to the
$\epsilon\Sigma\epsilon$ flavor structure, the induced kinetic mixing takes
different values in the $N$, $\Sigma$, and $\Xi$ channels, so that
this procedure would have to be carried out channel by channel, substantially
complicating the analysis. Since Eq.~\eqref{eq:kinmix} arises only at the
two-loop level and constitutes an $\mathcal{O}(\langle\Sigma\rangle)$
correction to the one-loop kinetic terms of Sec.~\ref{sec:mass}, we neglect it
in the present analysis as a higher-order correction to the kinetic sector.

As we stated at the end of Sec.~\ref{sec:composites}, 
we also omit the iso-singlet $\Lambda$ baryons included in the octets in the present analysis since they mix with the flavor singlet baryons.
It is interesting to study the mass spectra of $\Lambda$ baryons by including the effective interaction given in Eq.~(\ref{eq:LA}), which we leave as a future work.

Taken together, these results support a picture in which the part of the baryon mass that is insensitive to the chiral condensate is inherited from the
diquark sector---the invariant diquark mass together with the four-body quark--diquark interaction, both generated by gluon dynamics and chiral invariant---thereby providing a microscopic origin for the chiral invariant mass. A distinctive feature of the quark--diquark construction
is that this invariant mass is not a single number but is resolved into two
contributions, $m_{\cB0}$ and $m_{\psi0}$, tied to the
$\Rep{8}{1}\oplus\Rep{1}{8}$ and $\Rep{3}{\bar3}\oplus\Rep{\bar3}{3}$
representations, whose mixing shapes the physical spectrum in the vacuum and
whose separation is exposed in the restored phase.

Several directions deserve
further study. The chiral invariant mass generated here from the diquark makes
the model a natural starting point for the equation of state of dense matter
and its application to neutron stars, where a relatively large $m_0$ is
favored; the lightness of the restored hyperon doublets found above suggests,
in addition, that hyperonic degrees of freedom may play an enhanced role in
the vicinity of chiral restoration. On the spectroscopic side, the framework
can be extended to the flavor-singlet and decuplet baryons, and to spin-$1$
(``bad'') diquark correlations, which would test the diquark picture more
stringently. A systematic treatment of the $\Sigma$-induced kinetic mixing of
Eq.~\eqref{eq:kinmix}, including the channel-dependent rediagonalization 
that it entails, is another natural refinement. Finally, it would be valuable to
compute the invariant diquark mass directly from the underlying gluon
dynamics, rather than taking it as input, thereby closing the link between the
chiral invariant baryon mass and the color interactions from which we have
argued that  it originates. We leave these developments to future work.

\section*{Acknowledgement}
The work of B.G. is supported in part by JSPS KAKENHI Grant No. 26K17147. The work of M.H. is supported in part by JSPS KAKENHI Grant Nos. 23H05439 and 24K07045.

% ============================================================================
\appendix
% ============================================================================
 
\section{Discrete symmetries of the four-body interaction}
\label{app:PC}
 
In this appendix we show that the four-body interaction \eqref{eq:Lqd} is
invariant under parity ($P$) and charge conjugation ($C$), and that these
symmetries force the coupling constant to be real. We write $G\equiv G_{qd}$
throughout; the same argument applies verbatim to ${\mathcal L}_{qd}^{(2)}$ and
its coupling $G_{qd}^{(2)}$.
 
\subsection{Discrete transformations of the quark and diquark fields}
\label{app:PC1}
 
\paragraph{Parity.}
Under parity the chirality of a quark field is flipped,
\begin{align}
P:\quad q_{L,R}\to\gamma^0\,q_{R,L}, \qquad
\bar q_{L,R}\to\bar q_{R,L}\,\gamma^0.
\end{align}
Applying this to the diquark and using $(\gamma^0)^T=\gamma^0$ together with
$\gamma^0 C\gamma^0=-C$,
\begin{align}
P:\quad (d_R)_{a,i}=\epsilon_{abc}\,\epsilon_{ijk}\,(q_R^{b,j})^T C\,q_R^{c,k}
\;\to\;\epsilon_{abc}\,\epsilon_{ijk}\,(q_L^{b,j})^T\,\gamma^0 C\gamma^0\,q_L^{c,k}
=-(d_L)_{a,i},
\end{align}
and, in exactly the same way, $(d_L)_{a,i}\to-(d_R)_{a,i}$. Parity therefore
interchanges the two chiral diquarks up to a sign,
\begin{align}
P:\quad (d_L)_{a,i}\to-(d_R)_{a,i}, \qquad (d_R)_{a,i}\to-(d_L)_{a,i}.
\end{align}
 
\paragraph{Charge conjugation.}
Under charge conjugation the quark fields transform as
\begin{align}
C:\quad q\to C\bar q^T, \qquad \bar q\to q^T C.
\end{align}
Acting on the left-handed component and using $\gamma_5 C=C\gamma_5^T$ together
with $\bar q^T=\gamma^0 q^*$,
\begin{align}
q_L=\frac{1-\gamma_5}{2}\,q
\;\to\;\frac{1-\gamma_5}{2}\,C\bar q^T
=C\,\frac{1-\gamma_5^T}{2}\,\gamma^0 q^*
=C\left(\bar q\,\frac{1-\gamma_5}{2}\right)^{T}
=C\bar q_R^{\,T},
\end{align}
so that charge conjugation exchanges the chiralities. The full set of rules is
\begin{equation}
\begin{aligned}
C:\quad
q_L^{a,i}\to C(\bar q_R^T)_{a,i}&, \quad
q_R^{a,i}\to C(\bar q_L^T)_{a,i}, \quad \\
(\bar q_L)_{a,i}\to (q_R^{a,i})^T C &, \quad
(\bar q_R)_{a,i}\to (q_L^{a,i})^T C.
\end{aligned}
\end{equation}
Applying these to the diquark,
\begin{equation}
\begin{aligned}
(d_L)_{a,i}=\epsilon_{abc}\,\epsilon_{ijk}\,(q_L^{b,j})^T C\,q_L^{c,k} 
\xrightarrow{C} &
\epsilon_{abc}\,\epsilon_{ijk}\,(C\bar q_R^T)^T_{b,j}\,C\,(C\bar q_R^T)_{c,k}
\\ &=\epsilon^{abc}\epsilon^{ijk}\,(\bar q_R)_{b,j}\,C^T C C\,(\bar q_R^T)_{c,k}
\\ &=\epsilon^{abc}\epsilon^{ijk}\,(\bar q_R)_{b,j}\,C\,(\bar q_R^T)_{c,k},
\label{eq:dlC}
\end{aligned}
\end{equation}
where $C^T C C=C$. On the other hand, the hermitian conjugate of $d_R$ is
\begin{equation}
\begin{aligned}
(d_R^\dagger)^{a,i}
=\big[\epsilon_{abc}\,\epsilon_{ijk}\,(q_R^{b,j})^T C\,q_R^{c,k}\big]^\dagger
&=\epsilon^{abc}\epsilon^{ijk}\,(\bar q_R)_{c,k}\,\gamma^0 C^\dagger\gamma^0\,(\bar q_R^T)_{b,j}
\\&=\epsilon^{abc}\epsilon^{ijk}\,(\bar q_R)_{b,j}\,C\,(\bar q_R^T)_{c,k},
\label{eq:drHer}
\end{aligned}
\end{equation}
where we used $\gamma^0 C^\dagger\gamma^0=C$ and, in the last step, relabeled the
dummy indices $(b,j)\leftrightarrow(c,k)$, under which the product of the two
Levi-Civita symbols is unchanged. Comparing \eqref{eq:dlC} with
\eqref{eq:drHer},
\begin{align}
C:\quad (d_L)_{a,i}\to (d_R^\dagger)^{a,i}, \qquad
(d_R)_{a,i}\to (d_L^\dagger)^{a,i}.
\end{align}

\subsection{Invariance of the four-body interaction and reality of $G$}
\label{app:PC2}
 
It is convenient to define the operators
\begin{align}
\mathcal{O}_{L,R}
=\big\{(\bar q_{L,R})_{c,i}\,(d_{L,R}^\dagger)^{c,j}\big\}
 \big\{(i\partial_\mu d_{L,R})_{a,j}\,\gamma^\mu\,(q_{L,R})^{a,i}\big\},
\label{eq:Odef}
\end{align}
in terms of which the four-body interaction \eqref{eq:Lqd} takes the compact
form
\begin{align}
\mathcal{L}_4=-\sum_{\chi=L,R}\big(G\,\mathcal{O}_\chi+G^*\,\mathcal{O}_\chi^\dagger\big).
\label{eq:L4compact}
\end{align}
We allow $G$ to be complex a priori; hermiticity of the Lagrangian fixes only
the coefficient of $\mathcal{O}_\chi^\dagger$ to be $G^*$. We now show that
parity invariance holds for any $G$, whereas charge-conjugation invariance
forces $G$ to be real.
 
\paragraph{Parity.}
Using the transformations of Appendix~\ref{app:PC1},
\begin{align}
\mathcal{O}_L\xrightarrow{P}\;
&\big\{(\bar q_R)_{c,i}\,\gamma^0\,(-d_R^\dagger)^{c,j}\big\}
 \big\{(-i\partial^\mu d_R)_{a,j}\,\gamma^\mu\gamma^0\,(q_R)^{a,i}\big\}\notag\\
&=\big\{(\bar q_R)_{c,i}\,(d_R^\dagger)^{c,j}\big\}
  \big\{(i\partial^\mu d_R)_{a,j}\,\gamma_\mu\,(q_R)^{a,i}\big\}(\tilde x)
 =\mathcal{O}_R(\tilde x),
\end{align}
where the two minus signs from the diquark transformations cancel, and the sign
flip of the spatial derivatives, $\partial_\mu\to\partial^\mu$, is compensated
by $\gamma^0\gamma^\mu\gamma^0=\gamma_\mu$ ($\tilde x$ denotes the
parity-reflected argument). An identical computation gives
$\mathcal{O}_R\to\mathcal{O}_L(\tilde x)$, so parity simply interchanges the two
operators,
\begin{align}
P:\quad \mathcal{O}_L\to\mathcal{O}_R, \qquad \mathcal{O}_R\to\mathcal{O}_L.
\label{eq:Oparity}
\end{align}
Since $\mathcal{O}_L$ and $\mathcal{O}_R$ enter \eqref{eq:L4compact}
symmetrically and with the same coefficient, $\mathcal{L}_4$ is parity
invariant for any value of $G$.
 
\paragraph{Charge conjugation.}
Similarly,
\begin{align}
\mathcal{O}_L\xrightarrow{C}\;
&\big\{(q_R^{c,i})^T C\,(d_R)_{c,j}\big\}
 \big\{(i\partial_\mu d_R^\dagger)^{a,j}\,\gamma^\mu C\,(\bar q_R^T)_{a,i}\big\}\notag\\
&=\big\{(q_R^{c,i})^T\,(\gamma^\mu)^T\,(\bar q_R^T)_{a,i}\big\}
  \big\{(i\partial_\mu d_R^\dagger)^{a,j}\,(d_R)_{c,j}\big\}\notag\\
&=-\big\{(\bar q_R)_{a,i}\,\gamma^\mu\,(q_R)^{c,i}\big\}
   \big\{(i\partial_\mu d_R^\dagger)^{a,j}\,(d_R)_{c,j}\big\}\notag\\
&=\big\{(\bar q_R)_{a,i}\,\gamma^\mu\,(-i\partial_\mu d_R^\dagger)^{a,j}\big\}
  \big\{(d_R)_{c,j}\,(q_R)^{c,i}\big\}=\mathcal{O}_R^\dagger.
\end{align}
In the second line we used $C\gamma^\mu C=(\gamma^\mu)^T$ and regrouped the
(bosonic) diquark fields; in the third line we transposed the spinor bilinear,
which produces a minus sign because the quark fields anticommute. The final
expression is the hermitian conjugate of $\mathcal{O}_R$, as follows from
\eqref{eq:Odef} using $\gamma^0(\gamma^\mu)^\dagger\gamma^0=\gamma^\mu$. The
analogous computation for $\mathcal{O}_R$ then yields
\begin{align}
C:\quad \mathcal{O}_L\to\mathcal{O}_R^\dagger, \qquad
\mathcal{O}_R\to\mathcal{O}_L^\dagger.
\label{eq:Ocharge}
\end{align}
 
\paragraph{Reality of $G$.}
Applying \eqref{eq:Ocharge} to the Lagrangian \eqref{eq:L4compact},
\begin{equation}
\begin{aligned}
G\,\mathcal{O}_L+G^*\mathcal{O}_L^\dagger+G\,\mathcal{O}_R+G^*\mathcal{O}_R^\dagger
\xrightarrow{C}&
G\,\mathcal{O}_R^\dagger+G^*\mathcal{O}_R+G\,\mathcal{O}_L^\dagger+G^*\mathcal{O}_L
\\&=\sum_{\chi=L,R}\big(G^*\mathcal{O}_\chi+G\,\mathcal{O}_\chi^\dagger\big).
\end{aligned}
\end{equation}
Since $\mathcal{O}_L$, $\mathcal{O}_R$, $\mathcal{O}_L^\dagger$, and
$\mathcal{O}_R^\dagger$ are independent operator structures, invariance of
$\mathcal{L}_4$ under charge conjugation requires
\begin{align}
G=G^*.
\end{align}
The coupling constant $G$ is therefore real, and with real $G$ the four-body
interaction \eqref{eq:Lqd} is invariant under both $P$ and $C$, and hence under
$CP$ as well.
 
 % ============================================================================
\section{Evaluation of the two-point functions}
\label{app:2pt}
% ============================================================================

In this appendix we evaluate the one-loop quark--diquark two-point functions
for a general quark mass $M_q$ and external momentum $P$, and then take the
massless-quark and zero-momentum limits used in the matching of
Sec.~\ref{sec:mass}. The general expressions are also needed once the quarks
acquire the constituent mass $M_q=g_q\langle\sigma\rangle$ through the
coupling to $\Sigma$, and may be of use in future in-medium applications of
the model.

\subsection{Loop functions}
\label{app:2pt:loops}

In the rest frame $P=(W,\vec 0)$ the denominators factorize in $k^0$ as
\begin{align}
k^2-M_q^2 + i \epsilon  &=(k^0-E_q+i\epsilon)(k^0+E_q-i\epsilon),\notag\\
(P-k)^2-m_d^2 + i \epsilon  &=(k^0-W+E_d-i\epsilon)(k^0-W-E_d+i\epsilon),
\label{eq:app_factor}
\end{align}
with $E_q=\sqrt{\bm k^2+M_q^2}$ and $E_d=\sqrt{\bm k^2+m_d^2}$. Closing the
$k^0$ contour in the lower half plane picks up the poles $k^0=E_q-i\epsilon$
and $k^0=W+E_d-i\epsilon$, and
$\int\!\frac{d^3k}{(2\pi)^3}=\frac{1}{2\pi^2}\int_0^\Lambda dk\,k^2$ with the
three-momentum cutoff $\Lambda$. Integrands that fall off only linearly in
$k^0$ are defined by the symmetric (principal-value) $k^0$ integration; with
this prescription odd integrals vanish,
$i\!\int\!\frac{d^4k}{(2\pi)^4}\,\slashed k/[k^2-m^2]=0$, and equivalently
\begin{align}
i\!\int\!\frac{d^4k}{(2\pi)^4}\,
\frac{\slashed k}{(P-k)^2-m_d^2}=\slashed P\,T_d,
\label{eq:app_odd}
\end{align}
as follows from the shift $k\to P-k$, which leaves the three-momentum cutoff
unchanged in the rest frame.

\paragraph{Tadpoles.}
\begin{equation}
\begin{aligned}
T_d &= i\int\frac{d^4k}{(2\pi)^4}\frac{1}{(P-k)^2-m_d^2+i\epsilon}
\\&=i\int\frac{d^3k}{(2\pi)^3}\int\frac{dk^0}{2\pi}
\frac{1}{(W-k^0)^2-E_d^2+i\epsilon}
\\
&=\int\frac{d^3k}{(2\pi)^3}\frac{1}{2E_d}
=\frac{1}{2\pi^2}\int^\Lambda_0 dk\,\frac{k^2}{2E_d},
\end{aligned}
\label{eq:TdApp}
\end{equation}
independently of $W$, and analogously
\begin{align} 
T_q^{M_q} 
= i\int\frac{d^4k}{(2\pi)^4}\frac{1}{k^2-M_q^2+i\epsilon}
=\frac{1}{2\pi^2}\int^\Lambda_0 dk\,\frac{k^2}{2E_q}.
\label{eq:TqApp}
\end{align}

\paragraph{Scalar function $\PiS$.}
The two residues give
\begin{equation}
\begin{aligned}
\PiS &= i\int\frac{d^4k}{(2\pi)^4}\frac{1}{[k^2-M_q^2]\,[(P-k)^2-m_d^2]}
\\
&=\int\frac{d^3k}{(2\pi)^3}\left[
\frac{1}{2E_q\big[(W-E_q)^2-E_d^2\big]}
+\frac{1}{2E_d\big[(W+E_d)^2-E_q^2\big]}\right]\\
&=-\frac{1}{2\pi^2}\int^\Lambda_0 dk\,
\frac{k^2(E_q+E_d)}{2E_qE_d\big[(E_q+E_d)^2-W^2\big]}.
\end{aligned}
\label{eq:PiSApp}
\end{equation}
The last line uses, with $a\equiv E_q$, $b\equiv E_d$, and the common factor
$(W+b-a)$,
\begin{equation}
\begin{aligned}
&\frac{1}{2a\big[(W-a)^2-b^2\big]}+\frac{1}{2b\big[(W+b)^2-a^2\big]}
\\&=\frac{1}{2(W+b-a)}\left[\frac{-1}{a(a+b-W)}+\frac{1}{b(a+b+W)}\right]\\
&=\frac{1}{2(W+b-a)}\,
\frac{-b(a+b+W)+a(a+b-W)}{ab(a+b-W)(a+b+W)}
\\&=\frac{1}{2(W+b-a)}\,\frac{(a+b)(a-b-W)}{ab\big[(a+b)^2-W^2\big]}
=-\frac{a+b}{2ab\big[(a+b)^2-W^2\big]},
\end{aligned}
\label{eq:PFidentity}
\end{equation}
where the last step follows from $(a-b-W)/(W+b-a)=-1$.

\paragraph{Vector function $\PiV$.}
The vector loop function is defined by
\begin{align}
\slashed P\,\PiV=i\int\frac{d^4k}{(2\pi)^4}
\frac{\slashed k}{[k^2-M_q^2]\,[(P-k)^2-m_d^2]}.
\label{eq:PiVdef}
\end{align}
Contracting with $\slashed P$ and taking the Dirac trace,
$\mathrm{tr}[\slashed P\slashed k]=4\,P\!\cdot\!k$, and using
$2P\!\cdot\!k=k^2-(P-k)^2+P^2$,
\begin{equation}
\begin{aligned}
\PiV &=\frac{1}{4P^2}\,i\!\int\!\frac{d^4k}{(2\pi)^4}\,
\frac{\mathrm{tr}\!\left[\slashed P\slashed k\right]}
{[k^2-M_q^2]\,[(P-k)^2-m_d^2]}
\\&=\frac{1}{2P^2}\Big[\,T_d-T_q+\big(P^2+M_q^2-m_d^2\big)\,\PiS\,\Big],
\end{aligned}
\label{eq:PiVrel}
\end{equation}
where we used
\begin{equation}
\begin{aligned}
i\!\int\!\frac{d^4k}{(2\pi)^4}\,
\frac{k^2}{[k^2-M_q^2]\,[(P-k)^2-m_d^2]}&=T_d+M_q^2\,\PiS,\\
i\!\int\!\frac{d^4k}{(2\pi)^4}\,
\frac{(P-k)^2}{[k^2-M_q^2]\,[(P-k)^2-m_d^2]}&=T_q+m_d^2\,\PiS.
\end{aligned}
\label{eq:numerator_ids}
\end{equation}
Equivalently, the direct rest-frame evaluation of the $\gamma^0$ component of
Eq.~\eqref{eq:PiVdef} gives
\begin{align}
\PiV &=\frac{1}{W}\int\frac{d^3k}{(2\pi)^3}\left[
\frac{1}{2\big[(W-E_q)^2-E_d^2\big]}
+\frac{W+E_d}{2E_d\big[(W+E_d)^2-E_q^2\big]}\right]\nonumber\\
&=-\frac{1}{2\pi^2}\int^\Lambda_0 dk\,
\frac{k^2}{2E_d\big[(E_q+E_d)^2-W^2\big]},
\label{eq:PiVApp}
\end{align}
where the bracket is reduced by the same manipulation as in
Eq.~\eqref{eq:PFidentity}, now with numerators $\tfrac12$ and $W+b$, which
produce the numerator $W(a-b-W)$ and hence the factor $W$ that cancels the
$1/W$:
\begin{equation}
\frac{1}{2\big[(W-a)^2-b^2\big]}+\frac{W+b}{2b\big[(W+b)^2-a^2\big]}
=-\frac{W}{2b\big[(a+b)^2-W^2\big]}.
\end{equation}
Note that $\PiV$ is finite and negative at $W^2=0$.

\subsection{General two-point functions}
\label{app:2pt:general}

For a quark line of momentum $k$ and mass $M_q$ and a diquark line of momentum
$q_d=P-k$, the one-loop two-point function in vertex space reads
\begin{align}
\Pi_{ab}(P)=i\sum_{\chi_a,\chi_b=L,R}\int\frac{d^4k}{(2\pi)^4}\,
f_a\,g_a\,\hat P_{\chi_a}\,\frac{\slashed k+M_q}{k^2-M_q^2}\,
\hat P_{\chi_b}\,f_b\,g_b\,\frac{1}{(P-k)^2-m_d^2},
\qquad a,b=1,2,
\label{eq:PiabGen}
\end{align}
with the vertex factors $f_{1,2}$ and couplings of Eq.~\eqref{eq:f12}. In the
chirality-conserving components evaluated below, the two chiral projectors
enforce a chirality flip through the propagator, so that the $M_q$ piece of
the numerator drops out; the chirality-violating components, proportional to
$M_q$, are the origin of the $\Sigma$-induced mixings of
Sec.~\ref{sec:sigma}.

\paragraph{$\Pi_{22}$ component.}
With the trivial vertex $f_2$ on both sides, Eq.~\eqref{eq:PiVdef} directly
gives
\begin{align}
\Pi_{22}=\slashed P\,\PiV.
\label{eq:Pi22Gen}
\end{align}

\paragraph{$\Pi_{12}=\Pi_{21}$ component.}
With the derivative vertex $f_1$ on one side, the numerator reduces via
$\slashed q_d\,\slashed k=(\slashed P-\slashed k)\slashed k
=\slashed P\,\slashed k-k^2$, and Eqs.~\eqref{eq:PiVdef} and
\eqref{eq:numerator_ids} yield
\begin{align}
\Pi_{12}
=C_{qd}G_{qd}\,\big(P^2\,\PiV-T_d-M_q^2\,\PiS\big).
\label{eq:Pi12Gen}
\end{align}

\paragraph{$\Pi_{11}$ component.}
With $f_1$ on both sides, we use
$\slashed a\,\slashed b\,\slashed a=2(a\!\cdot\!b)\,\slashed a-a^2\slashed b$
with $a=q_d$, $b=k$, together with $q_d^2=P^2-2P\!\cdot\!k+k^2$ and
$q_d\!\cdot\!k=P\!\cdot\!k-k^2$, to obtain
\begin{equation}
\slashed q_d\,\slashed k\,\slashed q_d
=2\,(P\!\cdot\!k-k^2)\,\slashed P+(k^2-P^2)\,\slashed k.
\label{eq:numred}
\end{equation}
The first structure integrates, using
$i\!\int\!\tfrac{d^4k}{(2\pi)^4}\,(P\!\cdot\!k)/[\cdots]=P^2\PiV$ [which
follows from Eq.~\eqref{eq:PiVrel}] and Eq.~\eqref{eq:numerator_ids}, to
\begin{equation}
2\slashed P\,i\!\int\!\frac{d^4k}{(2\pi)^4}\,
\frac{P\!\cdot\!k-k^2}{[k^2-M_q^2]\left[(P-k)^2-m_d^2\right]}
=2\slashed P\big[\,P^2\PiV-T_d-M_q^2\PiS\,\big].
\end{equation}
For the second structure, splitting
$k^2-P^2=\left[(P-k)^2-m_d^2\right]+M_q^2-P^2$ \big[valid inside the
$1/(k^2-M_q^2)$ integrand after adding and subtracting $M_q^2$\big] and using
Eq.~\eqref{eq:app_odd},
\begin{equation}
i\!\int\!\frac{d^4k}{(2\pi)^4}\,
\frac{(k^2-P^2)\,\slashed k}{[k^2-M_q^2]\left[(P-k)^2-m_d^2\right]}
=\slashed P\,T_d+(M_q^2-P^2)\,\slashed P\,\PiV.
\end{equation}
Adding the two contributions,
\begin{align}
\Pi_{11}
=(C_{qd}G_{qd})^2\,\slashed P\,
\big[(P^2+M_q^2)\,\PiV-2M_q^2\,\PiS-T_d\big].
\label{eq:Pi11Gen}
\end{align}

\subsection{Massless-quark limit and zero-momentum matching}
\label{app:2pt:limits}

For $M_q=0$ one has $E_q=k$, and the loop functions reduce to
\begin{align} 
T_q = T_q^{M_q}\big|_{M_q=0} 
=\frac{1}{2\pi^2}\int_0^\Lambda dk\,\frac{k}{2}
=\frac{\Lambda^2}{8\pi^2},
\qquad
\PiV\big|_{M_q=0}=-\frac{1}{2\pi^2}\int^\Lambda_0 dk\,
\frac{k^2}{2E_d\big[(k+E_d)^2-W^2\big]},
\end{align}
while $T_d$ is unchanged. Since $\PiS$ remains finite as $M_q\to0$, all terms
proportional to $M_q^2$ in Eqs.~\eqref{eq:Pi12Gen} and \eqref{eq:Pi11Gen}
drop out,
\begin{align}
\Pi_{11}\to(C_{qd}G_{qd})^2\,\slashed P\,\big(P^2\PiV-T_d\big),
\qquad
\Pi_{12}\to C_{qd}G_{qd}\,\big(P^2\PiV-T_d\big),
\qquad
\Pi_{22}=\slashed P\,\PiV,
\end{align}
and the further limit $P^2\to0$ reproduces the values used in the matching of
Sec.~\ref{sec:mass},
\begin{align}
\Pi_{11}\to-(C_{qd}G_{qd})^2\,T_d\,\slashed P,
\qquad
\Pi_{12}\to-C_{qd}G_{qd}\,T_d,
\qquad
\Pi_{22}\to\slashed P\,\PiV\big|_{W^2=0}.
\end{align}

Finally, we collect the auxiliary integrals appearing in the $\Sigma$-induced
mixings of Sec.~\ref{sec:sigma}. The diquark bubble $L_d$ of
Eq.~\eqref{eq:Ld_loop} follows from the tadpole by differentiation,
\begin{align}
L_d=i\int\frac{d^4k}{(2\pi)^4}\left(\frac{1}{k^2-m_d^2}\right)^2
=\frac{\partial T_d}{\partial m_d^2}
=-\frac{1}{2\pi^2}\int_0^\Lambda dk\,\frac{k^2}{4E_d^3},
\end{align}
and the combination entering Eq.~\eqref{eq:effB2Lpsi2R} is
\begin{align}
i\int\frac{d^4k}{(2\pi)^4}\frac{1}{k^2\,[k^2-m_d^2]}
=\frac{T_d-  T_q  }{m_d^2}
=\frac{1}{2\pi^2}\int_0^\Lambda \frac{dk}{m_d^2}
\left(\frac{k^2}{2E_d}-\frac{k}{2}\right),
\end{align}
which is only logarithmically divergent, the quadratic divergences of the two
tadpoles cancelling in the difference.

\bibliographystyle{unsrt}  
\bibliography{ref}

\begin{thebibliography}{10}

\bibitem{Manohar:1983md}
Aneesh Manohar and Howard Georgi.
\newblock {Chiral Quarks and the Nonrelativistic Quark Model}.
\newblock {\em Nucl. Phys. B}, 234:189--212, 1984.

\bibitem{Kugo:1991da}
Taichiro Kugo.
\newblock {Basic concepts in dynamical symmetry breaking and bound state
  problems}.
\newblock In {\em {Nagoya Spring School on Dynamical Symmetry Breaking}}, 7
  1991.

\bibitem{Klevansky:1992qe}
S.~P. Klevansky.
\newblock {The Nambu-Jona-Lasinio model of quantum chromodynamics}.
\newblock {\em Rev. Mod. Phys.}, 64:649--708, 1992.

\bibitem{Roberts:1994dr}
Craig~D. Roberts and Anthony~G. Williams.
\newblock {Dyson-Schwinger equations and their application to hadronic
  physics}.
\newblock {\em Prog. Part. Nucl. Phys.}, 33:477--575, 1994.

\bibitem{Hatsuda:1994pi}
Tetsuo Hatsuda and Teiji Kunihiro.
\newblock {QCD phenomenology based on a chiral effective Lagrangian}.
\newblock {\em Phys. Rept.}, 247:221--367, 1994.

\bibitem{Miransky:1994vk}
V.~A. Miransky.
\newblock {\em {Dynamical symmetry breaking in quantum field theories}}.
\newblock 1994.

\bibitem{DeTar:1988kn}
Carleton~E. Detar and Teiji Kunihiro.
\newblock {Linear $\sigma$ Model With Parity Doubling}.
\newblock {\em Phys. Rev. D}, 39:2805, 1989.

\bibitem{Jido:1999hd}
D.~Jido, T.~Hatsuda, and T.~Kunihiro.
\newblock {Chiral symmetry realization for even parity and odd parity baryon
  resonances}.
\newblock {\em Phys. Rev. Lett.}, 84:3252, 2000.

\bibitem{Jido:2001nt}
Daisuke Jido, Makoto Oka, and Atsushi Hosaka.
\newblock {Chiral symmetry of baryons}.
\newblock {\em Prog. Theor. Phys.}, 106:873--908, 2001.

\bibitem{Aarts:2015mma}
Gert Aarts, Chris Allton, Simon Hands, Benjamin J{\"a}ger, Chrisanthi Praki,
  and Jon-Ivar Skullerud.
\newblock {Nucleons and parity doubling across the deconfinement transition}.
\newblock {\em Phys. Rev. D}, 92(1):014503, 2015.

\bibitem{Aarts:2017rrl}
Gert Aarts, Chris Allton, Davide De~Boni, Simon Hands, Benjamin J{\"a}ger,
  Chrisanthi Praki, and Jon-Ivar Skullerud.
\newblock {Light baryons below and above the deconfinement transition: medium
  effects and parity doubling}.
\newblock {\em JHEP}, 06:034, 2017.

\bibitem{Aarts:2018glk}
Gert Aarts, Chris Allton, Davide De~Boni, and Benjamin J{\"a}ger.
\newblock {Hyperons in thermal QCD: A lattice view}.
\newblock {\em Phys. Rev. D}, 99(7):074503, 2019.

\bibitem{Kim:2020zae}
Jisu Kim and Su~Houng Lee.
\newblock {Vector meson mass in the chiral symmetry restored vacuum}.
\newblock {\em Phys. Rev. D}, 103(5):L051501, 2021.

\bibitem{Kim:2021xyp}
Jisu Kim and Su~Houng Lee.
\newblock {Masses of hadrons in the chiral symmetry restored vacuum}.
\newblock {\em Phys. Rev. D}, 105(1):014014, 2022.

\bibitem{Lee:2023ofg}
Su~Houng Lee.
\newblock {Chiral Symmetry Breaking and the Masses of Hadrons: A Review}.
\newblock {\em Symmetry}, 15(4):799, 2023.

\bibitem{Hatsuda:1988mv}
T.~Hatsuda and M.~Prakash.
\newblock {Parity Doubling of the Nucleon and First Order Chiral Transition in
  Dense Matter}.
\newblock {\em Phys. Lett. B}, 224:11--15, 1989.

\bibitem{Zschiesche:2006zj}
D.~Zschiesche, L.~Tolos, Jurgen Schaffner-Bielich, and Robert~D. Pisarski.
\newblock {Cold, dense nuclear matter in a SU(2) parity doublet model}.
\newblock {\em Phys. Rev. C}, 75:055202, 2007.

\bibitem{Dexheimer:2007tn}
V.~Dexheimer, S.~Schramm, and D.~Zschiesche.
\newblock {Nuclear matter and neutron stars in a parity doublet model}.
\newblock {\em Phys. Rev. C}, 77:025803, 2008.

\bibitem{Sasaki:2010bp}
Chihiro Sasaki and Igor Mishustin.
\newblock {Thermodynamics of dense hadronic matter in a parity doublet model}.
\newblock {\em Phys. Rev. C}, 82:035204, 2010.

\bibitem{Mukherjee:2016nhb}
A.~Mukherjee, J.~Steinheimer, and S.~Schramm.
\newblock {Higher-order baryon number susceptibilities: interplay between the
  chiral and the nuclear liquid-gas transitions}.
\newblock {\em Phys. Rev. C}, 96(2):025205, 2017.

\bibitem{Suenaga:2017wbb}
Daiki Suenaga.
\newblock {Examination of $N^*(1535)$ as a probe to observe the partial
  restoration of chiral symmetry in nuclear matter}.
\newblock {\em Phys. Rev. C}, 97(4):045203, 2018.

\bibitem{Marczenko:2017huu}
Micha{\l} Marczenko and Chihiro Sasaki.
\newblock {Net-baryon number fluctuations in the Hybrid Quark-Meson-Nucleon
  model at finite density}.
\newblock {\em Phys. Rev. D}, 97(3):036011, 2018.

\bibitem{Marczenko:2018jui}
Micha{\l} Marczenko, David Blaschke, Krzysztof Redlich, and Chihiro Sasaki.
\newblock {Chiral symmetry restoration by parity doubling and the structure of
  neutron stars}.
\newblock {\em Phys. Rev. D}, 98(10):103021, 2018.

\bibitem{Eser:2023oii}
J{\"u}rgen Eser and Jean-Paul Blaizot.
\newblock {Thermodynamics of the parity-doublet model: Symmetric nuclear matter
  and the chiral transition}.
\newblock {\em Phys. Rev. C}, 109(4):045201, 2024.

\bibitem{Brodie:2025nww}
Liam Brodie and Robert~D. Pisarski.
\newblock {Parity-Doubled Nucleons Can Rapidly Cool Neutron Stars}.
\newblock {\em Phys. Rev. Lett.}, 135(15):152702, 2025.

\bibitem{Negreiros:2026ode}
Rodrigo Negreiros, Liam Brodie, Jan Steinheimer, Veronica Dexheimer, and
  Robert~D. Pisarski.
\newblock {Enhanced Neutrino Cooling from Parity-Doubled Nucleons in Neutron
  Star Cooling Simulations}.
\newblock 3 2026.

\bibitem{Yamazaki:2019tuo}
Takahiro Yamazaki and Masayasu Harada.
\newblock {Constraint to chiral invariant masses of nucleons from GW170817 in
  an extended parity doublet model}.
\newblock {\em Phys. Rev. C}, 100(2):025205, 2019.

\bibitem{Minamikawa:2020jfj}
Takuya Minamikawa, Toru Kojo, and Masayasu Harada.
\newblock {Quark-hadron crossover equations of state for neutron stars:
  constraining the chiral invariant mass in a parity doublet model}.
\newblock {\em Phys. Rev. C}, 103(4):045205, 2021.

\bibitem{Gao:2024chh}
Bikai Gao, Yan Yan, and Masayasu Harada.
\newblock {Reconciling constraints from the supernova remnant HESS J1731-347
  with the parity doublet model}.
\newblock {\em Phys. Rev. C}, 109(6):065807, 2024.

\bibitem{Kong:2025dwl}
Yuk-Kei Kong, Bikai Gao, and Masayasu Harada.
\newblock {Chiral Invariant Mass Constraints from HESS J1731{\textendash}347 in
  an Extended Parity Doublet Model with Isovector Scalar Meson}.
\newblock {\em Universe}, 11(10):345, 2025.

\bibitem{Gao:2025nkg}
Bikai Gao, Xiang Liu, Masayasu Harada, and Yong-Liang Ma.
\newblock {Implication of neutron star observations to the origin of nucleon
  mass}.
\newblock {\em Sci. China Phys. Mech. Astron.}, 69(3):232011, 2026.

\bibitem{Gao:2025vdc}
Bikai Gao, Yuk-Kei Kong, and Yong-Liang Ma.
\newblock {Origin of nucleon mass in the light of PSR J0614-3329 with
  quark-hadron crossover}.
\newblock {\em Phys. Rev. D}, 112(8):083041, 2025.

\bibitem{Chen:2009sf}
Hua-Xing Chen, V.~Dmitrasinovic, and Atsushi Hosaka.
\newblock {Baryon fields with U(L)(3) X U(R)(3) chiral symmetry II: Axial
  currents of nucleons and hyperons}.
\newblock {\em Phys. Rev. D}, 81:054002, 2010.

\bibitem{Chen:2010ba}
Hua-Xing Chen, V.~Dmitrasinovic, and Atsushi Hosaka.
\newblock {Baryon Fields with $U_L(3) times U_R(3)$ Chiral Symmetry III:
  Interactions with Chiral $(3,\bar{3})+ (\bar{3},3)$ Spinless Mesons}.
\newblock {\em Phys. Rev. D}, 83:014015, 2011.

\bibitem{Nishihara:2015fka}
Hiroki Nishihara and Masayasu Harada.
\newblock {Extended Goldberger-Treiman relation in a three-flavor parity
  doublet model}.
\newblock {\em Phys. Rev. D}, 92(5):054022, 2015.

\bibitem{Sasaki:2017glk}
Chihiro Sasaki.
\newblock {Parity doubling of baryons in a chiral approach with three flavors}.
\newblock {\em Nucl. Phys. A}, 970:388--397, 2018.

\bibitem{Minamikawa:2023ypn}
Takuya Minamikawa, Bikai Gao, Toru Kojo, and Masayasu Harada.
\newblock {Parity doublet model for baryon octets: Diquark classifications and
  mass hierarchy based on the quark-line diagram}.
\newblock {\em Phys. Rev. D}, 108(7):076017, 2023.

\bibitem{Gao:2024mew}
Bikai Gao, Toru Kojo, and Masayasu Harada.
\newblock {Parity doublet model for baryon octets: Ground states saturated by
  good diquarks and the role of bad diquarks for excited states}.
\newblock {\em Phys. Rev. D}, 110(1):016016, 2024.

\bibitem{Gao:2025eax}
Bikai Gao and Atsushi Hosaka.
\newblock {Linear realization of an SU(3) parity doublet model for octet
  baryons with a bad diquark}.
\newblock {\em Phys. Rev. D}, 113(3):036008, 2026.

\bibitem{Gao:2026scv}
Bikai Gao.
\newblock {Chiral symmetry restoration and hyperon suppression in neutron
  stars}.
\newblock {\em Phys. Rev. D}, 113(8):083012, 2026.

\bibitem{Anselmino:1992vg}
Mauro Anselmino, Enrico Predazzi, Svante Ekelin, Sverker Fredriksson, and D.~B.
  Lichtenberg.
\newblock {Diquarks}.
\newblock {\em Rev. Mod. Phys.}, 65:1199--1234, 1993.

\bibitem{Maris:2003vk}
Pieter Maris and Craig~D. Roberts.
\newblock {Dyson-Schwinger equations: A Tool for hadron physics}.
\newblock {\em Int. J. Mod. Phys. E}, 12:297--365, 2003.

\bibitem{Barabanov:2020jvn}
M.~Yu. Barabanov et~al.
\newblock {Diquark correlations in hadron physics: Origin, impact and
  evidence}.
\newblock {\em Prog. Part. Nucl. Phys.}, 116:103835, 2021.

\bibitem{Harada:2019udr}
Masayasu Harada, Yan-Rui Liu, Makoto Oka, and Kei Suzuki.
\newblock {Chiral effective theory of diquarks and the $U_A(1)$ anomaly}.
\newblock {\em Phys. Rev. D}, 101(5):054038, 2020.

\bibitem{Buck:1992wz}
A.~Buck, Reinhard Alkofer, and H.~Reinhardt.
\newblock {Baryons as bound states of diquarks and quarks in the
  Nambu-Jona-Lasinio model}.
\newblock {\em Phys. Lett. B}, 286:29--35, 1992.

\bibitem{Ishii:1995bu}
N.~Ishii, W.~Bentz, and K.~Yazaki.
\newblock {Baryons in the NJL model as solutions of the relativistic Faddeev
  equation}.
\newblock {\em Nucl. Phys. A}, 587:617--656, 1995.

\bibitem{Gasser:1983yg}
J.~Gasser and H.~Leutwyler.
\newblock {Chiral Perturbation Theory to One Loop}.
\newblock {\em Annals Phys.}, 158:142, 1984.

\bibitem{FlavourLatticeAveragingGroupFLAG:2024oxs}
Y.~Aoki et~al.
\newblock {FLAG review 2024}.
\newblock {\em Phys. Rev. D}, 113(1):014508, 2026.

\bibitem{Durr:2015dna}
S.~Durr et~al.
\newblock {Lattice computation of the nucleon scalar quark contents at the
  physical point}.
\newblock {\em Phys. Rev. Lett.}, 116(17):172001, 2016.

\bibitem{Yang:2015uis}
Yi-Bo Yang, Andrei Alexandru, Terrence Draper, Jian Liang, and Keh-Fei Liu.
\newblock {$\pi$N and strangeness sigma terms at the physical point with chiral
  fermions}.
\newblock {\em Phys. Rev. D}, 94(5):054503, 2016.

\bibitem{Abdel-Rehim:2016won}
A.~Abdel-Rehim, C.~Alexandrou, M.~Constantinou, K.~Hadjiyiannakou, K.~Jansen,
  Ch. Kallidonis, G.~Koutsou, and A.~Vaquero Aviles-Casco.
\newblock {Direct Evaluation of the Quark Content of Nucleons from Lattice QCD
  at the Physical Point}.
\newblock {\em Phys. Rev. Lett.}, 116(25):252001, 2016.

\bibitem{Bali:2016lvx}
Gunnar~S. Bali, Sara Collins, Daniel Richtmann, Andreas Sch{\"a}fer, Wolfgang
  S{\"o}ldner, and Andr{\'e} Sternbeck.
\newblock {Direct determinations of the nucleon and pion $\sigma$ terms at
  nearly physical quark masses}.
\newblock {\em Phys. Rev. D}, 93(9):094504, 2016.

\bibitem{Yamanaka:2018uud}
Nodoka Yamanaka, Shoji Hashimoto, Takashi Kaneko, and Hiroshi Ohki.
\newblock {Nucleon charges with dynamical overlap fermions}.
\newblock {\em Phys. Rev. D}, 98(5):054516, 2018.

\bibitem{Gubler:2018ctz}
Philipp Gubler and Daisuke Satow.
\newblock {Recent Progress in QCD Condensate Evaluations and Sum Rules}.
\newblock {\em Prog. Part. Nucl. Phys.}, 106:1--67, 2019.

\bibitem{Hoferichter:2015dsa}
Martin Hoferichter, J.~Ruiz~de Elvira, Bastian Kubis, and Ulf-G. Mei{\ss}ner.
\newblock {High-Precision Determination of the Pion-Nucleon
  {\ensuremath{\sigma}} Term from Roy-Steiner Equations}.
\newblock {\em Phys. Rev. Lett.}, 115:092301, 2015.

\bibitem{Kawaguchi:2025cuf}
Mamiya Kawaguchi, Masayasu Harada, and Yong-Liang Ma.
\newblock {Origin of hadron mass from gravitational D-form factor and neutron
  star measurements}.
\newblock {\em Phys. Lett. B}, 876:140400, 2026.

\end{thebibliography}

\end{document}